\documentclass[a4paper,USenglish,cleveref,autoref,thm-restate]{lipics-v2019-arxiv}

\newif\iflong
\longtrue %

\usepackage{amsmath}
\usepackage{stmaryrd}
\usepackage{xspace}
\usepackage{amssymb}
\usepackage{graphicx}
\usepackage{color}
\usepackage{comment}
\usepackage{wrapfig}
\usepackage[normalem]{ulem}
\usepackage[numbers,sort]{natbib}

\newcommand{\nproofsync}{A}
\newcommand{\nconsensus}{B}
\newcommand{\nlinear}{C}
\newcommand{\nhotstuffsafety}{B.1}
\newcommand{\ntwohs}{B.2}
\newcommand{\noriginaltendermint}{B.5}

\usepackage[vlined,linesnumbered,noresetcount]{algorithm2e}
\SetKwBlock{SubAlgoBlock}{}{end}
\newcommand{\SubAlgo}[2]{#1 \SubAlgoBlock{#2}}
\let\oldnl\nl
\newcommand{\nonl}{\renewcommand{\nl}{\let\nl\oldnl}}

\newcommand{\ms}{\\[2pt]}

\def\be{\begin{equation}}
\def\ee{\end{equation}}

\def\squareforqed{\hbox{\rlap{$\sqcap$}$\sqcup$}}
\def\qed{\ifmmode\squareforqed\else{\unskip\nobreak\hfil
\penalty50\hskip1em\null\nobreak\hfil\squareforqed
\parfillskip=0pt\finalhyphendemerits=0\endgraf}\fi}

\newcommand{\ag}[1]{{{\bf AG:} {\em #1}}}
\newcommand{\mb}[1]{{{\bf MB:} {\em #1}}}
\newcommand{\gc}[1]{{{\bf GC:} {\em #1}}}
\renewcommand{\ag}[1]{}
\renewcommand{\mb}[1]{}
\renewcommand{\gc}[1]{}

\makeatletter
\newcommand{\removelatexerror}{\let\@latex@error\@gobble}
\makeatother

\SetKw{Trigger}{trigger}
\SetKw{WhenReceived}{when received}
\SetKw{WhenAccepted}{when accepted}
\SetKw{KwTo}{to}
\SetKw{Send}{send}
\SetKw{Broadcast}{broadcast}
\SetKw{KwFrom}{from}
\SetKw{Upon}{upon}
\SetKw{Fun}{function}
\SetKw{WhenTimerFires}{when the timer expires}

\newcommand{\SYNC}{\textsc{FastSync}\xspace}

\newcommand{\DECIDE}{{\tt DECIDE}}
\newcommand{\DECIDESLOW}{{\tt DECIDE\_SLOW}}
\newcommand{\DECIDEFAST}{{\tt DECIDE\_FAST}}

\newcommand{\NEWLEADER}{{\tt NEWLEADER}}

\newcommand{\PREPARE}{{\tt PROPOSE}}
\newcommand{\PREPARED}{{\tt PREPARED}}

\newcommand{\COMMITTED}{{\tt COMMITTED}}
\newcommand{\COMMIT}{{\tt COMMIT}}
\newcommand{\WISH}{{\tt WISH}}

\newcommand{\PRECOMMIT}{{\tt PRECOMMIT}}
\newcommand{\PRECOMMITTED}{{\tt PRECOMMITTED}}

\newcommand{\voted}{{\sf voted}}
\newcommand{\cmd}{{\sf prepared\_val}}
\newcommand{\ctxn}{{\sf curr\_val}}

\newcommand{\vote}{{\sf vote}}
\newcommand{\ballot}{{\sf curr\_view}}
\newcommand{\cballot}{{\sf prepared\_view}}
\newcommand{\vcballot}{\mathit{view}}

\newcommand{\vctxn}{\mathit{cur\_val}}
\newcommand{\vcmd}{\mathit{val}}

\newcommand{\leader}{{\sf leader}}

\newcommand{\cert}{{\sf cert}}

\newcommand{\vcert}{\mathit{cert}}
\newcommand{\thcert}{C}
\newcommand{\valid}{{\sf wf}}
\newcommand{\validity}{{\sf valid}}
\newcommand{\Combine}{{\sf combine}}

\newcommand{\ValidNewLeader}{{\sf ValidNewLeader}}
\newcommand{\ValidNewView}{{\sf SafeProposal}}

\newcommand{\quorum}{{\sf quorum}}
\newcommand{\accepted}{{\sf prepared}}

\newcommand{\hash}{{\sf hash}}

\newcommand{\view}{{\sf view}}
\newcommand{\viewp}{{\sf view}^+}
\newcommand{\prevv}{\textit{prev\_v}}
\newcommand{\prevvp}{\textit{prev\_v}^+}

\newcommand{\postGST}{{\sf postGST}}

\newcommand{\te}[2]{E_{#1}(#2)}
\newcommand{\tm}[1]{E_{\rm first}(#1)}
\newcommand{\tl}[1]{E_{\rm last}(#1)}
\newcommand{\tf}{S_{\rm first}}
\newcommand{\ts}{S_{\rm last}}
\newcommand{\tfk}[1]{S_{#1}}

\newcommand{\B}{\mathcal{V}}

\newcommand{\GST}{{\sf GST}}

\newcommand{\all}{\ {\bf all}}
\newcommand{\lballot}{{\sf locked\_view}}
\newcommand{\lcmd}{{\sf locked\_val}}

\newcommand{\preballot}{{\sf pre\_view}}
\newcommand{\vpreballot}{\mathit{pre\_view}}
\newcommand{\timerview}{{\sf timer\_view}}
\newcommand{\timernetwork}{{\sf timer\_newleader}}
\newcommand{\timertendermint}{{\sf timer\_lock}}
\newcommand{\timersbft}{{\sf timer\_fast\_path}}

\newcommand{\newview}{{\tt new\_view}}
\newcommand{\View}{{\sf View}}
\newcommand{\Time}{{\sf Time}}
\newcommand{\lastViews}{{\sf max\_views}}

\newcommand{\vfast}{v_{\rm fast}}
\newcommand{\vslow}{v_{\rm slow}}
\newcommand{\valfast}{\val_{\rm fast}}
\newcommand{\valslow}{\val_{\rm slow}}

\newcommand{\start}{{\tt start}}
\newcommand{\gvsym}{{\sf GV}}
\newcommand{\lvsym}{{\sf LV}}
\newcommand{\GV}[1]{{\sf GV}(#1)}
\newcommand{\LV}[2]{{\sf LV}_{#1}(#2)}

\renewcommand{\_}{\texttt{\textunderscore}}

\newcommand{\TRUE}{\textsc{true}}
\newcommand{\FALSE}{\textsc{false}}

\newcommand{\Proc}{\mathcal{P}}

\newcommand{\committed}{\ensuremath{{\sf committed}}}
\newcommand{\committedslow}{\ensuremath{{\sf committed\_slow}}}
\newcommand{\committedfast}{\ensuremath{{\sf committed\_fast}}}

\newcommand{\val}{x}

\newcommand{\threshold}[4]{\langle #1(#2, #3) \rangle_{#4}}
\newcommand{\partialthreshold}[2]{\langle #1 \rangle_{#2}}

\newcommand{\starttimer}{\ensuremath{{\tt start\_timer}}}
\newcommand{\stoptimer}{\ensuremath{{\tt stop\_timer}}}
\newcommand{\timeout}{\ensuremath{F}}
\newcommand{\timeoutV}{\ensuremath{F}}
\newcommand{\timeoutS}{\ensuremath{F}_{p}}
\newcommand{\timeoutF}{\ensuremath{F}_{f}}
\newcommand{\timeoutL}{\ensuremath{F}_{l}}
\newcommand{\myproposal}{{\tt myval}}

\newcommand{\tr}[2]{\iflong{}\S#1\else{}\cite[\S#2]{ext}\fi}

\title{Making Byzantine Consensus Live\\ (Extended Version) \smallskip \smallskip}

\titlerunning{Making Byzantine Consensus Live}

\author{Manuel Bravo}{IMDEA Software Institute, Spain}{}{}{}
\author{Gregory Chockler}{University of Surrey, UK}{}{}{}
\author{Alexey Gotsman}{IMDEA Software Institute, Spain}{}{}{}

\EventEditors{Hagit Attiya}
\EventNoEds{1}
\EventLongTitle{34rd International Symposium on Distributed Computing (DISC 2020)}
\EventShortTitle{DISC 2020}
\EventAcronym{DISC}
\EventYear{2020}
\EventDate{October 12--18, 2020}
\EventLocation{Virtual Conference}
\EventLogo{}
\SeriesVolume{179}
\ArticleNo{19}

\nolinenumbers

\authorrunning{M. Bravo, G. Chockler, and A. Gotsman}

\Copyright{Manuel Bravo, Gregory Chockler, and Alexey Gotsman}

\ccsdesc[500]{Theory of computation~Distributed computing models}

\keywords{Byzantine consensus, blockchain, partial synchrony, liveness}

\funding{This work was partially supported by an ERC Starting Grant RACCOON.}

\acknowledgements{We want to thank Giuliano Losa, Dahlia Malkhi, Dragos-Adrian
  Seredinschi, Lacramioara Astefanoaei and Eugen Zalinescu for comments that
  helped improve the paper.}

\iflong
\makeatletter
\renewcommand\paragraph{\@startsection{subparagraph}{5}{\z@}%
                                       {2ex \@plus1ex \@minus .2ex}%
                                       {-1em}%
                                      {\sffamily\normalsize\bfseries}}
\makeatother
\fi

\begin{document}

\maketitle

\begin{abstract}
  Partially synchronous Byzantine consensus protocols typically structure their
  execution into a sequence of {\em views}, each with a designated leader
  process. The key to guaranteeing liveness in these protocols is to ensure that
  all correct processes eventually overlap in a view with a correct leader for
  long enough to reach a decision. We propose a simple {\em view synchronizer}
  abstraction that encapsulates the corresponding functionality for Byzantine
  consensus protocols, thus simplifying their design. We present a formal
  specification of a view synchronizer and its implementation under partial
  synchrony, which runs in bounded space despite tolerating message loss during
  asynchronous periods. We show that our synchronizer specification is strong
  enough to guarantee liveness for single-shot versions of several well-known
  Byzantine consensus protocols, including HotStuff, Tendermint, PBFT and
  SBFT. We furthermore give precise latency bounds for these protocols when
  using our synchronizer. By factoring out the functionality of view
  synchronization we are able to specify and analyze the protocols in a uniform
  framework, which allows comparing them and highlights trade-offs.
\end{abstract}

\smallskip
\smallskip
\smallskip
\smallskip

\section{Introduction}

The popularity of blockchains has renewed interest in Byzantine consensus
protocols, which allow a set of processes to reach an agreement on a value
despite a fraction of the processes being malicious. Unlike proof-of-work or
proof-of-stake protocols underlying many blockchains, classic Byzantine
consensus assumes a fixed set of processes, but can in exchange provide hard
guarantees on the finality of decisions. Byzantine consensus protocols are now
used in blockchains with both closed membership~\cite{hyperledger,sbft} and open
one~\cite{casper,tendermint-arxiv,algorand}, in the latter case by running
Byzantine consensus inside a committee elected among blockchain
participants. These use cases have motivated a wave of new
algorithms~\cite{hotstuff,tendermint-arxiv,sbft} that improve on classical
solutions, such as DLS~\cite{dls} and PBFT~\cite{pbft}.

Designing Byzantine consensus protocols is challenging, as witnessed by a number
of bugs found in recent
protocols~\cite{wild,zyzzyva-bug,tendermint-opodis,casper-bug}. Historically,
researchers have paid more attention to safety of these protocols rather than
liveness: e.g., while PBFT came with a safety proof~\cite{castro-thesis}, the
nontrivial mechanism used to guarantee its liveness has never had one. However,
achieving liveness of Byzantine consensus is no less challenging than its
safety. The seminal FLP result shows that guaranteeing both properties is
impossible when the network is asynchronous~\cite{flp}. Hence, consensus
protocols aim to guarantee safety under all circumstances and liveness only when
the network is synchronous. The expected network behavior is formalized by the
{\em partial synchrony} model~\cite{dls}. In one of its more general
formulations~\cite{CT96}, the model guarantees that after some unknown {\em
  Global Stabilization Time (GST)} the system becomes synchronous, with message
delays bounded by an unknown constant $\delta$ and process clocks tracking real
time. Before GST, however, messages can be lost or arbitrarily delayed, and
clocks at different processes can drift apart without bound. This behavior
reflects real-world phenomena: in practice, the space for buffering
unacknowledged messages in the communication layer is bounded, and messages will
be dropped if this space overflows; also, clocks are synchronized by exchanging
messages (e.g., using NTP), so network asynchrony will make clocks diverge.

Byzantine consensus protocols usually achieve liveness under partial synchrony
by dividing execution into {\em views} (aka rounds), each with a designated
leader process responsible for driving the protocol towards a decision. If a
view does not reach a decision (e.g., because its leader is faulty), processes
switch to the next one. To ensure liveness, the protocol needs to guarantee that
all correct processes will eventually enter the same view with a correct leader
and stay there long enough to complete the communication required for a
decision. Achieving such {\em view synchronization} is nontrivial, because
before GST, clocks that could measure the duration of a view can diverge, and
messages that could be used to bring processes into the same view can get lost
or delayed. Thus, by GST processes may end up in wildly different views, and the
protocol has to bring them back together, despite any disruption caused by
Byzantine processes. Some of the Byzantine consensus protocols integrate the
functionality required for view synchronization with the core consensus
protocol, which complicates their design~\cite{pbft,tendermint-arxiv}. 
In contrast, both the seminal DLS work on consensus under partial
synchrony~\cite{dls} and some of the more recent work~\cite{ADDNR19,hotstuff,lumiere}
suggest separating the complex functionality required for view synchronization
into a distinct component~-- {\em view synchronizer}, or simply {\em
synchronizer}. This approach allows designing Byzantine
protocols modularly, with mechanisms for ensuring liveness reused among
different protocols. 

However, to date there has been no rigorous analysis showing which properties of
a synchronizer would be sufficient for modern Byzantine consensus
protocols. Furthermore, the existing implementations of synchronizer-like
abstractions are either expensive or do not handle partial synchrony in its full
generality.
In particular, DLS~\cite{dls} implements view synchronization by constructing
clocks from program counters of processes.
Since these counters drift apart on every step, processes need to frequently
synchronize their local clocks. This results in prohibitive communication
overheads and makes this solution impractical.
Abraham et al.~\cite{ADDNR19} address this inefficiency by assuming hardware
clocks with a bounded drift, but only give a solution for a synchronous
system. Finally, recent synchronizers by Naor et al.~\cite{lumiere} only handle
a simplified variant of partial synchrony which disallows clock drift and
message loss before GST. 

In this paper we make several contributions that address the above limitations:
\begin{itemize}
\item We propose a simple and precise specification of a synchronizer
  abstraction sufficient for single-shot consensus (\S\ref{sec:sync}). The
  specification ensures that from some point on after GST, all correct processes
  go through the same sequence of views, overlapping for some time in each one
  of them. It precisely characterizes the duration of the overlap and gives
  bounds on how quickly correct processes switch between views.
\item We propose a synchronizer implementation, called \SYNC, and rigorously
  prove that it satisfies our specification. \SYNC handles the general version
  of the partial synchrony model~\cite{dls}, allowing for an unknown $\delta$
  and -- before GST -- unbounded clock drift and message loss
  (\S\ref{sec:sync-impl}). Despite the latter, the synchronizer runs in bounded
  space -- a key feature under Byzantine failures, because the absence of a
  bound on the required memory opens the system to denial-of-service
  attacks. Our synchronizer also does not use digital signatures, relying only
  on authenticated point-to-point links.
\item We show that our synchronizer specification is strong enough to guarantee
  liveness under partial synchrony for single-shot versions of a number of
  Byzantine consensus protocols. All of these protocols can thus achieve
  liveness using a single synchronizer -- \SYNC. In the paper we consider in
  detail HotStuff~\cite{hotstuff} (\S\ref{sec:hotstuff}) and its two-phase
  version similar to Tendermint~\cite{tendermint-arxiv}
  (\S\ref{sec:tendermint}); \iflong{}in an appendix
  (\S\ref{app:consensus})\else{}in an extended
  version~\cite[\S\nconsensus]{ext}\fi{} we also analyze PBFT~\cite{pbft},
  SBFT~\cite{sbft} and Tendermint itself. The precise guarantees about the
  timing of view switches provided by our specification are key to handle such a
  wide range of protocols.
\item We provide a precise latency analysis of \SYNC, showing that it quickly
  converges to a synchronized view (\S\ref{sec:sync-bounds}).
  Building on this analysis, we prove worst-case latency bounds for the above
  consensus protocols when using \SYNC. Our bounds consider both favorable and
  unfavorable conditions: if the protocol executes during a synchronous period,
  they determine how quickly all correct processes decide; and if the protocol
  starts during an asynchronous period, how quickly the processes decide after
  GST.
\item Most of the protocols we consider were originally presented in a form
  optimized for solving consensus repeatedly. By specializing them to the
  standard single-shot consensus problem and factoring out the functionality
  required for view synchronization, we are able to succinctly capture their
  core ideas in a uniform
  framework. %
  This allows us to easily compare the protocols and to shed light on trade-offs
  between them.
\end{itemize}

\section{System Model}

We assume a system of $n = 3f+1$ processes, out of which at most $f$ can be
Byzantine, i.e., can behave arbitrarily. In the latter case the process is {\em
  faulty}; otherwise it is {\em correct}. We call a set $Q$ of $2f+1$ processes
a {\em quorum} and write $\quorum(Q)$ in this case. Processes communicate using
authenticated point-to-point links and, when needed, can sign messages using
digital signatures. We denote by $\langle m \rangle_i$ a message $m$ signed by
process $p_i$. We sometimes use a cryptographic hash function $\hash()$, which
must be collision-resistant: the probability of an adversary producing inputs
$m$ and $m'$ such that $\hash(m) = \hash(m')$ is negligible. Processes are
equipped with clocks to measure timeouts. We denote the set of time points by
$\Time$ (ranged over by $t$) and assume that local message processing takes zero
time.

We consider a generalized {\em partial synchrony} model~\cite{dls,CT96}, where
after some time $\GST$ message delays between correct processes are bounded by a
constant $\delta$, and both $\GST$ and $\delta$ are unknown to the
protocol. Before $\GST$ messages can get arbitrarily delayed or lost (although
for simplicity we assume that self-addressed messages are never lost). Assuming
that both $\GST$ and $\delta$ are unknown to the protocol (as in~\cite{CT96})
reflects the requirements of practical systems, whose designers cannot
accurately predict when network problems leading to asynchrony will stop and
what the latency will be during the following synchronous period. We also assume
that the processes are equipped with hardware clocks that can drift unboundedly
from real time before $\GST$, but do not drift thereafter (our results can be
trivially adjusted to handle bounded clock drift after $\GST$, but we omit this
for conciseness).

\section{Synchronizer Specification and Implementation}
\label{sec:sync}

We now define a {\em view synchronizer} interface sufficient for single-shot Byzantine
consensus, and present its specification and implementation.
Let $\View = \{1, 2, \ldots\}$ be the set of {\em views}, ranged over by $v$; we
sometimes use $0$ to denote an invalid view. The job of the synchronizer is to
produce notifications $\newview(v)$ at each correct process, telling it to {\em
  enter} view $v$. A process can ensure that the synchronizer has started
operating by calling a special ${\tt start}()$ function. We assume that each
correct process eventually calls ${\tt start}()$.

For a consensus protocol to terminate, its processes need to stay in the same
view for long enough to complete the message exchange leading to a
decision. Since the message delay $\delta$ after GST is unknown to the protocol,
we need to increase the view duration until it is long enough.
To this end, the synchronizer is parameterized by a function defining this
duration -- $\timeout : \View \cup \{0\} \to \Time$, which is monotone,
satisfies $\timeout(0) = 0$, and increases unboundedly:
\begin{equation}
\label{prop:increasing}
\forall \theta.\,\exists v.\,\forall v'.\, v'\ge v \implies F(v')>\theta.
\end{equation}

\begin{figure}[t]
\center
\includegraphics[width=.95\textwidth]{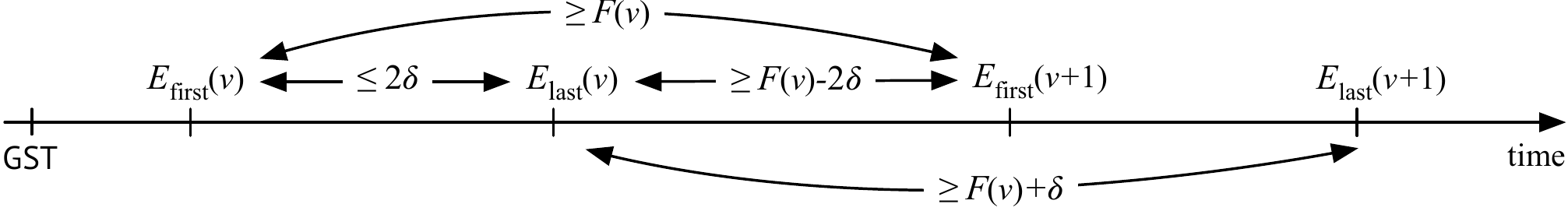}

\vspace{10pt}

\begin{tabular}{@{}l@{\!\!\!\!\!\!}|@{}l@{}}
\begin{minipage}[t]{6.75cm}
\removelatexerror
\vspace*{-8pt}
\setlength{\leftmargini}{14pt}
\renewcommand{\labelenumi}{\theenumi.}
\begin{enumerate}
\item %
  \label{prop:local-order} $\forall i, v, v'.\, (\te{i}{v} \mbox{ and }
  \te{i}{v'} \mbox{ are defined}) \,{\wedge}\, $\\[2pt]
 $ v < v' {\implies} \te{i}{v} < \te{i}{v'}$
\item %
  \label{prop:after-t} $\tm{\B} \ge \GST$
\item %
  \label{prop:no-skip-sync:1}
  $\forall i.\, \forall v\ge \B.\, p_i \text{~is~correct} {\implies}
  p_i \text{~enters~} v$
\item %
  \label{prop:no-skip-sync:2}
  $\forall v \ge \B.\, \tl{v} \le \tm{v}+d$
\item %
\label{prop:no-skip-sync:3}
  $\forall v \ge \B.\, \tm{v+1} \ge \tm{v}+\timeout(v)$
\end{enumerate}
\end{minipage}
&
\begin{minipage}[t]{8cm}
\removelatexerror
\vspace*{-8pt}
\setlength{\leftmargini}{20pt}
\renewcommand{\theenumi}{\Alph{enumi}}
\renewcommand{\labelenumi}{\theenumi.}
\begin{enumerate}
\item %
  \label{prop:last-entry}
  $\forall v \ge \B.\, \tl{v+1} \le \tl{v} + \timeout(v) + \delta$

\item %
  \label{prop:view1} $\tf \geq \GST \wedge
  F(1) > 2\delta {\implies} {}$\\[2pt]
$\B=1 \wedge \tl{1}\leq \ts+\delta$

\item
  \label{prop:convergence}
  $F(\GV{\GST+\rho}+1) > 2\delta \wedge \tfk{f+1} \le \GST + \rho {\implies}$\\[2pt]
  $\B = \GV{\GST+\rho}+1 \wedge {}$\\[2pt]
  $\tl{\B} \le \GST + \rho + F(\B-1) + 3\delta$
\end{enumerate}
\end{minipage}
\end{tabular}
\caption{Synchronizer properties (holding for some $\B \in \View$) and their
  visual illustration. Properties on the left specify the synchronizer
  abstraction, sufficient to ensure consensus liveness. Properties on the right
  give latency bounds specific to our \SYNC synchronizer
  (\S\ref{sec:sync-impl}). The latter satisfies
  Property~\ref{prop:no-skip-sync:2} for $d = 2\delta$. The parameter $\rho$ is
  the retransmission interval used by \SYNC.}
\label{fig:sync-properties}
\end{figure}

The properties on the left of Figure~\ref{fig:sync-properties} define our
synchronizer specification (ignore the properties on the right for the time
being). The specification strikes a balance between usability and
implementability. On one hand, it is sufficient to prove the liveness of a range
of consensus protocols (as we show in \S\ref{sec:consensus}). On the other hand,
it can be efficiently implemented under partial synchrony by our \SYNC
synchronizer (\S\ref{sec:sync-impl}).

Ideally, a synchronizer should ensure that all correct processes overlap in each
view $v$ for a duration determined by $F(v)$. However, achieving this before GST
is impossible due to network and clock asynchrony. Therefore, we require a
synchronizer to provide nontrivial guarantees only after GST and starting from
some view $\B$. To formulate the guarantees we use the following notation. Given
a view $v$ that was entered by a correct process $p_i$, we denote by
$\te{i}{v}$ the time when this happens; we let $\tm{v}$ and $\tl{v}$ denote
respectively the earliest and the latest time when some correct process enters
$v$. We let $\tf$ and $\ts$ be respectively the earliest and the latest time
when some correct process calls $\start()$, and $\tfk{k}$ the earliest
time by which $k$ correct processes do so.
Thus, a synchronizer must guarantee that views may only increase at a given
process (Property~\ref{prop:local-order}), and ensure view synchronization
starting from some view $\B$, entered after GST
(Property~\ref{prop:after-t}). Starting from $\B$, correct processes do not skip
any views (Property~\ref{prop:no-skip-sync:1}), enter each view $v \ge \B$
within at most $d$ of each other (Property~\ref{prop:no-skip-sync:2}) and stay
there for a determined amount of time: until $\timeout(v)$ after the first
process enters $v$ (Property~\ref{prop:no-skip-sync:3}). Our \SYNC
implementation satisfies Property~\ref{prop:no-skip-sync:2} for $d = 2\delta$.
Properties~\ref{prop:no-skip-sync:2} and~\ref{prop:no-skip-sync:3} imply a lower
bound on the overlap between the time intervals during which all correct
processes execute in view $v$:
\begin{equation}
  \forall v \ge \B.\, \tm{v + 1} - \tl{v} \ge (\tm{v} + \timeout(v)) - (\tm{v} +
  d) = \timeout(v) - d.
\label{overlap}
\end{equation}
Due to~(\ref{prop:increasing}), the overlap increases unboundedly as processes
keep switching views. Byzantine consensus protocols are often leader-driven, with
leaders rotating round-robin across views. Hence,~(\ref{overlap}) allows us to
prove their liveness by showing that there will eventually be a view with a
correct leader (due to Property~\ref{prop:no-skip-sync:1}) where all correct
processes will overlap for long enough. %
Having separate Properties~\ref{prop:no-skip-sync:2}
and~\ref{prop:no-skip-sync:3} instead of a single property in~(\ref{overlap}) is
required to prove the liveness of some protocols, e.g., two-phase HotStuff
(\S\ref{sec:tendermint}) and Tendermint (\S\ref{sec:pbft-sbft}).

\subsection{\SYNC: a Bounded-Space Synchronizer for Partial Synchrony}
\label{sec:sync-impl}

In Figure~\ref{alg:sync-lossy:bounded} we present our \SYNC synchronizer, which
satisfies the synchronizer specification on the left of
Figure~\ref{fig:sync-properties} for $d = 2\delta$. Despite tolerating message
loss before GST, \SYNC only requires bounded space; it also does not rely on
digital signatures.

\SYNC measures view duration using a timer $\timerview$: when the synchronizer
tells the process to enter a view $v$, it sets the timer for the duration
$F(v)$. When the timer expires, the synchronizer does not immediately move to
the next view $v'$; instead, it disseminates a special $\WISH(v')$ message,
announcing its intention. Each process maintains an array
$\lastViews : \{1,\ldots, n\} \rightarrow \View \cup \{0\}$, whose $j$-th entry
stores the maximal view received in a $\WISH$ message from process $p_j$
(initially $0$, updated in line~\ref{line:update-maxviews}). Keeping track of
only the maximal views allows the synchronizer to run in bounded space. The
process also maintains two variables, $\view$ and $\viewp$, derived from
$\lastViews$ (initially $0$, updated in lines~\ref{line:update-view}
and~\ref{line:update-viewp}): $\viewp$ (respectively, $\view$) is equal to the
maximal view such that at least $f+1$ processes (respectively, $2f+1$ processes)
wish to switch to a view no lower than this. The two variables monotonically
increase and we always have $\view \le \viewp$.

The process enters the view determined by the $\view$ variable
(line~\ref{line:trigger-newview}) when the latter increases ($\view > \prevv$ in
line~\ref{line:enter-condition}; we explain the extra condition later). At this
point the process also resets its $\timerview$
(line~\ref{line:start-timer}). Thus, a process enters a view only if it receives
a quorum of $\WISH$es for this view or higher, and a process may be forced to switch
views even if its $\timerview$ has not yet expired. The latter helps
lagging processes to catch up, but poses another challenge. Byzantine
processes may equivocate, sending $\WISH$ messages to some processes but not
others. In particular, they may send $\WISH$es for views $\ge v$ to some correct
process, helping it to form a quorum of $\WISH$es sufficient for entering
$v$. But they may withhold the same $\WISH$es from another correct process, so
that it fails to form a quorum for entering $v$, as necessary, e.g., for
Property~\ref{prop:no-skip-sync:2}. To deal with this, when a
process receives a $\WISH$ that makes its $\viewp$ increase, the process sends
$\WISH(\viewp)$ (line~\ref{line:send5}). By the definition of $\viewp$, at least
one correct process has wished to move to a view no lower than $\viewp$. The
$\WISH(\viewp)$ message replaces those that may have been omitted by
Byzantine processes and helps all correct processes to quickly form the
necessary quorums of $\WISH$es.

An additional guard on entering a view is $\viewp = \view$ in
line~\ref{line:enter-condition}, which ensures that a process does not enter a
``stale'' view such that another correct process already wishes to enter a
higher one.
Similarly,
when the timer of the current view expires (line~\ref{line:timer-exp1}), the
process sends a $\WISH$ for the maximum of $\view+1$ and $\viewp$. In other
words, if $\view = \viewp$, so that the values of the two variables have not
changed since the process entered the current view, then the process sends a
$\WISH$ for the the next view ($\view+1$).
Otherwise, $\view < \viewp$, and the process sends a $\WISH$ for the higher view
$\viewp$.

To deal with message loss before GST, a process retransmits the highest $\WISH$
it sent every $\rho$ units of time, according to its local clock
(line~\ref{line:retransmit-start}). Depending on whether $\timerview$ is
enabled, the $\WISH$ is computed as in lines~\ref{line:send5}
or~\ref{line:send2}.
Finally, the ${\tt start}$ function ensures that the synchronizer has started
operating at the process by sending $\WISH(1)$, unless the process has already
done so in line~\ref{line:send5} due to receiving $f+1$ $\WISH$es from other
processes.

\begin{figure}[t]
\begin{tabular}{@{}l@{\!\!\!\!\!}|@{\ \ }l@{}}
\begin{minipage}[t]{6.8cm}
\removelatexerror
\vspace*{-12pt}
\begin{algorithm*}[H]
  \setcounter{AlgoLine}{0}

  \SubAlgo{\Fun $\texttt{start}$()}{\label{line:init1}
    \If{$\viewp = 0$}{\label{line:init-check}
      \Send $\WISH(1)$ \KwTo \all;\label{line:send1}
    }
  }

  \smallskip

  \SubAlgo{\textbf{when \timerview\ expires}}{\label{line:timer-exp1}  
    \Send $\WISH(\max(\view+1, \viewp))$ \label{line:send2}\\
    \nonl \quad \KwTo \all\;
  }

  \smallskip

 \SubAlgo{\textbf{periodically}}{\label{line:retransmit-start}
    \uIf{$\timerview$ \upshape is enabled } {
      \Send $\WISH(\viewp)$ \KwTo \all\;\label{line:send3}
    }
    \ElseIf{$\lastViews[i] > 0$}{\label{line:start-guard}
      \Send $\WISH(\max(\view+1, \viewp))$\label{line:send4}\\
      \nonl \quad \KwTo \all\;
    }
  }

\end{algorithm*}
\vspace*{-5pt}
\end{minipage}
&
\begin{minipage}[t]{8.1cm}
\removelatexerror
\vspace*{-12pt}
\begin{algorithm*}[H]
  \SubAlgo{\WhenReceived $\WISH(v)$ {\bf from $p_j$}}{
    $\prevv, \prevv^+ \leftarrow \view, \viewp$\;
    \lIf{$v > \lastViews[j]$}{$\lastViews[j] \leftarrow v$\label{line:update-maxviews}}
    $\view \ \ \leftarrow \max\{v \mid \exists k.\, \lastViews[k] = v \wedge
    {}$\label{line:update-view}\\
        \nonl\hspace{1.32cm}$|\{j \mid \lastViews[j] \ge v\}| \ge 2f+1\}$\;
    $\viewp  \leftarrow \max\{v \mid \exists k.\, \lastViews[k] = v \wedge{}$\label{line:update-viewp}\\
       \nonl\hspace{1.32cm}$|\{j \mid \lastViews[j] \ge v\}| \ge f+1\}$\; 
    \If{$\viewp = \view
          \wedge \view > \prevv$}{\label{line:enter-condition}
      $\stoptimer(\timerview)$\; \label{line:timer-stop}
      $\starttimer(\timerview, \timeout(\view))$\; \label{line:start-timer}
      \Trigger $\newview(\view)$\; \label{line:trigger-newview}
    }

    \If{$\viewp > \prevv^+$\label{line:cond-send5}}
    {\Send $\WISH(\viewp)$ \KwTo \all; \label{line:send5}}
  } 
\end{algorithm*}
\vspace*{-5pt}
\end{minipage}
\end{tabular}
\caption{The \SYNC synchronizer. The periodic handler is invoked every $\rho$ units of time.
}
\label{alg:sync-lossy:bounded}
\end{figure}

\subparagraph*{Discussion.} \SYNC requires only $O(n)$ variables for storing
views. When proving its correctness, we establish that every view is entered by
some correct process \iflong(Lemma~\ref{lem:gv-noskip} in
\S\ref{sec:proof-sync})\else{}\cite[\S\nproofsync, Lemma 18]{ext}\fi, and
eventually, correct processes do not skip views
(Property~\ref{prop:no-skip-sync:1}). %
These two properties limit the power of the adversary to exhaust the value space
for views, similarly to~\cite{BazziD04}.

The basic mechanisms we use in our synchronizer --
entering views supported by $2f+1$ $\WISH$es and relaying views supported by
$f+1$ $\WISH$es -- are similar to the ones used in Bracha's algorithm for
reliable Byzantine broadcast~\cite{Bra87}. However, Bracha's algorithm only
makes a step upon receiving a set of {\em identical} messages. Thus, its naive
application to view synchronization~\cite[\S{}A.2]{lumiere} requires unbounded
space to store the views $v$ for which the number of received copies of
$\WISH(v)$ still falls below the threshold required for delivery or relay.
Moreover, tolerating message loss would require a process to retain a copy of
every message it has broadcast, to enable retransmissions. \SYNC can be viewed
as specializing the mechanisms of Bracha broadcast to take advantage of the
particular semantics of $\WISH$ messages, by keeping track of only the highest
$\WISH$ received from each process and by acting on sets of $\WISH$es for
non-identical views. This allows tolerating message loss before GST in bounded
space and without compromising liveness, as illustrated by the following
example.

We first show that, before $\GST$, we may end up in the situation where
processes are split as follows: a set $P_1$ of $f$ correct processes entered
$v_1$, a set $P_2$ of $f$ correct processes entered $v_2>v_1$, a correct process
$p_i$ entered $v_2+1$, and $f$ processes are faulty. To reach this state, assume
that all correct processes manage to enter view $v_1$ and then all messages
between $P_1$ and $P_2 \cup \{p_i\}$ start getting lost. The $f$ faulty
processes help the processes in $P_2 \cup \{p_i\}$ to enter all views between
$v_1$ and $v_2$, by providing the required $\WISH$es
(line~\ref{line:enter-condition}), while the processes in $P_1$ get stuck in
$v_1$. After the processes in $P_2 \cup \{p_i\}$ time out on $v_2$, they start
sending $\WISH(v_2+1)$ (line~\ref{line:send2}), but all messages directed to
processes other than $p_i$ get lost, so that the processes in $P_2$ get stuck in
$v_2$.  The faulty processes then help $p_i$ gather $2f+1$ messages
$\WISH(v_2+1)$ and enter $v_2+1$ (line~\ref{line:enter-condition}).

Assume now that $\GST$ occurs, the faulty processes go silent and the correct
processes time out on the views they are in. Thus, the $f$ processes in $P_1$
send $\WISH(v_1+1)$, the $f$ processes in $P_2$ send $\WISH(v_2+1)$, and $p_i$
sends $\WISH(v_2+2)$ (line~\ref{line:send4}). The processes in $P_1$ eventually
receive the $\WISH$es from $P_2 \cup \{p_i\}$, so that they set $\viewp = v_2+1$
and send $\WISH(v_2+1)$ (line~\ref{line:send5}). Note that here processes act on
$f+1$ mismatching $\WISH$es, unlike in Bracha broadcast. Eventually, the
processes in $P_1 \cup P_2$ receive $2f$ copies of $\WISH(v_2+1)$ and one
$\WISH(v_2+2)$, which causes them to set $\view =v_2+1$ and enter $v_2+1$
(line~\ref{line:enter-condition}). Note that here processes act on $2f+1$
mismatching $\WISH$es, again unlike in Bracha broadcast. Finally, the processes
$P_1 \cup P_2$ time out and send $\WISH(v_2+2)$ (line~\ref{line:send2}), which
allows all correct processes to enter $v_2+2$. Acting on sets of mismatching
$\WISH$es is crucial for liveness in this example: if processes only accepted
matching sets, like in Bracha broadcast, message loss before GST would cause
them to get stuck, and they would never converge to the same view.

\subsection{Correctness and Latency Bounds of \SYNC}
\label{sec:sync-bounds}

As we demonstrate shortly, the synchronizer specification given by
Properties~\ref{prop:local-order}-\ref{prop:no-skip-sync:3} in
Figure~\ref{th:main} serves to prove that consensus {\em eventually} reaches a
decision. However, \SYNC also satisfies some additional properties that allow us
to quantify {\em how quickly} this happens under both favorable and unfavorable
conditions. We list these properties on the right of
Figure~\ref{fig:sync-properties}.

\begin{theorem}
  {\em \SYNC} satisfies all properties in Figure~\ref{fig:sync-properties} for
  $d = 2\delta$.
\label{th:main}
\end{theorem}

Due to space constraints, we defer the proof to
\tr{\ref{sec:proof-sync}}{\nproofsync}. Property~\ref{prop:last-entry} allows us to quantify the
cost of switching between several views (e.g., due to faulty leaders). This is
formalized by the following proposition, easily proved using
Property~\ref{prop:last-entry} by induction on $v'$.
\begin{proposition}
$\forall v, v'.\, \B \le v \le v' {\implies}
\tl{v'} \le \tl{v} + \sum_{k=v}^{v'-1}(\timeoutV(k)+\delta)$.
\label{lem:last-entry-sum}
\end{proposition}

Property~\ref{prop:view1} guarantees that, when the synchronizer starts after
$\GST$ ($\tf \geq \GST$) and the initial timeout is long enough
($F(1) > 2\delta$), processes synchronize in the very first view ($\B = 1$) and
enter it within $\delta$ of the last correct process calling $\start()$.

Let the {\em global view}\/ at time $t$, denoted $\GV{t}$, be the maximum view
entered by a correct process at or before $t$, or $0$ if no view was entered by
a correct process. Property~\ref{prop:convergence} quantifies the latency of
view synchronization in a more general case when the synchronizer may be started
before $\GST$. The property depends on the interval $\rho$ at which the
synchronizer periodically retransmits its internal messages to deal with
possible message loss. The property considers the highest view $\GV{\GST+\rho}$
a correct process has at time $\GST+\rho$ and ensures that all correct processes
synchronize in the immediately following view within at most
$\rho + F(\B-1) + 3\delta$ after $\GST$. This is guaranteed under an assumption
that the timeout of this view exceeds $2\delta$
and $f+1$ correct processes call $\start()$ early enough. Since $\GST$ can be
arbitrary, in principle, so can be the view $\B$ and, thus due
to~(\ref{prop:increasing}), the timeout $F(\B-1)$. However, practical
implementations usually stop increasing timeouts when they exceed a reasonable
value. Hence, Property~\ref{prop:convergence} guarantees that to reach $\B$,
processes need to wait for at most a single maximal timeout.

\section{Liveness and Latency of Byzantine Consensus Protocols}
\label{sec:consensus}

We show that our synchronizer abstraction allows ensuring liveness and
establishing latency bounds for several consensus protocols. The protocols solve
a variant of Byzantine consensus problem that relies on an application-specific
$\validity()$ predicate to indicate whether a value is
valid~\cite{cachin-crypto01,dbft}. In the context of blockchain systems a value
represents a block, which may be invalid if it does not include correct
signatures authorizing its transactions. Assuming that each correct process
proposes a valid value, each of them has to decide on a value so that:
\vspace{2pt}
\begin{itemize}
\item
{\bf Agreement.} No two correct processes decide on different values.

\item
{\bf Validity.} A correct process decides on a valid value, i.e., satisfying
$\validity()$.

\item {\bf Termination.} Every correct process eventually decides on a value.
\end{itemize}

\subsection{Single-Shot HotStuff}
\label{sec:hotstuff}

We first consider the HotStuff protocol~\cite{hotstuff}, 
underlying the upcoming Libra cryptocurrency~\cite{libra}. The protocol
was originally presented as solving an inherently multi-shot problem, agreeing
on a hash-chain of blocks. In Figure~\ref{fig:hotstuff} we present its
single-shot version
that concisely expresses the key idea and allows comparing the protocol with
others. For brevity, we %
eschew the use of threshold signatures, which makes the communication complexity
of a leader change $O(n^2)$ rather than $O(n)$, like in the original
HotStuff. This complexity is still better than that of PBFT, which is
$O(n^3)$. We handle 
linear versions of the protocols we consider %
in \tr{\ref{sec:linear}}{\nlinear}.  HotStuff delegated view synchronization to a separate
component~\cite{hotstuff}, but did not provide its practical implementation or
analyze how view synchronization affects the protocol latency. We show that our
single-shot version of HotStuff is live when used with a synchronizer satisfying
the specification in \S\ref{sec:sync} and give precise bounds on its latency. We
also show that the protocol requires only bounded space when using our
synchronizer \SYNC.

The protocol in Figure~\ref{fig:hotstuff} works in a succession of views
produced by the synchronizer. Each view $v$ has a fixed leader
$\leader(v) = p_{((v-1)\ \mathrel{\rm mod}\ n) + 1}$ that is responsible for
proposing a value to the other processes, which vote on the proposal. A correct
leader needs to choose its proposal carefully so that, if %
a value was decided in a previous view, the leader will propose the same value. To enable
the leader to do this, when a process receives a notification 
to move to a view $v$ (line~\ref{hotstuff:newview}), it sends a
$\NEWLEADER$ message to the leader of $v$ with information about the latest
value it %
accepted in a previous view (as described in the following). The process also
stores the view $v$ in a variable $\ballot$, and sets a flag $\voted$ to
$\FALSE$, to record that it has not yet received any proposal from the leader in
the current view. The leader computes its proposal (as described in the
following) based on a quorum of $\NEWLEADER$ messages
(line~\ref{hotstuff:receive-newleader}) and sends the proposal, 
along with some supporting information, in a $\PREPARE$ message to all processes
(for uniformity, including itself).

The leader's proposal is processed in three phases. 
A process receiving a proposal $\val$ from the leader of its view $v$
(line~\ref{hotstuff:receive-prepare}) first checks that $\voted$ is $\FALSE$, so
that it has not yet accepted a proposal in $v$. It also checks that $x$
satisfies a $\ValidNewView$ predicate (explained later), which ensures that a
faulty leader cannot reverse decisions reached in previous views. The process
then sets $\voted$ to $\TRUE$ and stores $\val$ in $\ctxn$.

Since a faulty leader may send different proposals to different processes, the
process next communicates with others to check that they received the same
proposal. To this end, the process disseminates a $\PREPARED$ message with the
hash of the proposal it received. The process then waits until it gathers a set
$C$ of $\PREPARED$ messages from a quorum with a hash matching the proposal
(line~\ref{hotstuff:receive-prepared}); we call this set of messages a {\em
  prepared certificate} for the value and check it using the $\accepted$
predicate. The process stores the proposal in $\cmd$, the view in which it
formed the prepared certificate in $\cballot$, and the certificate itself in
$\cert$. At this point we say that the process {\em prepared} the value. Since a
certificate consists of at least $2f+1$ $\PREPARED$ messages and there are
$3f+1$ replicas in total, it is impossible to prepare different values in the
same view: this would require some correct process to send two $\PREPARED$
messages with different values in the same view, which is impossible due to the
check on the $\voted$ flag in line~\ref{hotstuff:safety-check}. Formally, let us
write $\valid(C)$ (for {\em well-formed}) if the set of correctly signed
messages $C$ have been sent in the execution of the protocol.
\begin{proposition}
\label{lemma:hotstuff:singlecmd}
$\forall v, C, C', \val, \val '.\, \accepted(C, v, \hash(\val)) \wedge
\accepted(C', v, \hash(\val'))  \wedge {}$
\\
\hspace*{2.83cm}$\valid(C) \wedge \valid(C') {\implies} \val = \val'$.
\end{proposition}

\begin{figure}[t]
\begin{tabular}{@{}l@{\!\!\!\!\!}|@{\ \ }l@{}}
\begin{minipage}[t]{7.8cm}
\removelatexerror
\vspace*{-12pt}
\begin{algorithm*}[H]
  \setcounter{AlgoLine}{0}
  \SubAlgo{\Upon $\newview(v)$\label{hotstuff:newview}}{
    $\ballot \leftarrow v$\;
    $\voted \leftarrow \FALSE$\;
    \Send $\langle \NEWLEADER(\ballot, \cballot,$\label{hotstuff:send-newleader}\\
    \nonl \quad $ \cmd, \cert) \rangle_i$ \KwTo
    $\leader(\ballot)$\; 
  }

  \smallskip

  \SubAlgo{\WhenReceived $\{\langle \NEWLEADER(v, \vcballot_j, \vcmd_j,$
    \label{hotstuff:receive-newleader}\\  
    \nonl \quad$\vcert_j) \rangle_j \mid p_j \in Q\} = M$
  {\bf for a quorum $Q$}}{
    \textbf{pre:} $\ballot = v \wedge p_i = \leader(v) \wedge {}$\\
    \nonl \hspace{.76cm}$ (\forall m \in M.\, \ValidNewLeader(m))$\;
    \uIf{$\exists j.\hspace{1pt} \vcballot_j \,{=}\, \max\{\vcballot_{k} \,{\mid}\, p_k \,{\in}\, Q\}
      \,{\not=}\, 0$}{
      \Send $\langle \PREPARE(v, \vcmd_j, \vcert_j) \rangle_i$ \KwTo \all\; 
    }\Else{
      \Send $\langle \PREPARE(v, \myproposal(), \bot) \rangle_i$ \KwTo \all\;\label{hotstuff:send-proposal}
    }
  }

  \smallskip

  \SubAlgo{\WhenReceived $\langle \PREPARE(v, \val, \_) \rangle_j =
    m$\label{hotstuff:receive-prepare}}{ 
    \textbf{pre:}
    $\ballot = v \wedge \voted = \FALSE \wedge {}$\label{hotstuff:safety-check}\\
    \nonl \hspace{.76cm}$\ValidNewView(m)$\;
    $\ctxn \leftarrow \val$\;
    $\voted \leftarrow \TRUE$\;
    \Send $\langle \PREPARED(v, \hash(\ctxn)) \rangle_i$ \KwTo  \all\; \label{hotstuff:send-prepared}
  }

\end{algorithm*}
\vspace*{-5pt}
\end{minipage}
&
\begin{minipage}[t]{7.6cm}
\removelatexerror
\vspace*{-12pt}
\begin{algorithm*}[H]
  \SubAlgo{\WhenReceived $\{\langle \PREPARED(v, h)\rangle_j \mid {}$\label{hotstuff:receive-prepared}\\
    \nonl \quad  $p_j \in Q\} = C$  {\bf for a quorum $Q$}}{
    \textbf{pre:} $\ballot = v \wedge \voted = \TRUE \wedge{}$\\
    \nonl \hspace{.77cm}$\hash(\ctxn) = h$\;
    $\cmd \leftarrow \ctxn$\; \label{hotstuff:send-prepared1} 
    $\cballot \leftarrow \ballot$\; \label{hotstuff:send-prepared2}
    $\cert \leftarrow C$\; \label{hotstuff:send-prepared3}
    \Send $\langle \PRECOMMITTED(v, h) \rangle_i$ \KwTo
    \all; \label{hotstuff:send-precommitted} 
  }

  \smallskip

  \SubAlgo{\WhenReceived $\{\langle \PRECOMMITTED(v,h) \rangle_j \,{\mid}$\\
    \nonl \quad  $ p_j \in
    Q\}$
    {\bf for a quorum $Q$}\label{hotstuff:receive-precommitted}}{
    \textbf{pre:} $\ballot \,{=}\, \cballot \,{=}\, v \wedge {}$\\
    \nonl \hspace{.77cm}$\hash(\ctxn) = h$\;
    $\lballot \leftarrow \cballot$\;\label{hotstuff:lock}
    \Send $\langle \COMMITTED(v, h)\rangle_i$ \KwTo \all\; \label{hotstuff:send-committed}
  }

  \smallskip

  \SubAlgo{\WhenReceived $\{\langle \COMMITTED(v, h)\rangle_j \mid$\\
    \nonl \quad $p_j \in Q\}$
    {\bf for a quorum $Q$}\label{hotstuff:receive-committed}}{ 
    \textbf{pre:} $\ballot= \lballot = v \wedge {}$\\
    \nonl \hspace{.77cm}$\hash(\ctxn) = h$\;
    ${\tt decide}(\ctxn)$;
  }
\end{algorithm*}
\vspace*{-5pt}
\end{minipage}
\end{tabular}

\vspace{10pt}

{\small
\centerline{$
\begin{array}{@{}l@{}}
\accepted(C, v, h)
\iff
\exists Q.\, 
\quorum(Q) \wedge C = \{\langle \PREPARED(v, h) \rangle_j \mid p_j \in Q\}
\\[5pt]
\ValidNewLeader(\langle \NEWLEADER(v', v, \val, C) \rangle_{\_})
  \iff 
v < v' \wedge
({v \not= 0} {\implies} \accepted(C, v, \hash(\val)))
\\[5pt]
\ValidNewView(\langle \PREPARE(v, \val, C) \rangle_i) \iff 
p_i = \leader(v) \wedge \validity(\val) \wedge {}
\\[2pt]
\quad (\lballot \not \not= 0 {\implies}
\val = \cmd \vee 
(\exists v'.\, v > v'>\lballot \wedge \accepted(C, v', \hash(\val))))
\end{array}
$}}

\caption{Single-shot HotStuff. All variables storing views are
  initially set to $0$ and others to $\bot$.}
\label{fig:hotstuff}
\end{figure}

Preparing a value is a prerequisite for deciding on it. Hence, by
Proposition~\ref{lemma:hotstuff:singlecmd} a prepared certificate for a value
$\val$ and a view $v$ guarantees that $\val$ is the only value that can be
possibly decided in $v$. For this reason, it is this certificate, together with
the corresponding value and view, that the process sends upon a view change to
the new leader in a $\NEWLEADER$ message
(line~\ref{hotstuff:send-newleader}). The leader makes its proposal based on a
quorum of $\NEWLEADER$ messages with prepared certificates formed in lower views
than the one it is in (line~\ref{hotstuff:receive-newleader}), as checked by
$\ValidNewLeader$. Similarly to Paxos~\cite{paxos} and PBFT~\cite{pbft}, the
leader selects as its proposal the value prepared in the highest view, or, if
there are no such values, its own proposal given by $\myproposal()$. In the
former case, the leader sends the corresponding certificate in its $\PREPARE$
message, to justify its choice; in the latter case this is replaced by $\bot$.

Once a process prepares a value $\val$, it participates in the next message
exchange: it disseminates a $\PRECOMMITTED$ message with the hash of the value and
waits until it gathers a quorum of $\PRECOMMITTED$ messages matching the
prepared value (line~\ref{hotstuff:receive-precommitted}). This ensures that at
least $f+1$ correct processes have prepared the value $\val$. Since the leader
of the next view will gather prepared commands from at least $2f+1$ processes,
at least one correct process will tell the leader about the value $\val$, and
thus the leader will be aware of this value as a potential decision in the
current view.

Having gathered a quorum of $\PRECOMMITTED$ messages for a value, the process
becomes {\em locked} on this value, which is recorded by setting a special
variable $\lballot$ to the current view. From this point on, the process will
not accept a proposal of a different value from a leader of a future view,
unless the leader can convince the process that no decision was reached in the
current view. This is ensured by the $\ValidNewView$ check the process does on a
$\PREPARE$ message from a leader (line~\ref{hotstuff:safety-check}). This checks
that the value is valid and that, if the process has previously locked on a
value, then either the leader proposes the same value, or its proposal is
justified by a prepared certificate from a higher view than the lock. In the
latter case the process can be sure that no decision was reached in the view it
is locked on.

Having locked a value, the process participates in the final message exchange:
it disseminates a $\COMMITTED$ message with the hash of the value and waits until
it gathers a quorum of matching $\COMMITTED$ messages for the locked value
(line~\ref{hotstuff:receive-committed}). Once this happens, the process decides
on this value. Gathering a quorum of $\COMMITTED$ messages on a value $\val$
ensures that at least $f+1$ correct processes are locked on the same value. This
guarantees that a leader in a future view cannot get processes to decide on a
different value: this would require $2f+1$ processes to accept the leader's
proposal; but at least one correct process out of these would be locked on
$\val$ and would refuse to accept a different value due to the $\ValidNewView$
check.
Thus, while the exchange of $\PRECOMMITTED$ messages ensures that a future
correct leader will be aware of the value being decided and will be able to make
a proposal passing $\ValidNewView$ checks (liveness), the exchange of
$\COMMITTED$ ensures that a faulty leader cannot revert the decision (safety).

Since processes transition through increasing views
(Property~\ref{prop:local-order} in Figure~\ref{fig:sync-properties}), we get
\begin{proposition}
\label{lemma:hotstuff:increase}
The variables $\lballot$, $\cballot$ and $\ballot$ at a correct process never
decrease and we always have $\lballot \le \cballot \le \ballot$.
\end{proposition}

Note that, when a process enters view $1$, it trivially knows that no decision
could have been reached in prior views. Hence, the leader of view $1$ can send
its proposal immediately, without waiting to receive a quorum of $\NEWLEADER$
messages (line~\ref{hotstuff:send-proposal}), and processes can avoid sending
these messages to this leader. For brevity, we omit this optimization from the
pseudocode, even though we take it into account in our latency analysis.

Since the synchronizer is not guaranteed to switch processes between views all
at the same time, a process in a view $v$ may receive a message from a higher
view $v' > v$, which needs to be stored in case the process finally switches to
$v'$. If implemented naively, this would require a process to store unboundedly
many messages. Instead, we allow a process to store, for each message type and
sender, only the message of this type received from this sender that has the
highest view. As we show below (Theorem~\ref{thm:livehotstuff}), this does not
violate liveness. Thus, assuming consensus proposals of bounded size, the
protocol Figure~\ref{fig:hotstuff} runs in bounded space, and so does the
overall consensus protocol with the \SYNC synchronizer.

We defer the proof that the protocol satisfies Validity and Agreement to
\tr{\ref{sec:hotstuff-safety}}{\nhotstuffsafety} and focus on our core
contribution: proving its liveness and analyzing its latency.

\subparagraph*{\textbf{Protocol liveness.}}
Assume that the protocol is used with a synchronizer satisfying
Properties~\ref{prop:local-order}-\ref{prop:no-skip-sync:3} on the left of
Figure~\ref{fig:sync-properties}; to simplify the following latency analysis, we
assume $d = 2\delta$, as for \SYNC. The next theorem states requirements on a
view sufficient for the protocol to reach a decision and quantifies the
resulting latency.
\begin{theorem}
  Let $v \ge \B$ be a view such that $\timeout(v) > 7\delta$ and $\leader(v)$ is
  correct. Then in single-shot HotStuff all correct processes decide in view $v$
  by $\tl{v}+5\delta$.
\label{thm:livehotstuff}
\end{theorem}
\noindent {\sf \textbf{Proof.}} 
By Property~\ref{prop:after-t} we have $\tm{v} \ge \GST$, so that all messages
sent by correct processes after $\tm{v}$ get delivered to all correct processes
within $\delta$. Once a correct process enters $v$, it sends its $\NEWLEADER$
message, so that $\leader(v)$ will receive a quorum of such messages by
$\tl{v} + \delta$. When this happens, the leader will send its proposal in a
$\PREPARE$ message, which correct processes will receive by $\tl{v}+2\delta$.
If they deem the proposal safe, it takes them at most $3\delta$ to exchange the
sequence of $\PREPARED$, $\PRECOMMITTED$ and $\COMMITTED$
messages. By~(\ref{overlap}), all correct processes will stay in $v$ until at
least $\tl{v}+(\timeout(v) - d) > \tl{v}+5\delta$, and thus will not send a
message with a view $>v$ until this time. Thus, none of none of the above
messages will be discarded at correct processes before this time, and assuming
the safety checks pass, the sequence of message exchanges will lead to decisions
by $\tl{v}+5\delta$.

It remains to show that the proposal $\leader(v)$ makes in view $v$
(line~\ref{hotstuff:receive-newleader}) will satisfy $\ValidNewView$ at all
correct processes (line~\ref{hotstuff:receive-prepare}). It is easy to show that 
the proposal satisfies $\validity$,
so we now need to prove the last conjunct of $\ValidNewView$. This trivially
holds if no correct process is locked on a value when receiving the $\PREPARE$
message from the leader.

We now consider the case when some correct process is locked on a value when
receiving the $\PREPARE$ message, and let $p_i$ be a process that is locked on
the highest view among correct processes. Let $\val = p_i.\cmd$ be the value
locked and $v_0 = p_i.\lballot < v$ be the corresponding view. Since $p_i$
locked $\val$ at $v_0$, it must have previously received messages
$\PRECOMMITTED(v_0, \hash(\val))$ from a quorum of processes
(line~\ref{hotstuff:receive-precommitted}), at least $f+1$ of which have to be
correct. The latter processes must have prepared the value $\val$ at view $v_0$
(line~\ref{hotstuff:receive-prepared}). By
Proposition~\ref{lemma:hotstuff:increase}, when each of these $f+1$ correct
processes enters view $v$, it has $\cballot \ge v_0$ and thus sends the
corresponding value and its prepared certificate in the $\NEWLEADER(v, \ldots)$
message to $\leader(v)$. The leader is guaranteed to receive at least one of
these messages before making a proposal, since it only does this after receiving
at least $2f+1$ $\NEWLEADER$ messages
(line~\ref{hotstuff:receive-newleader}). Hence, the leader proposes a value
$\val'$ with a prepared certificate formed at some view $v' \ge v_0$ no lower
than any view that a correct process is locked on when receiving the leader's
proposal. Furthermore, if $v' = v_0$, then by
Proposition~\ref{lemma:hotstuff:singlecmd} we have that $\val' = \val$ and
$\val$ is the only value that can be locked by a correct process at
$v_0$. Hence, the leader's proposal will satisfy $\ValidNewView$ at each correct
process.\qed

\smallskip
\smallskip

Since by Property~\ref{prop:no-skip-sync:1} correct processes enter every view
starting from $\B$ and, by the definition of $\leader()$, leaders rotate
round-robin, we are always guaranteed to encounter a correct leader after at
most $f$ view changes. Then Theorem~\ref{thm:livehotstuff} implies that the
protocol is live when using a timeout function $\timeoutV$ that grows without
bound.
\begin{corollary}
  Let $\timeoutV$ be such that~(\ref{prop:increasing}) holds. Then in
  single-shot HotStuff all correct processes eventually decide.
\end{corollary}

\subparagraph*{Protocol latency.} 
When single-shot HotStuff is used with the \SYNC synchronizer, rather than an
arbitrary one, we can use
Properties~\ref{prop:last-entry}-\ref{prop:convergence} on the right of
Figure~\ref{fig:sync-properties} to bound how quickly the protocol reaches a
decision after $\GST$. To this end, we combine Theorem~\ref{thm:livehotstuff}
with Property~\ref{prop:convergence}, which bounds the latency of view
synchronization, and Proposition~\ref{lem:last-entry-sum}, which bounds the
latency of going through up to $f$ views with faulty leaders.
\begin{corollary}
  Let $v = \GV{\GST+\rho}+1$ and assume that $\timeoutV(v) > 7\delta$ and
  $\tfk{f+1} \le \GST + \rho$. Then in single-shot HotStuff all correct
  processes decide by
  $\GST + \rho + \sum_{k=v-1}^{v+f-1} (\timeoutV(k)+\delta) + 7\delta$.
\label{thm:hotstuff:worst}
\end{corollary}
\vspace{-\baselineskip} 
We can also quantify the latency of the protocol under favorable conditions,
when it is started after $\GST$. In this we rely on Property~\ref{prop:view1},
which gives conditions under which processes synchronize in view $1$. The
following corollary of Theorem~\ref{thm:livehotstuff} exploits this property to
bound the latency of HotStuff when it is started after $\GST$
and the initial timeout is set appropriately, but the protocol may still go
through a sequence of up to $f$ faulty leaders. The summation in the bound
(coming from Proposition~\ref{lem:last-entry-sum}) quantifies the overhead in
the latter case.
\begin{corollary}
  Assume that $\tf\geq\GST$ and $\timeout(1) > 7\delta$.
  Then in single-shot HotStuff all correct processes decide
  no later than $\ts+\sum_{k=1}^{f} (\timeoutV(k)+\delta) + 6\delta$.
\label{thm:hotstuff:latencyfromVequal1}
\end{corollary}
Finally, the next corollary bounds the latency when additionally the leader of
view $1$ is correct, in which case the protocol can benefit from the optimized
execution of this view noted earlier. The corollary follows from
Property~\ref{prop:view1} and an easy strengthening of
Theorem~\ref{thm:livehotstuff} for the special case of $v = \B =1$.
\begin{corollary}
  Assume that $\tf \geq\GST$, $\timeout(1) > 6\delta$, 
  and $\leader(1)$ is correct. Then in single-shot HotStuff
  all correct processes decide no later than $\ts + 5\delta$.
 \label{cor:Vequal1andcorrectleader}
\end{corollary}

\subsection{Two-Phase HotStuff}
\label{sec:tendermint}

We next consider a {\em two-phase} variant of HotStuff~\cite{hotstuff}, which
processes the leader's proposals in two phases instead of three. In exchange,
it uses timeouts not just for view synchronization, but also in the core
consensus protocol to delimit different stages of a single view. This
demonstrates that our synchronizer specification is strong enough to deal with
interactions between the timeouts in different parts of the overall protocol.
When used with our \SYNC synchronizer, the protocol furthermore requires only
bounded space.  Two-phase HotStuff is similar to
Tendermint~\cite{tendermint-arxiv} and Casper~\cite{casper}, which use timeouts
for the same purposes. We chose this protocol for conciseness of presentation,
but in \tr{\ref{sec:original-tendermint}}{\noriginaltendermint} we also present
a variant of the original Tendermint consensus based on our synchronizer (see
\S\ref{sec:pbft-sbft}).

Due to space constraints, we describe the changes to the protocol in
Figure~\ref{fig:hotstuff} required to get its two-phase version informally and
defer the pseudocode to \tr{\ref{sec:2hs}}{\ntwohs}. In two-phase HotStuff, a
process handles a proposal from the leader in the same way as in the three-phase
one, by sending a $\PREPARED$ message (line~\ref{hotstuff:receive-prepare} in
Figure~\ref{fig:hotstuff}). Upon assembling a quorum of matching $\PREPARED$
messages (line~\ref{hotstuff:receive-prepared}), the process updates its
variables as per
lines~\ref{hotstuff:send-prepared1}-\ref{hotstuff:send-prepared3}, but in
addition immediately becomes locked on the prepared value $\cmd$, without
exchanging $\PRECOMMITTED$ messages: the process assigns $\lballot$ to the
current view and sends a $\COMMITTED$ message with the hash of the value. As
before, assembling a quorum of such messages causes the process to decide on the
value (line~\ref{hotstuff:receive-committed}).  Upon entering a new view
(line~\ref{hotstuff:newview}), a process sends to the leader a $\NEWLEADER$
message with the information about the last value it prepared (and therefore
locked, line~\ref{hotstuff:send-newleader}). The leader chooses its proposal in
the same way as in three-phase HotStuff (line~\ref{hotstuff:receive-newleader}).

The two-phase version of HotStuff is safe for the same reasons as the
three-phase one: %
the exchange of $\PRECOMMITTED$ messages, %
omitted from the current
protocol, is only needed for liveness, not safety. However, ensuring liveness in
two-phase HotStuff requires a different mechanism: since a correct process $p_i$
gets locked on a value immediately after preparing it, gathering prepared values
from an arbitrary quorum of processes is not enough for the leader to ensure it
will make a proposal that will pass the $\ValidNewView$ check at $p_i$: the
quorum may well exclude this process. To solve this problem, the leader waits
before making a proposal so that eventually in some view it will receive
$\NEWLEADER$ messages from {\em all correct processes}. This ensures the leader
will eventually make a proposal that will pass the $\ValidNewView$ checks at all
of them. In more detail, when a process enters a view where it is the leader, it
sets a special timer $\timernetwork$ for the duration determined by a function
$\timeoutS$. The leader makes a proposal by executing the handler in
line~\ref{hotstuff:receive-newleader} only after the timer expires.

For the leader to make an acceptable proposal, the duration of $\timernetwork$
needs to be long enough for all $\NEWLEADER$ messages for this view from correct
processes to reach the leader. For the protocol to decide, after $\timernetwork$
expires, processes also need to stay in the view long enough to complete the
necessary message exchanges. The following theorem characterizes these
requirements formally, again assuming $d = 2\delta$ in
Property~\ref{prop:no-skip-sync:2}. Note that in the proof of the theorem we
rely on the guarantees about the timing of correct processes entering a view
(Property~\ref{prop:no-skip-sync:2}) to show that $\timernetwork$ fulfills its
intended function.
\begin{theorem}
  Let $v \ge \B$ be a view such that $\timeoutS(v) > 3\delta$,
  $\timeoutV(v) - \timeoutS(v) > 5\delta$ %
  and $\leader(v)$ is correct. Then in two-phase HotStuff all correct processes
  decide at $v$ by $\tl{v}+\timeoutS(v)+3\delta$.
\label{thm:livetendermint}
\end{theorem}
\vspace{-\baselineskip}
\noindent {\sf \textbf{Proof.}} 
Once a correct process enters $v$, it sends its $\NEWLEADER$ message, so that
$\leader(v)$ is guaranteed to receive such messages from all correct processes
by $\tl{v} + \delta$. By Property~\ref{prop:no-skip-sync:2}, the leader enters
$v$ by $\tl{v}-2\delta$ at the earliest. Since the leader starts its
$\timernetwork$ when it enters $v$ and $\timeoutS(v) > 3\delta$, $\timernetwork$
can only expire after $\tl{v}+\delta$. Thus, the leader is guaranteed to receive
$\NEWLEADER$ messages from all correct processes before $\timernetwork$
expires. When $\timernetwork$ expires, which happens no later than
$\tl{v}+\timeoutS(v)$, the leader will send its proposal in a $\PREPARE$
message, which correct processes will receive by
$\tl{v}+\timeoutS(v)+\delta$. If they deem the proposal safe, it takes them at
most $2\delta$ to exchange the sequence of $\PREPARED$ and $\COMMITTED$ messages
leading to decisions.
By~(\ref{overlap}), all correct processes will stay in $v$ until
at least $\tl{v}+(\timeout(v) - d) > \tm{v}+\timeoutS(v)+3\delta$. By then the
above sequence of message exchanges will complete, and all correct processes
will decide.

It remains to show that the proposal $\leader(v)$ makes in view $v$
will satisfy $\ValidNewView$ at all correct processes.
It is easy to show that this proposal is valid,
so we now need to prove the last conjunct of $\ValidNewView$. This trivially
holds if no correct process is locked on a value when receiving the $\PREPARE$
message from the leader.
We now consider the case when some correct process is locked on a value when
receiving the $\PREPARE$ message, and let $p_i$ be a process that is locked on
the highest view among correct processes. Let $\val = p_i.\cmd$ be the value
locked and $v_0 = p_i.\lballot < v$ be the corresponding view. Since
$\leader(v)$ receives all of the $\NEWLEADER$ messages sent by correct processes
before making its proposal, it proposes a value $\val'$ with a prepared
certificate formed at some view $v' \ge v_0$. 
Also, if $v' = v_0$, then by Proposition~\ref{lemma:hotstuff:singlecmd},
$\val' = \val$ and $\val$ is the only value that can be locked by a correct
process at $v_0$. Hence, the leader's proposal will satisfy $\ValidNewView$ at
each correct process.\qed

\smallskip
\smallskip

Since leaders rotate round-robin, Theorem~\ref{thm:livetendermint} implies that
the protocol is live, provided the functions $\timeoutV$ and $\timeoutS$, {\em
  as well as the difference between them}, grow without bound. This can be
satisfied, e.g., by letting $\timeoutV(v) = 2v$ and $\timeoutS(v) = v$.
\begin{corollary}\label{thm:tendermint-live}
  Let $\timeoutV$ and $\timeoutS$ be such that~(\ref{prop:increasing}) holds and
  $\forall \theta.\,\exists v.\,\forall v'.\, v' \ge v {\implies}
  \timeoutV(v')-\timeoutS(v')>\theta$. Then in two-phase HotStuff all correct
  processes eventually decide.
\end{corollary}

\subparagraph*{Protocol latency.} 
Similarly to \S\ref{sec:hotstuff}, when the protocol is used with the \SYNC
synchronizer, we can quantify its latency in both unfavorable scenarios (when
starting before $\GST$) and favorable scenarios (when starting after
$\GST$). The first corollary of Theorem~\ref{thm:livetendermint} below uses
Property~\ref{prop:convergence} and Proposition~\ref{lem:last-entry-sum}, and
the following two corollaries, Property~\ref{prop:view1}.
\begin{corollary}
  Let $v = \GV{\GST+\rho}+1$ and assume that $\tfk{f+1} \le \GST + \rho$,
  $\timeoutS(v) > 3\delta$ and $\timeoutV(v) - \timeoutS(v) > 5\delta$. Then in
  two-phase HotStuff all correct processes decide no later than
  $\GST + \rho + \sum_{k=v-1}^{v+f-1} (\timeoutV(k)+\delta) + \timeoutS(v+f) +
  5\delta$.
\label{thm:tendermint:worst}
\end{corollary}
\begin{corollary}
  Assume that $\tf\geq\GST$, $\timeoutS(1) > 3\delta$ and
  $\timeoutV(1) - \timeoutS(1) > 5\delta$.
  Then in two-phase HotStuff all correct processes decide no later than
  $\ts+\sum_{k=1}^{f} (\timeoutV(k)+\delta) + \timeoutS(f+1) + 4\delta$.
\label{thm:tendermint:latencyfromVequal1}
\end{corollary}
\vspace{-20pt}
\begin{corollary}
  Assume that $\tf\geq\GST$, $\timeout(1) > 5\delta$
  and $\leader(1)$ is correct. Then in two-phase HotStuff all correct processes
  decide no later than $\ts + 4\delta$.
 \label{cor:tendermint:Vequal1andcorrectleader}
\end{corollary}

Like in \S\ref{sec:hotstuff}, the last corollary takes into account the
optimized execution of view $1$. 
The above latency bounds allow us to compare the two-phase and three-phase
versions of HotStuff (\S\ref{sec:hotstuff}). In the ideal case when the timeouts
are set optimally and the leader of view $1$ is correct, two-phase HotStuff has
a lower latency than three-phase one: $4\delta$ in
Corollary~\ref{cor:tendermint:Vequal1andcorrectleader} vs $5\delta$ in
Corollary~\ref{cor:Vequal1andcorrectleader}. When the initial leader is faulty,
both protocols incur the overhead of switching through several views until they
encounter a correct leader (Corollaries~\ref{thm:tendermint:latencyfromVequal1}
and~\ref{thm:hotstuff:latencyfromVequal1}). In this case, the latency of
deciding in the first view with a correct leader is at most $6\delta$ for
three-phase HotStuff and $\timeoutS(f+1) + 4\delta$ for two-phase one. Even when
$\timeoutS(f+1)$ is the optimal $3\delta$, the two-phase HotStuff bound yields
$7\delta$ -- a higher latency than for three-phase HotStuff. The latency bounds
for the case of starting before $\GST$ relate similarly
(Corollaries~\ref{thm:tendermint:worst} and~\ref{thm:hotstuff:worst}). The
higher latency of two-phase HotStuff in these cases are caused by the inclusion
of the timeout $\timeoutS(f+1)$, which reflects the lack of ``optimistic
responsiveness'' of this protocol~\cite{hotstuff}.

\subsection{Single-Shot PBFT, SBFT and Tendermint}
\label{sec:pbft-sbft}

Using our synchronizer specification, we have also proved the correctness and
analyzed the latency of single-shot versions of PBFT~\cite{pbft},
SBFT~\cite{sbft} and Tendermint~\cite{tendermint-arxiv}, thus demonstrating the
wide applicability of the specification. Due to space constraints we defer the
details to \tr{\ref{app:consensus}}{\nconsensus}. Our analysis of PBFT is
similar to that of HotStuff. SBFT is a recent improvement of PBFT that adds a
fast path for cases when all processes are correct, and our analysis quantifies
the latency of both paths.

Tendermint is similar to two-phase HotStuff; in particular, it also
uses timeouts both for view synchronization and to delimit different stages of a
single view. %
However, the protocol never sends messages with certificates, and thus, like
\SYNC, does not need digital signatures.
Tendermint integrates the functionality required for view synchronization with
the core consensus protocol, breaking its control flow in multiple places. We
consider its variant that delegates this functionality to the synchronizer, thus
simplifying the protocol. Our analysis of the resulting protocol is similar to
the one of two-phase HotStuff in \S\ref{sec:tendermint}. Apart from deriving
latency bounds for the protocol, our analysis exploits the synchronizer
specification to give a proof of its liveness that is more rigorous than the
existing ones~\cite{tendermint-arxiv,tendermint-netys}, which lacked a detailed
correctness argument for the view synchronization mechanism used in the
protocol.

\section{Related Work}
\label{sec:related}

Most Byzantine consensus protocols are based on the concept of views (aka
rounds), and thus include a mechanism for view synchronization. This mechanism
is typically integrated with the core consensus protocol, which complicates the
design~\cite{pbft,tendermint-arxiv,sbft}. Subtle view synchronization mechanisms
have often come without a proof of liveness (e.g., PBFT~\cite{castro-thesis}) or
had liveness bugs (e.g., Tendermint~\cite{tendermint-opodis} and
Casper~\cite{casper-bug}). Furthermore, liveness proofs have not usually given
concrete bounds on the latency of reaching a decision (exceptions
are~\cite{prime,bounded-delay}).

Several papers suggested separating the functionality of view synchronization
into a distinct component, starting with the seminal DLS paper on consensus
under partial synchrony~\cite{dls}. DLS specified the guarantees provided by
view synchronization indirectly, by proving that its implementation simulated an
abstract computational model with a built-in notion of rounds. Unlike us, DLS
did not give a specification determining how long processes stay in a round and
how quickly they switch between rounds; as we have demonstrated, such properties
are needed to reason about modern Byzantine consensus protocols. DLS implemented
rounds using a distributed protocol that synchronizes process-local clocks
obtained by counting state transitions of each process. This protocol has to
synchronize local clocks on every step of the consensus algorithm, which results
in prohibitive communication overheads and makes this solution impractical.

Abraham et al.~\cite{ADDNR19} build upon ideas from fault-tolerant clock
synchronization~\cite{SWL86,DHSS95} to implement view synchronization assuming
that processes have access to hardware clocks with bounded drift. But this work
only gives a solution for a synchronous system. Our \SYNC\/ synchronizer also
assumes hardware clocks but removes the assumption of bounded drift before GST,
thus making them compatible with partial synchrony. We note that, although the
problems of clock and view synchronization are different, they are closely
related at the algorithmic level. We therefore believe that our view
synchronization techniques can in the future be adapted to obtain an efficient
partially synchronous clock synchronization protocol.

The HotStuff protocol~\cite{hotstuff} delegated the functionality of view
synchronization to a separate component, called a pacemaker. But it did not
provide a formal specification of this component or a practical
implementation. To address this, Naor et al. have recently formalized view
synchronization as a separate problem~\cite{lumiere,NK20}. Unlike us, they did
not provide a comprehensive study of the applicability of their specifications
to a wide range of modern Byzantine consensus protocols. In particular, their
specifications do not expose bounds on how quickly processes switch views
(Property~\ref{prop:no-skip-sync:2} in Figure~\ref{fig:sync-properties}), which
are necessary for protocols such as two-phase HotStuff (\S\ref{sec:tendermint})
and Tendermint (\S\ref{sec:pbft-sbft}).

Naor et al. also proposed synchronizer implementations in a simplified variant
of partial synchrony where $\delta$ is known a priori, and messages sent before
$\GST$ are guaranteed to arrive by $\GST +
\delta$~\cite{lumiere,NK20}. These implementations focus on optimizing
communication complexity, making it linear in best-case scenarios~\cite{lumiere}
or in expectation~\cite{NK20}. They achieve linearity by relying on digital
signatures (more precisely, threshold signatures), which \SYNC eschews. Unlike
\SYNC, they also require unbounded space (for the reasons explained in
\S\ref{sec:sync-impl}). Finally, we give exact latency bounds for \SYNC under
both favorable and unfavorable conditions whereas~\cite{lumiere,NK20} only
provide expected latency analysis. It is interesting to investigate whether the
benefits of the two approaches can be combined to tolerate message loss before
GST with both bounded space and a low communication complexity.

LibraBFT~\cite{libra} extends HotStuff with a view synchronization mechanism,
integrated with the core protocol; the protocol assumes reliable
channels. LibraBFT is optimized to solve repeated consensus, whereas in this
paper we focus on single-shot one. We leave investigating synchronizer
abstractions optimized for the multi-shot case to future work.

The original idea of using synchronizers to simulate a round-based
synchronous system on top  of an asynchronous one is due to Awerbuch~\cite{Awe85}. 
This work however, 
did not consider failures. 
Augmented round models
to systematically study properties of distributed consensus under various failure
and environment assumptions were proposed in~\cite{Gaf98,heardof,KS06,heardof-bft}.
These papers however, 
do not deal with implementing the proposed models under partial synchrony.
Upper bounds for deciding after $\GST$ in round-based crash fault-tolerant 
consensus algorithms were studied in~\cite{DGL05,AGGT08}. 
While we derive similar bounds for Byzantine failures, it remains open 
if these are optimal or can be further improved. 
Failure detectors~\cite{CT96,CHT96}, which abstract away
the timeliness guarantees of the environment, have been extensively used 
for developing and analyzing  consensus algorithms~\cite{CT96,MR99} 
in the presence of benign failures.
However, since capturing all possible faulty behaviors is algorithm-specific,
the classical notion of a failure detector does not naturally
generalize to Byzantine settings. 
As a result, the existing work on Byzantine failure detectors 
either limits the types of failures being addressed (e.g.,~\cite{MR97}),
or focuses on other means (such as accountability~\cite{HK09}) to
mitigate faulty behavior.

\bibliographystyle{abbrv}
\bibliography{biblio}

\iflong

\clearpage
\appendix

\makeatletter
    \def\@evenhead{\large\sffamily\bfseries
                   \llap{\hbox to0.5\oddsidemargin{\thepage\hss}}APPENDIX\hfil}%
    \def\@oddhead{\large\sffamily\bfseries APPENDIX\hfil
                  \rlap{\hbox to0.5\oddsidemargin{\hss\thepage}}}%
\makeatother

\section{Correctness of the Synchronizer Algorithm}
\label{sec:proof-sync}

The {\em local view}\/ of a process $p_i$ at time $t$, denoted $\LV{i}{t}$, is
the latest view entered by $p_i$ at or before $t$, or $0$ if $p_i$ has not
entered any views by then. Thus, $\GV{t} = \max\{\LV{i}{t} \mid p_i \text{~is correct}\}$.
We say that a process $p_i$ {\em attempts to advance}\/ from a view $v \ge 0$ at time $t$
if at this time $p_i$ executes the code in either
line~\ref{line:send1} or line~\ref{line:send2}, and $\LV{i}{t} = v$.

\begin{lemma}%
For all times $t$ and views $v>0$, if a correct process sends
$\WISH(v)$ at $t$, then there exists a time $t' \le t$
such that some correct process attempts to advance from $v-1$ at $t'$.
\label{lem:wish-attempt}
\end{lemma}

\paragraph{Proof.}
We first prove the following auxiliary proposition:
\begin{multline} 
\forall p_i.\, \forall v.\, 
p_i \text{~is~correct} \wedge p_i \text{~sends~} \WISH(v)
\text{~at~} t \implies\\
\exists t' \le t.\, \exists v' \ge v-1.\, \exists p_j.\,
p_j \text{~is~correct} \wedge p_j 
\text{~attempts~to~advance~from~} v' \text{~at~} t'.
\label{eq:attempt-gtv}
\end{multline}
By contradiction, assume that a correct process $p_i$ sends $\WISH(v)$ at $t$,
but for all $t' \le t$ and all $v' \ge v-1$, no correct process attempts to
advance from $v'$ at $t'$. Consider the earliest time $t_k$ when some correct
process $p_k$ sends a $\WISH(v_k)$ with $v_k \ge v$, so that $t_k \le t$. Since
at $t_k$ process $p_k$ does not attempt to advance from $v_k$, it has to
execute the code in either one of the following lines: \ref{line:send2},
\ref{line:send3}, \ref{line:send4}, or \ref{line:send5}.

Since $p_k$ sends $\WISH(v_k)$ at $t_k$, then either 
$v_k = p_k.\viewp(t_k)$ or $p_k.\view(t_k) = p_k.\viewp(t_k) = v_k - 1$, and in
the latter case $p_k$ executes either line~\ref{line:send2} or line~\ref{line:send4}.
If $p_k.\viewp(t_k) = v_k \ge v$, then $p_k.\lastViews(t_k)$ includes $f+1$ entries 
$\ge v_k \ge v$, and therefore, there exists a correct process $p_l$ that 
sent $\WISH(v')$ with $v' \ge v$ at $t_l < t_k$, contradicting the
assumption that $t_k$ is the earliest time when this can happen.
Suppose that $p_k.\view(t_k) = p_k.\viewp(t_k) = v_k - 1$ and 
at $t_k$, $p_k$ executes either line~\ref{line:send2}
or line~\ref{line:send4}. Then $\LV{k}{t_k} = v_k - 1$. If $p_k$ executes 
line~\ref{line:send2} at $t_k$, then since $\LV{k}{t_k} = v_k - 1$,
$p_k$ attempts to advance from $v_k-1 \ge v-1$ at $t_k\le t$, contradicting our
assumption that no such attempt can occur. 

Suppose now that $p_k$ executes the code in
line~\ref{line:send4} at $t_k$ and
$p_k.\view(t_k) = p_k.\viewp(t_k) = v_k - 1$. 
Consider first the case when $v_k = 1$. Since $\lastViews[k] > 0$,
$p_k$ has already sent $\WISH(v_k')$ for some view $v_k' \ge 1$
at time $s_k < t_k$. Since $v_k' \ge v_k \ge v$, this is a contradiction
to our assumption that no $\WISH$ messages with views $\ge v$ can 
be sent before $t_k$.
It remains to consider the case when $v_k > 1$. Then
$\te{k}{v_k-1}$ is defined and satisfies $\te{k}{v_k-1} < t_k$.
Thus, $p_k.\view(\te{k}{v_k-1}) = p_k.\viewp(\te{k}{v_k-1}) = v_k - 1$.
Since $p_k$ starts $p_k.\timerview$ at $\te{k}{v_k-1}$, and
$p_k.\timerview(t_k)$ is not enabled, there exists a time 
$\te{k}{v_k-1} < t_k' < t_k$ such that $p_k.\timerview$
expires at $t_k'$, triggering the execution of the $\timerview$
expiration handler. Since both $p_k.\view$ and $p_k.\viewp$ are non-decreasing,
and both are equal to $v_k - 1$ at 
$\te{k}{v_k-1}$ as well as $t_k$, $p_k.\view(t_k') = p_k.\viewp(t_k') = v_k -
1$. Thus, $\LV{k}{t_k'} = v_k-1$, which implies that at $t_k' < t_k \le t$,
$p_k$ attempts to advance from $v_k-1 \ge v - 1$,
contradicting our assumption that no such attempt can happen. 
We conclude that~(\ref{eq:attempt-gtv}) holds.

We now prove the lemma. Let $t$ be a time and $v$ be a view such that
some correct process sends $\WISH(v)$ at $t$.
By~(\ref{eq:attempt-gtv}), there exists a correct process
that attempts to advance from a view $\ge v-1$ at or before $t$.
Let $t'$ be the earliest time when some correct process attempts to advance from 
a view $\ge v-1$, and let $p_j$ be this process and $v' \ge v-1$ be the view
from which $p_j$ attempts to advance at $t'$.
Thus, at $t'$, $p_j$ executes the code
in either line~\ref{line:send1} or line~\ref{line:send2}, 
and $\LV{j}{t'} = v' \ge v - 1$.
Suppose first that $p_j$ executes the code in line~\ref{line:send2} at $t'$.
Since $\LV{j}{t'} = v'$, there exists an earlier time at which 
$p_j.\viewp = p_j.\view = v'$. Since $p_j.\viewp$ is non-decreasing,
$p_j.\viewp(t') \ge v'$. If $p_j.\viewp(t') > v'$, then given that
$v' \ge v - 1$, $p_j.\viewp(t') \ge v$. Thus, there exists a correct
process $p_k$ and time $t'' < t'$ such that $p_k$ sent 
$\WISH(v'')$ with $v'' \ge v$ to $p_j$ at $t''$. By~(\ref{eq:attempt-gtv}),
there exists a time $\le t'' < t'$ at which some correct process
attempts to advance from a view $\ge v''-1 \ge v-1$, which is impossible.
Thus, $p_j.\viewp(t') = v'$. Since $\LV{j}{t'} = v'$, we have
$p_j.\view(t') = p_j.\viewp(t') = v'$. 
Suppose now that $p_j$ executes the code in line~\ref{line:send1}. Then 
$p_j.\viewp(t') = p_j.\view(t') = 0 = \LV{j}{t'} = v'$.
Hence, in both cases
$$
p_j.\view(t') = p_j.\viewp(t') = v' \ge v - 1.
$$
By the definitions of $\view$ and $\viewp$,
$v'$ is both the lowest view among the highest $2f+1$ views
in $p_j.\lastViews(t')$, and the lowest view among the highest
$f+1$ views in $p_j.\lastViews(t')$. 
Hence,  $p_j.\lastViews(t')$ includes $f+1$ entries equal to $v'$, and 
therefore, there exists a correct process $p_k$ such that
\begin{equation}\label{entry-eq-v-1}
p_j.\view(t') = p_j.\viewp(t') = p_j.\lastViews[k](t') = v'
\ge v-1.
\end{equation}
Also, for all correct processes $p_l$, $p_j.\lastViews[l](t') < v$
for otherwise, some correct process sent $\WISH(v'')$ with $v'' \ge v$
at $t'' < t'$, and therefore, by~(\ref{eq:attempt-gtv}),
some correct process attempted to advance from a view $\ge v-1$ earlier
than $t'$, which is impossible. Thus, 
$$
p_j.\view(t') = p_j.\viewp(t') = p_j.\lastViews[k](t') < v.
$$
Together with~(\ref{entry-eq-v-1}), this implies
$$
p_j.\view(t') = p_j.\viewp(t') = v - 1.
$$
Hence, $\LV{j}{t'} = v - 1$, and therefore, $p_j$ attempts to advance from $v-1$ at $t'$. Thus,
$v' = v-1$ and $t' \le t$, as required.\qed

\begin{lemma}
  If a correct process $p_i$ enters a view $v$, then there exists a time
  $t < \te{i}{v}$ at which some correct process attempts to advance from $v-1$.
\label{lem:desu-noto-main}
\end{lemma}

\paragraph{Proof.}
Since $p_i$ enters a view $v$, we have
$p_i.\view(\te{i}{v}) = p_i.\viewp(\te{i}{v})=v$. By the definitions of $\view$
and $\viewp$, $v$ is both the lowest view among the highest $2f+1$ views in
$p_i.\lastViews(\te{i}{v})$, and the lowest view among the highest $f+1$ views
in $p_i.\lastViews(\te{i}{v})$. Hence, $p_i.\lastViews(\te{i}{v})$ includes
$f+1$ entries equal to $v$. Then there exists a time $t' < \te{i}{v}$ at which
some correct process sends $\WISH(v)$. Hence, by Lemma~\ref{lem:wish-attempt},
there exists a time $t \le t' < \te{i}{v}$ at which some correct process
attempts to advance from $v-1$.\qed

\begin{lemma}
For all times $t$ and views $v>0$, if a correct process sends
$\WISH(v)$ at $t$, then there exists a time $t' \le t$
such that some correct process calls $\start$ at $t'$.
\label{lem:wish-start}
\end{lemma}

\paragraph{Proof.}
Consider the earliest time $t_k \le t$ at which some
correct process $p_k$ sends $\WISH(v_k)$ for some view $v_k$. 
By Lemma~\ref{lem:wish-attempt}, there exists a time $t_j \le t_k$ 
at which some correct process attempts to advance from 
$v_k - 1 \ge 0$, and therefore, sends $\WISH(v_k)$ at $t_j$.
Since $t_k$ is the earliest time when this could happen, 
we have $t_j = t_k$. Also, if $v_k - 1 > 0$, 
then $\te{k}{v_k-1}$ is defined, and hence,
by Lemma~\ref{lem:desu-noto-main}, some correct 
process attempts to advance from $v_k - 2$ by 
sending $\WISH(v_k - 1)$ earlier than $t_j = t_k$, which
cannot happen. Thus, at $t_k$, $p_k$ attempts to advance
from view $0$, so that $v_k = 1$ and $\LV{k}{t_k} = 0$.
Assume first that $p_k$ executes the code in line~\ref{line:send2} at $t_k$.
Then $p_k.\timerview$ expires at $t_k$, and hence,
there exists a time $s_k < t_k$ such that 
$p_k.\timerview$ is set at $s_k$. Thus, at $s_k$,
$p_k$ enters a view $>0$. Since $\lvsym$ is non-decreasing,
$\LV{k}{t_k} > 0$, which is a contradiction. Thus, $p_k$ cannot execute 
line~\ref{line:send2} at $t_k$, and has to call $\start$ at this time.\qed

\begin{lemma}%
Global view never skips values:
$\forall t.\, \GV{t} > 0 {\implies} \exists t' < t.\, \GV{t'} = \GV{t}-1$.
\label{lem:gv-noskip}
\end{lemma}

\vspace{-\baselineskip} 
\vspace{-\baselineskip} 

\paragraph{Proof.}
Assume by contradiction that there exists time $t$ such that
$$
\GV{t} > 0 \wedge \forall t' < t.\, \GV{t} \neq \GV{t'} + 1.
$$
Since $\gvsym$ is non-decreasing, the above implies
\begin{equation}
\GV{t} > 0 \wedge \forall t' < t.\, \GV{t} > \GV{t'} + 1.
\label{eq:gv-prev}
\end{equation}
Since $\GV{0} = 0$, (\ref{eq:gv-prev}) implies that $t>0$ and $\GV{t} > 1$.  By
the definition of $\gvsym$, there exists a correct process $p_i$ such that
$\te{i}{\GV{t}} \le t$. Then by Lemma~\ref{lem:desu-noto-main}, there exist a
time $t' < t$ at which some correct process $p_j$ attempted to advance from $\GV{t}-1$. 
Thus, $\LV{j}{t'} = \GV{t}-1$, which by the
definition of $\gvsym$, implies $\GV{t'} \ge \LV{j}{t'} = \GV{t}-1$.  Hence,
$\GV{t} \le \GV{t'} + 1$, which is a contradiction to~(\ref{eq:gv-prev}).\qed

\begin{lemma}
For all views $v$, if a correct process enters $v$,
then $\GV{\tm{v}} = v$.
\label{lem:efirst-gv}
\end{lemma}

\paragraph{Proof.}
By the definition of $\gvsym$, $\GV{\tm{v}} \ge v$. If $\GV{\tm{v}} > v$, then
there exists a view $v' > v$, and a time $t' < \tm{v}$ such that some correct 
process enters $v'$ at time $t'$. Thus, $\GV{t'} \ge v' > v$. 
By Lemma~\ref{lem:gv-noskip}, there exists a time $t'' < t'$, such that
$\GV{t''} = v$, and therefore, $\tm{v} \le t'' < t' < \tm{v}$, which is 
a contradiction. We conclude that $\GV{\tm{v}} = v$, as needed.\qed

\begin{corollary}
  For all views $v, v' > 0$ such that $v \le v'$, if some correct processes
  enter $v$ and $v'$, then $\tm{v} \le \tm{v'}$.
\label{cor:ef-monotone}
\end{corollary}

\paragraph{Proof.}
By Lemma~\ref{lem:efirst-gv}, $v < v'$ implies that $\GV{\tm{v}} \le \GV{\tm{v'}}$.
Since $\gvsym$ is non-decreasing, $\tm{v} \le \tm{v'}$, as needed.\qed

\begin{lemma}
If a correct process $p_i$ sends $\WISH(v)$ at $t$ and 
$\WISH(v')$ at $t' \ge t$, then $v' \ge v$.
\label{lem:wish-grow} 
\end{lemma}

\paragraph{Proof.}
We first state three simple facts that follow directly
from the structure of the code.
First, for all times $t$, 
if a correct process $p_i$ sends $\WISH(v)$ with $v > p_i.\viewp(t)$ 
by executing the code in lines~\ref{line:send1}, 
\ref{line:send2} or~\ref{line:send4} at $t$, 
and $v = p_i.\view(t) + 1 > p_i.\viewp(t)$, then given 
that $p_i.\view(t) \le p_i.\viewp(t)$, we have
\begin{multline}
\forall t.\, \forall v.\, 
(p_i \text{~sends~} \WISH(v) 
\text{~in~lines~\ref{line:send1},~\ref{line:send2},~or~\ref{line:send4}~at~} t)
{\implies}\\
(p_i.\viewp(t) \le v - 1 \iff v = p_i.\view(t) + 1 \iff
p_i.\view(t) = p_i.\viewp(t)).
\label{eq:lem:aux:1}
\end{multline}
Also, since $p_i$ never sends a $\WISH$ message with a view $< \viewp$, we have
\begin{equation}
\forall t.\, \forall v.\, 
(p_i \text{~sends~} \WISH(v) \text{~at~} t) {\implies} v \ge p_i.\viewp(t).
\label{eq:lem:aux:2}
\end{equation}
Finally, since a view sent in a $\WISH$ message is equal to $\viewp$
when either $\timerview$ is enabled or line~\ref{line:send5} is executed, we have
\begin{multline}
\forall t.\, \forall v.\, 
(p_i \text{~sends~} \WISH(v) \text{~at~} t) \wedge ((p_i.\timerview(t)
\text{~is~enabled}) \vee 
(p_i \text{~executes~line~\ref{line:send5}}))\\ 
{\implies}  v = p_i.\viewp(t).
\label{eq:lem:aux:3}
\end{multline}
We now prove the lemma. Suppose that at $t$,
a correct process $p_i$ sends $\WISH(v)$, and
consider a time $t' > t$ such that $p_i$ sends $\WISH(v')$ at $t'$.
We consider two cases:
\begin{itemize}
\item $p_i.\viewp(t) \le v - 1$. Then $v \ge p_i.\viewp(t) + 1 > p_i.\viewp(t)$,
and hence, by~(\ref{eq:lem:aux:3}), $p_i.\timerview(t)$ is disabled and
$p_i$ does not execute the code in line~\ref{line:send5} at $t$. Thus,
at $t$, $p_i$ executes the code in 
lines~\ref{line:send1}, \ref{line:send2}, or~\ref{line:send4}, which
by~(\ref{eq:lem:aux:1}), implies
$$
p_i.\view(t) = p_i.\viewp(t) \wedge v = p_i.\view(t) + 1. 
$$
Since $p_i.\viewp$ is non-decreasing,
$p_i.\viewp(t') \ge p_i.\viewp(t)$. If $p_i.\viewp(t') = p_i.\viewp(t)$,
then $p_i$ does not execute the code in line~\ref{line:send5} at $t$.
Also, since $p_i.\view$ is non-decreasing and 
$p_i.\view(t') \le p_i.\viewp(t')$, we have $p_i.\view(t') = p_i.\viewp(t')$
and $p_i.\view(t) = p_i.\view(t')$. Then since $p_i.\timerview(t)$ is disabled, 
$p_i.\timerview(t')$ is disabled as well. Hence, at $t'$ the process $p_i$ executes
the code in lines~\ref{line:send1}, \ref{line:send2}, or~\ref{line:send4},
which by~(\ref{eq:lem:aux:1}) implies 
$v' = p_i.\view(t') + 1 = p_i.\view(t) + 1 = v$, as needed. 
On the other hand, if $p_i.\viewp(t) < p_i.\viewp(t')$,
then by~(\ref{eq:lem:aux:2}), 
$$
v' \ge p_i.\viewp(t') > p_i.\viewp(t).
$$
Hence,
$$
v' \ge p_i.\viewp(t') \ge p_i.\viewp(t) + 1 = p_i.\view(t) + 1 = v,
$$
as needed.

\item $p_i.\viewp(t) > v - 1$. Then 
$p_i.\viewp(t) \ge v$. Since by~(\ref{eq:lem:aux:2}), 
$v' \ge p_i.\viewp(t')$ and $p_i.\viewp$ is non-decreasing, we have
$v' \ge p_i.\viewp(t') \ge p_i.\viewp(t) \ge v$, as needed.
\end{itemize}
\qed

\begin{lemma}%
If a correct process enters a view $v>0$ and $\tm{v} \ge \GST$,
then for all $v' > v$, no correct process attempts to advance from $v'-1$ before
$\tm{v} + \timeout(v)$.
\label{lem:i-wont-try}
\end{lemma}

\paragraph{Proof.}
Suppose by contradiction that there exists a time 
$t' < \tm{v} + \timeout(v)$ and a correct process $p_i$
such that $p_i$ attempts to advance from $v'-1 > v-1$ at $t'$.
If $p_i$ executes the code in line~\ref{line:send1} at $t'$, then
$\LV{i}{t'} = 0 = v'-1 > v - 1 \ge 0$, which is impossible. 
Thus, at $t'$, the process $p_i$ executes
the code in line~\ref{line:send2}, and 
$\LV{i}{t'} = v' - 1$.
Since $p_i.\timerview$ is not enabled at $t'$, $p_i$ must have
entered $v' - 1$ at least $\timeout(v)$ before $t'$ according
to its local clock. Since
$v' - 1 \ge v$, by Corollary~\ref{cor:ef-monotone},
we have $\tm{v' - 1} \ge \tm{v} \ge \GST$. Therefore,
given that the clocks of all correct processes progress
at the same rate as real time after $\GST$, we get
$$
\tm{v} \le \tm{v' - 1} \le t' - \timeout(v' - 1).
$$
Hence,
$$
t' \ge \tm{v} + \timeout(v' - 1).
$$
Since $\timeout$ is non-decreasing and $v' - 1 \ge v$, 
we have $\timeout(v' - 1) \ge \timeout(v)$ and
$$
t' \ge \tm{v} + \timeout(v' - 1) \ge \tm{v} + \timeout(v),
$$
which contradicts our assumption that $t' < \tm{v} + \timeout(v)$.
Thus, no correct process can attempt to advance from $v'-1$ before 
$\tm{v} + \timeout(v)$, as needed.\qed

\begin{corollary}
Assume a correct process enters a view $v>0$ and $\tm{v} \ge \GST$.
For all views $v' > v$, if there exists a correct process
that enters $v'$, then $\tm{v'} > \tm{v} + \timeout(v)$.
\label{lem:all-by-myself}
\end{corollary}

\vspace{-\baselineskip} 

\paragraph{Proof.}
Since a correct process enters a view $v'>0$, by Lemma~\ref{lem:desu-noto-main},
there exist a time $t < \tm{v'}$ at which some correct process attempts to advance from
$v'-1$. By Lemma~\ref{lem:i-wont-try}, we get $t \ge \tm{v} + \timeout(v)$, so
that $\tm{v'} > t \ge \tm{v} + \timeout(v)$, as required.\qed

\begin{corollary}
Consider a view $v$ and assume that $v$ is entered
by a correct process. If $\tm{v} \ge \GST$, then
a correct process cannot send a $\WISH(v')$ with $v' > v$ 
earlier than $\tm{v} + \timeout(v)$.
\label{lem:upper-efirst}
\end{corollary}

\paragraph{Proof.}
Assume a correct process sends a $\WISH(v')$ with $v' > v$ 
at time $t'$. By Lemma~\ref{lem:wish-attempt},
there exists a time $s \le t'$ such that 
some correct process $p_i$ attempts to advance from $v'-1 > v-1$ at $s$.
By Lemma~\ref{lem:i-wont-try}, $s \ge \tm{v} + \timeout(v)$, 
which implies that $t' \ge s \ge \tm{v} + \timeout(v)$, as
required.\qed

\bigskip

For an arbitrary time $t$, we let  $\postGST(t)$ be a predicate
defined as follows:
$$
\begin{array}{@{}l@{}}
\postGST(t) \iff
(\forall p_i.\, \forall v > 0.\, 
p_i \text{~is~correct~} \implies
\ms
(
\exists s \le t.\, p_i \text{~sends~} \WISH(v) \text{~at~} s 
\implies 
\exists t'.\, \exists v' \ge v.\, 
p_i \text{~sends~} \WISH(v') \text{~at~} t' \wedge 
\GST \le t' \le t
)).
\end{array}
$$

\begin{lemma}
$\tf \ge \GST \implies \forall t \ge \tf.\, \postGST(t)$.
\label{lem:postgst0}
\end{lemma}

\paragraph{Proof.}
Let $t \ge \tf$, and 
consider a correct process $p_i$ and view $v$
such that $p_i$ sends $\WISH(v)$ at time $s \le t$. 
By Lemma~\ref{lem:wish-start}, no correct process can send 
a $\WISH$ message before $\tf$, and therefore, 
$s \ge \tf \ge \GST$. Thus, $t' = s$ and $v' = v$ satisfy 
$p_i \text{~sends~} \WISH(v') \text{~at~} t' \wedge 
\GST \le t' \le t \wedge v' \ge v$, as needed.\qed

\begin{lemma}
$\forall t \ge \GST + \rho.\, \postGST(t)$.
\label{lem:postgst1}
\end{lemma}

\paragraph{Proof.}
Let $t \ge \GST + \rho$, and
consider a correct process $p_i$ and a view $v$
such that $p_i$ sends $\WISH(v)$ at $s \le t$.
If $s \ge \GST$, then choosing $t' = s$ and $v' = v$
validates $\postGST$. 
Suppose that $s < \GST$. Since after $\GST$
the local clock of $p_i$ advances at the same rate as real time,
there exists a time $t_i$ satisfying $\GST \le t_i \le t$ such that
$p_i$ executes the periodic retransmission code in 
lines~\ref{line:retransmit-start}-\ref{line:send4} at $t_i$.
Since $p_i$ already sent a $\WISH$ message at $s < \GST \le t_i$,
and every message sent by a correct process is instantaneously 
received by the sender,
$p_i.\lastViews[i](t_i) > 0$,
and therefore, the code sending $\WISH(v')$ for some view $v'$ is
guaranteed to be reached at $t_i$. 
Since $t_i > s$,  by Lemma~\ref{lem:wish-grow}, $v' \ge v$. Thus, 
choosing $t' = t_i$ validates $\postGST$.
Hence, we get that for all values of $s \le t$, we can find a time $t'$
and a view $v'$ validating $\postGST$, which implies the result.\qed

\begin{lemma}
Let $t \ge \GST$ be a time such that $\postGST(t)$ holds.
Then for all times $t' \ge t$ and views $v$, if a correct process $p_i$ 
sends $\WISH(v)$ at a time $s < t'$, then $p_i$ also sends $\WISH(v')$ with some
$v' \ge v$ at a time $s'$ such that $\GST \le s' \le t'$.
\label{lem:postgst2}
\end{lemma}

\paragraph{Proof.}
If $t \le s< t'$, then since $t\ge \GST$, we have $\GST \le s' < t'$. Thus, 
choosing $s' = s$ validates the lemma. On the other hand, if $s < t$, then
since $\postGST(t)$ holds, there exists a time $s'$ such that 
$\GST \le s' \le t \le t'$ and $p_i$ sends $\WISH(v')$ with $v' \ge v$ at
$s'$. Thus, $s'$ chosen in this way satisfies the required.\qed 

\begin{lemma}
For all $v$, if some correct process enters $v$, and
{\renewcommand{\labelenumi}{(\roman{enumi})}
\renewcommand{\theenumi}{(\roman{enumi})}
\setlength{\leftmargini}{22pt}
\begin{enumerate}
\item $\tm{v} \ge \GST$, \label{lem:GST-bound2-generic:1}
\item $\postGST(\tm{v})$ holds, and \label{lem:GST-bound2-generic:2}
\item $\timeout(v) > 2\delta$,
\end{enumerate}}
\noindent then all correct processes enter $v$ and
$\tl{v} \le \tm{v} + 2\delta$.
\label{lem:GST-bound2}
\end{lemma}

\paragraph{Proof.}
Since $\tm{v} \ge \GST$ and $\timeout(v) > 2\delta$, 
by Corollary~\ref{lem:upper-efirst}, we have:
\smallskip
\smallskip
{\renewcommand{\labelenumi}{(\roman{enumi})}
\renewcommand{\theenumi}{(\roman{enumi})}
\setlength{\leftmargini}{22pt}
\begin{enumerate}
\setcounter{enumi}{3}
\item no correct process sends $\WISH(v')$ with $v' > v$
until after $\tm{v} + 2\delta$. \label{lem:GST-bound2-generic:3}
\end{enumerate}}
\smallskip
\smallskip

Let $p_i$ be a correct process that enters $v$ at $\tm{v}$.
By the view entry condition,
$p_l.\view(\tm{v}) = v$, and therefore
$p_i.\lastViews(\tm{v})$ includes $2f+1$ entries $\ge v$.
At least $f+1$ of these entries belong to correct processes,
and by~\ref{lem:GST-bound2-generic:3},
none of them can be $> v$.
Hence, there exists a set $C$ of $f+1$ correct processes, each of which sends 
$\WISH(v)$ to all processes before $\tm{v}$.

Since $\postGST(\tm{v})$ holds, any $p_j \in C$ also sends a message
$\WISH(v')$ with $v' \ge v$ at some time $t'_j$ such that
$\GST \le t_j' < \tm{v}$.  Then by~\ref{lem:GST-bound2-generic:3} we have
$v'=v$. It follows that each $p_j \in C$ is guaranteed to send a copy of
$\WISH(v)$ to all correct processes between $\GST$ and $\tm{v}$. Since all
messages sent by correct processes after $\GST$ are guaranteed to be received by
all correct processes within $\delta$ of their transmission, by
$\tm{v} + \delta$ all correct processes will receive $\WISH(v)$ from at least
$f+1$ distinct correct processes.

Consider an arbitrary correct process $p_j$ and let $t_j \le \tm{v} + \delta$ be
the earliest time by which $p_j$ receives $\WISH(v)$ from $f+1$
correct processes. By~\ref{lem:GST-bound2-generic:3}, no correct process sends
$\WISH(v')$ with $v' > v$ before $t_j < \tm{v} + 2\delta$. Thus, $p_j.\lastViews(t_j)$
includes at least $f+1$ entries equal to $v$ and at most
$f$ entries $>v$, so that $p_j.\viewp(t_j) = v$.
If $p_j.\prevvp(t_j) < v$, then at $t_j$ the process $p_j$ sends $\WISH(v)$
to all processes by executing the code in line~\ref{line:send5}.
Otherwise, $p_j$ must have already sent $\WISH(v)$ to all processes at some time 
before $t_j \le \tm{v} + \delta$. Thus, in both cases $p_j$ sends $\WISH(v)$ with 
to all processes no later than $\tm{v} + \delta$.
Since $\postGST(\tm{v})$ holds and $\tm{v} \ge \GST$, 
by~\ref{lem:GST-bound2-generic:3} and Lemma~\ref{lem:postgst2}, 
$p_j$ also sends $\WISH(v)$ to all processes
sometime between $\GST$ and $\tm{v} + \delta$.
Hence, all correct processes are guaranteed to send $\WISH(v)$ to all
correct processes between $\GST$ and $\tm{v} + \delta$. 

Consider an arbitrary correct process $p_k$ and let $t_k \le \tm{v} + 2\delta$
be the earliest time by which $p_k$ receives $\WISH(v)$ from all correct
processes. Then by~\ref{lem:GST-bound2-generic:3}, at $t_k$ the array
$p_k.\lastViews$ will contain include at least $2f+1$ entries equal to $v$, and
at most $f$ entries greater than $v$. Therefore,
$p_k.\view(t_k) = p_k.\viewp(t_k) = v$, so that $p_k$ enters $v$ no later than 
$t_k \le \tm{v}+2\delta$. We have thus shown that by $\tm{v}+2\delta$, all
correct processes will enter $v$, as required.\qed

\bigskip

Since by Lemma~\ref{lem:postgst1}, $\postGST(t)$ holds for all
$t > \GST + \rho$, from Lemma~\ref{lem:GST-bound2} we get
\begin{corollary}
For all views $v$, if a correct process enters $v$, 
$\tm{v} > \GST + \rho$, and $\timeout(v) > 2\delta$, 
then all correct processes enter $v$ and
$\tl{v} \le \tm{v} + 2\delta$.
\label{lem:GST-bound2-final}
\end{corollary}

\begin{lemma}
Let $t \ge \GST$ be a time such that $\postGST(t)$ holds,
and $T\ge t$ be a time such that:
{\renewcommand{\labelenumi}{(\roman{enumi})}
\renewcommand{\theenumi}{(\roman{enumi})}
\setlength{\leftmargini}{22pt}
\begin{enumerate}
\item no correct process sends $\WISH(v)$ with $v > \GV{t} + 1$ before
  $T + 2\delta$; and \label{lem:simple-bound-generic:item2}
\item there exists a time $s \le T+\delta$ such that all correct processes send
  $\WISH(\GV{t}+1)$ to all processes no later than at $s$.
  \label{lem:simple-bound-generic:item3}
\end{enumerate}}
\noindent Then all correct processes enter $\GV{t}+1$ and
$\tl{\GV{t} + 1} \le s + \delta$.
\label{lem:simple-bound-generic}
\end{lemma}

\paragraph{Proof.}
Fix an arbitrary correct process $p_i$ that sends
$\WISH(\GV{t}+1)$ to all 
processes at time $t_i \le s \le T+\delta$, as stipulated by~\ref{lem:simple-bound-generic:item3}.
Since $t \ge \GST$, $\postGST(t)$, and $T+\delta \ge t$, by 
Lemma~\ref{lem:postgst2} and~\ref{lem:simple-bound-generic:item2}
there exists a time $t_i'$ such that $\GST \le t_i' \le s$ and at $t_i'$ the process
$p_i$ sends $\WISH(\GV{t} + 1)$ to all processes.
Since $t_i' \ge \GST$, all correct processes 
receive $\WISH(\GV{t}+1)$ from all correct processes 
no later than at $t_i' + \delta \le s + \delta$. 

Consider an arbitrary correct process $p_j$ and let 
$t_j \le s + \delta \le T + 2\delta$ be the earliest time by which 
$p_j$ receives receives $\WISH(\GV{t}+1)$ from all correct processes.
Thus, at $t_j$, the entries of all correct processes 
in $p_j.\lastViews$ are occupied by views $\ge \GV{t} + 1$.
By~\ref{lem:simple-bound-generic:item2}
none of the entries in $p_j.\lastViews(t_j)$ belonging to correct processes
are occupied by views $> \GV{t} + 1$. Thus, each such entry
in $p_j.\lastViews(t_j)$ stores $\GV{t} +1$.
Since at least $2f+1$ entries in $p_j.\lastViews$ belong to correct processes,
$p_j.\view(t_j) = p_j.\viewp(t_j) = \GV{t} + 1$. Therefore,
$p_j$ enters $\GV{t} + 1$ no later than $t_j \le s + \delta$.
Thus, we conclude that all correct processes enter $\GV{t} + 1$ and
$\tl{\GV{t} + 1} \le s + \delta$, as needed.\qed

\begin{lemma}
Let $t\ge \GST$ be a time such that $\postGST(t)$ holds,
$t_1 = \max(\tfk{f+1}, t)$, 
$t_2 = \max(\ts, t)$, and 
$T = \max(\min(t_1, t_2 - \delta), t)$.
Assume that
{\renewcommand{\labelenumi}{(\roman{enumi})}
\renewcommand{\theenumi}{(\roman{enumi})}
\setlength{\leftmargini}{22pt}
\begin{enumerate}
\item $\GV{t} = 0$, and
\item no correct process can send $\WISH(v)$ with $v > 1$ before
$T + 2\delta$. \label{item:upper}
\end{enumerate}}
\noindent Then all correct processes enter view $1$ and
$\tl{1} \le \min(t_1+2\delta, t_2+\delta)$.
\label{lem:bound0}
\end{lemma}

\paragraph{Proof.}
We consider three cases:

\begin{itemize}

\item $t \le \tfk{f+1} \le \ts$, i.e., $t_1 = \tfk{f+1}$ and $t_2 = \ts$.
Hence $T = \max(\min(\tfk{f+1}, \ts - \delta), t)$. 
We consider two cases:

\begin{itemize}

\item $\tfk{f+1} + \delta < \ts$. Hence, $\tfk{f+1} < \ts - \delta$, 
and therefore,
\begin{equation}
T = \max(\tfk{f+1}, t) = \tfk{f+1}.
\label{eq:upper-1}
\end{equation}
Let $C$ be the set of the 
$f+1$ correct processes $p_i$ calling $\start()$ at $t_i \le \tfk{f+1}$.
If $p_i.\viewp(t_i) = 0$, then at $t_i$, $p_i$ sends $\WISH(1)$ to all
processes by executing the code in line~\ref{line:send1}. Otherwise,
by~(\ref{eq:upper-1}) and~\ref{item:upper}, $p_i.\viewp(t_i) = 1$, and
$p_i$ sent $\WISH(1)$ when $p_i.\viewp$ first 
became equal to $1$ at some time $s_i < t_i \le \tfk{f+1}$.
Since $t_i \le \tfk{f+1}$, in both cases, by Lemma~\ref{lem:postgst2},
$p_i$ sends $\WISH(v)$ with $v \ge 1$ sometime between 
$\GST$ and $\tfk{f+1}$. By~(\ref{eq:upper-1}) and~\ref{item:upper},
we have $v=1$. Thus, we get that all processes in $C$ send $\WISH(1)$ to all 
processes in-between $\GST$ and $\tfk{f+1}$. It follows that 
all correct processes receive all these $\WISH(1)$ messages
no later than $\tfk{f+1}+\delta$. Consider a correct process
$p_j$, and let $\GST \le t_j \le \tfk{f+1}+\delta$ be the earliest time by which 
$p_j$ receives the $\WISH(1)$ 
messages sent by the processes in $C$ in-between $\GST$ and $\tfk{f+1}$.
By~(\ref{eq:upper-1}) and~\ref{item:upper}, 
$p_j.\lastViews[k] = 1$ for all $p_k \in C$, and
there are at most $f$ entries in $p_j.\lastViews(t_j)$ 
occupied by views $>1$. Thus, $p_j.\viewp(t_j) = 1$. 
If $p_j.\viewp(t_j) > p_j.\prevvp(t_j)$, then $p_j$ sends $\WISH(1)$ to 
all processes at $t_j\ge \GST$ by executing the code in line~\ref{line:send5}.
Otherwise, $p_j$ sent $\WISH(1)$ when $p_j.\viewp$ first became equal 
to $1$ sometime before $t_j$. Thus, we get that $p_j$ sends $\WISH(1)$ to all 
processes no later than $\tfk{f+1} + \delta \le T + \delta$. 
By~Lemma~\ref{lem:simple-bound-generic},
all correct processes enter view $1$, and 
$\tl{1} \le \tfk{f+1} + 2\delta = t_1 + 2\delta$.
Since $t_1 + \delta < t_2$, we also have 
$\tl{1} \le \min(t_1 + 2\delta, t_2 + \delta)$, as needed.

\item $\tfk{f+1} + \delta \ge \ts$. Hence, $\tfk{f+1} \ge \ts - \delta$, 
and therefore,
$$
T = \max(\min(\tfk{f+1}, \ts - \delta), t) = \max(\ts - \delta, t).
$$
Hence,
\begin{equation}
T + 2\delta \ge \ts + \delta
\label{eq:upper-2}
\end{equation}
and
\begin{equation}
\ts \le T + \delta.
\label{eq:upper-22}
\end{equation}
Let $p_i$ be a correct process calling $\start()$ at $t_i \le \ts$.
If $p_i.\viewp(t_i) = 0$, then at $t_i$, $p_i$ sends $\WISH(1)$ to all
processes by executing the code in line~\ref{line:send1}. 
Otherwise, by~(\ref{eq:upper-2}) and~\ref{item:upper}, 
$p_i.\viewp(t_i) = 1$, and
$p_i$ sent $\WISH(1)$ when $p_i.\viewp$ first 
became equal to $1$ at some time $s_i < t_i \le \ts$.
Since by~(\ref{eq:upper-22}), $\ts \le T + \delta$, we have
that all correct processes send $\WISH(1)$ to all processes
no later than $\ts \le T+\delta$. Thus, by Lemma~\ref{lem:simple-bound-generic},
all correct processes enter view $1$, and
$\tl{1} \le \ts + \delta = t_2 + \delta$. 
Since $t_1 + \delta \ge t_2$, 
we also have $\tl{1} \le \min(t_1 + 2\delta, t_2 + \delta)$,
as needed.
\end{itemize}

\item $\tfk{f+1} < t \le \ts$. Hence, $t_1 = t$ and $t_2 = \ts$,
and therefore, $T = \max(\min(t, \ts - \delta), t)$. We consider two cases:

\begin{itemize}

\item $t + \delta < \ts$. Hence, $t < \ts - \delta$, and therefore
\begin{equation}
T = t.
\label{eq:upper-3}
\end{equation}
Let $C$ be the set of the 
$f+1$ correct processes $p_i$ calling $\start()$ at $t_i < t$.
If $p_i.\viewp(t_i) = 0$, then at $t_i$, $p_i$ sends $\WISH(1)$ to all
processes by executing the code in line~\ref{line:send1}. Otherwise,
by~(\ref{eq:upper-3}) and~\ref{item:upper},
$p_i.\viewp(t_i) = 1$, and $p_i$ sent $\WISH(1)$ when $p_i.\viewp$ first 
became equal to $1$ at sometime before $t_i$. Since $t_i < t$, and
$\postGST(t)$ holds, there exists a time $s_i$ such that
$\GST \le s_i < t$ and at $s_i$, $p_i$ sends
$\WISH(v)$ with $v\ge 1$ to all processes.
By~(\ref{eq:upper-3}) and~\ref{item:upper}, we have
$v=1$. Thus, we get that all processes in $C$ send
$\WISH(1)$ to all processes in-between $\GST$ and $t$.
It follows that all correct processes receive all these $\WISH(1)$ messages
no later than $t+\delta$. Consider a correct process
$p_j$, and let $\GST \le t_j \le t+\delta$ be the earliest time by which 
$p_j$ receives the $\WISH(1)$ 
messages sent by the processes in $C$ in-between $\GST$ and $t$.
By~(\ref{eq:upper-3}) and~\ref{item:upper},
$p_j.\lastViews[k] = 1$ for all $p_k \in C$, and
there are at most $f$ entries in $p_j.\lastViews(t_j)$ 
occupied by views $>1$. Thus, $p_j.\viewp(t_j) = 1$. 
If $p_j.\viewp(t_j) > p_j.\prevvp(t_j)$, then $p_j$ sends $\WISH(1)$ to 
all processes at $t_j\ge \GST$ by executing the code in line~\ref{line:send5}.
Otherwise, $p_j$ sent $\WISH(1)$ when $p_j.\viewp$ first became equal 
to $1$ sometime before $t_j$. 
Since $t+\delta \ge t_j$, by
Thus, we get that $p_j$ sends $\WISH(1)$ to all 
processes no later than $t + \delta \le T + \delta$. 
From Lemma~\ref{lem:simple-bound-generic},
all correct processes enter view $1$ and 
$\tl{1} \le t + 2\delta = t_1 + 2\delta$.
Since $t_1 + \delta < t_2$, 
we also have $\tl{1} \le \min(t_1 + 2\delta, t_2+\delta)$, as needed.

\item $t + \delta \ge \ts$. Hence, $t \ge \ts - \delta$, and therefore
$$
T = \max(\min(t, \ts - \delta), t) = \max(\ts - \delta, t).
$$
Hence,
\begin{equation}
T + 2\delta \ge \ts + \delta
\label{eq:upper-4}
\end{equation}
and
\begin{equation}
\ts \le T + \delta.
\label{eq:upper-42}
\end{equation}
Let $p_i$ be a correct process calling $\start()$ at $t_i \le \ts$.
If $p_i.\viewp(t_i) = 0$, then at $t_i$, $p_i$ sends $\WISH(1)$ to all
processes by executing the code in line~\ref{line:send1}. Otherwise,
by~(\ref{eq:upper-4}) and~\ref{item:upper},
$p_i.\viewp(t_i) = 1$, and
$p_i$ sent $\WISH(1)$ when $p_i.\viewp$ first 
became equal to $1$ at some time $s_i < t_i \le \ts$.
Since by~(\ref{eq:upper-42}), $\ts \le T + \delta$, we have
that all correct processes send $\WISH(1)$ to all processes
no later than $\ts \le T+\delta$. Thus, by Lemma~\ref{lem:simple-bound-generic},
all correct processes enter view $1$, and
$\tl{1} \le \ts + \delta = t_2 + \delta$. 
Since $t_2 \le t_1 + \delta$, we also have
$\tl{1} \le \min(t_1 + 2\delta, t_2 + \delta)$, as needed.
\end{itemize}

\item $\tfk{f+1} \le \ts < t$. Hence, $t_1 = t_2 = t$, 
and therefore,
\begin{equation}
T = \max(\min(t_1, t_2 - \delta), t) = t.
\label{eq:upper-5}
\end{equation}
Let $p_i$ be a correct process calling $\start()$ at $t_i \le \ts < t$.
If $p_i.\viewp(t_i) = 0$, then at $t_i$, $p_i$ sends $\WISH(1)$ to all
processes by executing the code in line~\ref{line:send1}. Otherwise,
by~(\ref{eq:upper-5}) and~\ref{item:upper},
$p_i.\viewp(t_i) = 1$, and
$p_i$ sent $\WISH(1)$ when $p_i.\viewp$ first 
became equal to $1$ at some time $s_i < t_i \le \ts < t$.
Since $\postGST(t)$ holds, in both cases, 
$p_i$ sends $\WISH(v)$ with $v \ge 1$ sometime between 
$\GST$ and $t$.
By~(\ref{eq:upper-4}) and~\ref{item:upper}, we have $v=1$.
Thus, we get that all correct processes send $\WISH(1)$ to all 
processes in-between $\GST$ and $t$. By 
Lemma~\ref{lem:simple-bound-generic}, this implies that
all correct processes enter view $1$
and $\tl{1} \le t + \delta = t_2 + \delta$.
Since $t_1 = t_2$, we also have
$\tl{1} \le \min(t_1 + 2\delta, t_2 + \delta)$, as needed.
\end{itemize}
Thus, we get that in all three cases above, all correct processes
enter view $1$, and $\tl{1} \le \min(t_1 + 2\delta, t_2 + \delta)$,
as required.\qed

\begin{lemma}
Let $t \ge \GST$ be a time such that $\postGST(t)$ holds,
$T = t + \timeout(\GV{t}) + \delta$,
and assume $\GV{t} > 0$, and
no correct process sends $\WISH(v)$ with $v > \GV{t} + 1$
before $T + \delta$.
Then all correct processes send $\WISH(\GV{t} + 1)$ 
to all processes no later than at $T+\delta$.
\label{lem:all-send-wish1-generic}
\end{lemma}

\paragraph{Proof}
Since $\GV{t} > 0$, 
the definition of $\gvsym$ implies that there exists a correct process $p_l$ such 
that $p_l$ entered $\GV{t}$ and $\te{l}{\GV{t}} \le t$. By the view entry condition,
$p_l.\view(\te{l}{\GV{t}}) = \GV{t}$, and therefore
$p_l.\lastViews(\te{l}{\GV{t}})$ includes $2f+1$ entries 
$\ge \GV{t}$. Since $f+1$ of these entries belong to correct processes, 
there exists a set $C$ of $f+1$ correct processes, each of which sends 
$\WISH(v)$  with $v\ge \GV{t}$ to all processes before 
$\te{l}{\GV{t}} \le t$. 
Since  $\postGST(t)$ holds, $p_i$ sends $\WISH(v')$ with $v'\ge \GV{t}$ sometime
between $\GST$ and $t$. Since no correct process sends $\WISH(v')$ 
with $v' > \GV{t}+1$ before $t <  t + \timeout(\GV{t}) + 2\delta = T + \delta$, 
we have:
$$
\forall p_i \in C.\,
\exists t_i.\, \exists v_i'.\, 
p_i \text{~sends~} \WISH(v_i') \text{~at~} t_i \wedge 
\GST \le t_i < t \wedge v_i' \in \{\GV{t}, \GV{t}+1\}.
$$
Since after $\GST$ every message sent by a correct process is received
by all correct processes within $\delta$ of its transmission, 
the above implies that by $t + \delta$ every correct process receives a
$\WISH(v)$ with $v \in \{\GV{t}, \GV{t}+1\}$ from each process in $C$. 

Consider an arbitrary correct process $p_j$ and let $t_j \le t + \delta$ be the
earliest time by which $p_j$ receives $\WISH(v)$ with $v \in \{\GV{t},
\GV{t}+1\}$ from each process in $C$. 
Since $t_j \le t + \delta < T+\delta$ and 
no correct process sends $\WISH(v)$ with $v > \GV{t} + 1$
before $T + \delta$, we get that
for all processes $p_i\in C$, 
$p_j.\lastViews[i](t_j) \in \{\GV{t}, \GV{t}+1\}$, and
for all correct processes $p_k$, 
$p_j.\lastViews[k](t_j) \le \GV{t}+1$. Since $|C|=f+1$, this
implies $p_j.\viewp(t_j)\in \{\GV{t}, \GV{t}+1\}$.
If $p_j.\viewp(t_j) > p_j.\prevvp(t_j)$, 
then at $t_j$ the process $p_j$ sends $\WISH(v)$
with $v = p_j.\viewp(t_j) \in \{\GV{t}, \GV{t}+1\}$ 
by executing the code in line~\ref{line:send5}. 
Otherwise, $p_k$ must have already sent $\WISH(v)$ with 
$v = p_j.\viewp(t_j) \in \{\GV{t}, \GV{t}+1\}$ to all processes at some time
before $t_j \le t + \delta$. Thus, in both cases $p_j$ sends $\WISH(v)$ with 
$v \in \{\GV{t}, \GV{t}+1\}$ to all processes no later than $t + \delta$.
Then Lemma~\ref{lem:postgst2} implies that 
$p_j$ also sends $\WISH(v')$ with $v'\in \{\GV{t}, \GV{t}+1\}$ to all processes
sometime between $\GST$ and $t+\delta$, inclusive.

Consider an arbitrary correct process $p_k$ and let $t_k$ be the
earliest time by which $p_k$ receives $\WISH(v)$ with $v \in
\{\GV{t}, \GV{t}+1\}$ from each correct process, with the message
being sent between $\GST$ and $t+\delta$. Then 
\begin{equation}
\label{ineq-t-k}
t_k \le t + 2\delta = (t + \timeout(\GV{t}) + \delta) +
\delta - \timeout(\GV{t}) = T + \delta - \timeout(\GV{t}).
\end{equation}
Since no correct process sends $\WISH(v)$ with $v > \GV{t} + 1$
before $T + \delta$, for all correct processes
$p_j$, we have $p_k.\lastViews[j](t_k)\in \{\GV{t}, \GV{t}+1\}$. Since 
there are $2f+1$ correct processes, by the definitions
of $\view$ and $\viewp$, 
$p_k.\view(t_k) \in \{\GV{t}, \GV{t}+1\}$ and
$p_k.\viewp(t_k) \in \{\GV{t}, \GV{t}+1\}$. 
Thus, given that $p_k.\view(t_k) \le p_k.\viewp(t_k)$, there are three cases
to consider: 
\emph{(i)} $p_k.\view(t_k) = \GV{t} \wedge p_k.\viewp(t_k) = \GV{t} +1$;
\emph{(ii)} $p_k.\view(t_k) = p_k.\viewp(t_k) = \GV{t} + 1$; and
\emph{(iii)} $p_k.\view(t_k) = p_k.\viewp(t_k) = \GV{t}$;

Suppose first that either \emph{(i)} or \emph{(ii)} holds.
Then $p_k.\viewp(t_k) = \GV{t} + 1$ and therefore, $p_k$
either sends $\WISH(\GV{t}+1)$ by executing line~\ref{line:send5}
at $t_k$, or sent it when $p_k.\viewp$ first became equal to $\GV{t} + 1$ 
sometime before $t_k$. Then $p_k$ sends $\WISH(\GV{t}+1)$
to all correct processes no later than $t_k$, and by~(\ref{ineq-t-k}) we have 
$t_k \le T + \delta$, as required.

Suppose now that \emph{(iii)} holds. Then $p_k$ enters $\GV{t}$ at
$\te{k}{\GV{t}} \le t_k$ and starts $p_k.\timerview$ for the duration of
$\timeout(\GV{t})$. Since $t_k > \GST$, and the clocks of the correct processes
advance at the same rate as real time after $\GST$, $p_k.\timerview$ cannot last
past $t_k + \timeout(\GV{t})$. Let $s_k$ be the time at which $p_k.\timerview$
either expires or is stopped prematurely by executing the code in
line~\ref{line:timer-stop}; then
$\te{k}{\GV{t}} < s_k \le t_k + \timeout(\GV{t})$.
We consider two cases.
\begin{itemize}
\item
$p_k.\timerview$ expires at $s_k$, so that at this
time $p_k$ executes the code in lines~\ref{line:timer-exp1}-\ref{line:send2}. 
Since~(\ref{ineq-t-k}) implies $t_k + \timeout(\GV{t}) \le T+\delta$, we get
\begin{equation}
s_k \le T+\delta.
\label{eq:sktd}
\end{equation}
We consider two cases:
\begin{itemize}
\item $s_k < t_k$. 
We have 
\begin{equation} 
\label{inbetween-enter-t-k}
p_k.\view(\te{k}{\GV{t}}) = p_k.\viewp(\te{k}{\GV{t}}) =
p_k.\view(t_k) = p_k.\viewp(t_k) = \GV{t}.
\end{equation} 
Since $p_k.\view$ and $p_k.\viewp$ are non-decreasing, this implies
$p_k.\view(s_k) = p_k.\viewp(s_k) = \GV{t}$. Then
$\max(p_k.\view(s_k)+1, p_k.\viewp(s_k)) = p_k.\view(s_k)+1 = \GV{t}+1$,
and thus, $p_k$ sends $\WISH(\GV{t} + 1)$ to all processes at
$s_k$. By~(\ref{eq:sktd}) this implies the required.
\item $t_k \le s_k \le t_k + \timeout(\GV{t})$. 
Since no correct process sends $\WISH(v)$ with $v > \GV{t} + 1$ 
before $T + \delta$, from~(\ref{eq:sktd}) we get 
$p_k.\view(s_k) \le p_k.\viewp(s_k) \le \GV{t} + 1$. 
Since $p_k.\view(t_k) = p_k.\viewp(t_k) = \GV{t}$
and both $p_k.\view$ and $p_k.\viewp$ are non-decreasing,
$t_k \le s_k$ implies 
$$
\begin{array}{@{}l@{}}
p_k.\view(s_k) = \GV{t} \wedge p_k.\viewp(s_k) \in \{\GV{t}, \GV{t} + 1\}
\vee{}
\ms
p_k.\view(s_k) = p_k.\viewp(s_k) = \GV{t}+1.
\end{array}
$$
If the first disjunct holds, then $\max(p_k.\view(s_k)+1, p_k.\viewp(s_k)) = \GV{t}+1$,
and therefore $p_k$ sends $\WISH(\GV{t}+1)$ to all processes at $s_k$.
Otherwise, $p_k$ enters $\GV{t}+1$ prior to the expiration
of $p_k.\timerview$ at $s_k$, which is impossible. Hence, 
in both cases $p_k$ sends $\WISH(\GV{t}+1)$ to all processes at $s_k$, 
which by~(\ref{eq:sktd}) implies the required.
\end{itemize}
\item
$p_k.\timerview$ is stopped prematurely at $s_k$,
by executing the code in line~\ref{line:timer-stop}. Then
the condition in line~\ref{line:enter-condition} is true at $s_k$,
so that
\begin{equation}
p_k.\viewp(s_k) = p_k.\view(s_k) \wedge p_k.\view(s_k) > p_k.\prevv(s_k).
\label{eq:entry}
\end{equation}
We consider two cases.
\begin{itemize}
\item
$s_k < t_k$. We again have~(\ref{inbetween-enter-t-k}), so that 
$p_k.\view(s_k) = p_k.\viewp(s_k) = \GV{t}$. But
since $p_k$ enters $\GV{t}$ prior to $s_k$, we have $p_k.\prevv(s_k) = \GV{t}$,
contradicting~(\ref{eq:entry}). Hence, this case is impossible.
\item
$s_k \ge t_k$. Since $p_k.\view(t_k) = p_k.\viewp(t_k) = \GV{t}$ and both
$p_k.\view$ and $p_k.\viewp$ are non-decreasing, (\ref{eq:entry}) implies
that $p_k.\view(s_k) = p_k.\viewp(s_k) \ge \GV{t}+1$. 
Since no correct process 
sends $\WISH(v)$ with $v > \GV{t} + 1$ 
before $T + \delta$, from~(\ref{eq:sktd}) we get
$p_k.\view(s_k) = p_k.\viewp(s_k) = \GV{t}+1$. 
If $p_k.\viewp(s_k) > p_k.\prevvp(s_k)$, then 
the condition in line~\ref{line:send5} is true at $s_k$, and therefore,
$p_k$ sends $\WISH(p_k.\viewp(s_k)) = \WISH(\GV{t}+1)$ to all processes at $s_k$.
Otherwise, $p_k.\viewp(s_k) = p_k.\prevvp(s_k)$, which implies
that $p_k$ sent $\WISH(\GV{t}+1)$ when $p_k.\viewp$ first became
equal to $\GV{t}+1$ sometime before $s_k$.
Thus, in both cases $p_k$ sends $\WISH(\GV{t}+1)$ by $s_k$, and the required
follows from~(\ref{eq:sktd}).
\end{itemize}
\end{itemize}
\qed

\begin{lemma}%
Global view keeps increasing
$\forall t.\, \exists t' > t.\, \GV{t'} > \GV{t}$.
\label{lem:gv-live}
\end{lemma}

\paragraph{Proof.}
Assume by contradiction that there exists a time $t$ such that
for all $t' \ge t$, $\GV{t'} \le \GV{t}$. Since $\gvsym$ is non-decreasing,
this implies that for all $t' \ge t$, $\GV{t'} = \GV{t}$, and
for all times $t'' < t$, $\GV{t''} \le \GV{t}$. Thus, we have
\begin{equation}
\forall t.\, \forall v \ge \GV{t} + 1.\, 
\neg (p_i \text{~enters~} v \text{~at~} t \wedge 
p_i \text{~is~correct}).
\label{eq:enter-never}
\end{equation}
Furthermore, if there is a correct process that sends 
$\WISH(v)$ with $v > \GV{t} + 1$ at any time $s$, then 
by Lemma~\ref{lem:wish-attempt},
a correct process $p_i$ attempts to advance from $v - 1 > \GV{t}$ at 
some time $s' \le s$. 
Thus, $\LV{i}{s} = \LV{i}{s'} = v - 1 \ge \GV{t} + 1$, and therefore,
$\GV{s} \ge \GV{t} + 1$. Hence, by Lemma~\ref{lem:gv-noskip},
some correct process must enter $\GV{t} + 1$,
contradicting~(\ref{eq:enter-never}).
Thus, we have 
\begin{equation}
\forall t.\, \forall v > \GV{t} + 1.\, 
\neg (p_i \text{~sends~} \WISH(v) \text{~at~} t \wedge 
p_i \text{~is~correct}).
\label{eq:upper-never}
\end{equation}
Since we assume that all correct processes eventually call $\start()$,
there exists a time $t^* = \max\{t, \GST + \rho, \ts\}$. 
Our choice of $t^*$ implies that $t^* \ge \GST + \rho$,  
and therefore, by Lemma~\ref{lem:postgst1} we have:
\begin{equation}
t^* \ge \GST \wedge \postGST(t^*).
\label{eq:live-precon}
\end{equation}
Since $t^* \ge t$, we also have $\GV{t^*} = \GV{t}$. 

Suppose first that $\GV{t} = 0$. Let 
$$
\begin{array}{@{}l@{}}
t_1 = \max(\tfk{f+1}, t^*) = t^*;
\ms
t_2 = \max(\ts, t^*) = t^*;
\ms
T = \max(\min(t_1, t_2 - \delta), t^*) = 
\max(\min(t^*, t^* -\delta), t^*) = t^*.
\end{array}
$$
By~(\ref{eq:upper-never}), no correct process
can send $\WISH(v)$ with $v > 1$ before $T + 2\delta$.
Since~(\ref{eq:live-precon}) holds, by Lemma~\ref{lem:bound0}, 
all correct processes enter
view $1$, which is a contradiction to~(\ref{eq:enter-never}).

Suppose that $\GV{t} > 0$.
Let $T = t^* + \timeout(\GV{t^*}) + \delta$. 
By~(\ref{eq:upper-never}),
no correct process sends  $\WISH(v)$ with $v > \GV{t^*} + 1$
before $T + 2\delta$. Thus, (\ref{eq:live-precon}), and
Lemma~\ref{lem:all-send-wish1-generic} imply
that all correct processes send $\WISH(\GV{t^*} + 1)$ 
to all processes no later than $T+\delta$. 
Since~(\ref{eq:live-precon}) holds,
by Lemma~\ref{lem:simple-bound-generic}, 
all correct processes enter $\GV{t^*}+1$ by $T+2\delta$,
which is a contradiction to~(\ref{eq:enter-never}).\qed

\begin{lemma}
Assume a correct process enters a view $v$, 
$\tm{v} \ge \GST$, $\postGST(\tm{v})$ holds, and
$\timeout(v) > 2\delta$. Then 
all correct processes enter the view $v+1$ and
$\tl{v+1} \le \tl{v}+\timeout(v) + \delta$.
\label{lem:after-GST-bound-final-gen}
\end{lemma}

\paragraph{Proof.}
If some correct process enters the view 
$v+1$ before $T = \tl{v} + \timeout(v) -\delta$, then by Lemma~\ref{lem:GST-bound2}, 
all correct processes enter the view $v+1$ and
$$
\tl{v+1} \le \tm{v+1} + 2\delta \le T + 2\delta = 
\tl{v} + \timeout(v) -\delta + 2\delta = 
\tl{v} + \timeout(v) +\delta,
$$
as required.

Suppose that no correct process enters $v+1$ before $T$.
We have $T = \tm{v} + \timeout(v) - \delta > \tm{v} \ge \GST$.
By Lemmas~\ref{lem:gv-live} and~\ref{lem:gv-noskip}, some correct process eventually enters
$v+1$, and therefore, by Corollary~\ref{lem:upper-efirst}, $T \ge \GST$ implies
that no correct process can send $\WISH(v')$ for any $v' > v + 1$
earlier than $T + \timeout(v+1)$.
Thus, given that $\timeout(v+1) \ge \timeout(v) > 2\delta$, 
we get:
\begin{equation}
\mbox{no correct process sends $\WISH(v')$ with $v' > v + 1$
before $T + 2\delta$.} \label{lem:after-GST-bound-final-gen:1}
\end{equation}

By Lemma~\ref{lem:GST-bound2}, all correct
processes enter $v$.
Let $p_i$ be a correct process that enters $v$ at
$\te{i}{v}$; at this moment $p_i$ starts $p_i.\timerview$ for the duration of
$\timeout(v)$. Since $\te{i}{v} \ge \tm{v} \ge \GST$,
and the clocks of the correct processes
advance at the same rate as real time after $\GST$, $p_i.\timerview$ cannot last
past $\te{i}{v} + \timeout(v) \le \tl{v} + \timeout(v)$. 
Let $s_i$ be the time at which $p_i.\timerview$
either expires or is stopped prematurely by executing the code in
line~\ref{line:timer-stop}; then
$\te{i}{v} < s_i \le \tl{v} + \timeout(v)$, and therefore,
\begin{equation}
s_i \le \tl{v} + \timeout(v) =  (\tl{v} + \timeout(v) -\delta) + \delta = 
T + \delta.
\label{eq:sitd}
\end{equation}
We consider two cases.
\begin{itemize}
\item
$p_i.\timerview$ expires at $s_i$, so that at this
time $p_i$ executes the code in 
lines~\ref{line:timer-exp1}-\ref{line:send2}. 
Since from~(\ref{lem:after-GST-bound-final-gen:1}), 
no correct process sends $\WISH(v')$ with $v' > v + 1$ 
before $T + \delta$, from~(\ref{eq:sitd}) 
we get $p_i.\view(s_i) \le p_i.\viewp(s_i) \le v + 1$. 
Since $p_i.\view(\te{i}{v}) = p_i.\viewp(\te{i}{v}) = v$,
both $p_i.\view$ and $p_i.\viewp$ are non-decreasing,
and $p_i.\view \le p_i.\viewp$, $\te{i}{v} \le s_i$ implies 
$$
p_i.\view(s_i) = v \wedge p_i.\viewp(s_i) \in \{v, v + 1\}
\vee
p_i.\view(s_i) = p_i.\viewp(s_i) = v+1.
$$
If the first disjunct holds, then 
$\max(p_i.\view(s_i)+1, p_i.\viewp(s_i)) = v+1$,
and therefore $p_i$ sends $\WISH(v+1)$ to all processes at $s_i$.
Otherwise, $p_i$ enters $v+1$ prior to the expiration
of $p_i.\timerview$ at $s_i$, which is impossible. Hence, 
$p_i$ sends $\WISH(v+1)$ to all processes at $s_i$. 

\item
$p_i.\timerview$ is stopped prematurely at $s_i$,
by executing the code in line~\ref{line:timer-stop}. Then
the condition in line~\ref{line:enter-condition} is true at $s_i$,
so that
\begin{equation}
p_i.\viewp(s_i) = p_i.\view(s_i) \wedge p_i.\view(s_i) > p_i.\prevv(s_i).
\label{eq:entry-si}
\end{equation}
Since $p_i.\view(\te{i}{v}) = p_i.\viewp(\te{i}{v}) = v$ and both
$p_i.\view$ and $p_i.\viewp$ are non-decreasing, (\ref{eq:entry-si}) implies
that $p_i.\view(s_i) = p_i.\viewp(s_i) \ge v+1$. 
Since by~(\ref{lem:after-GST-bound-final-gen:1}), 
no correct process sends $\WISH(v')$ with $v' > v + 1$ 
before $T + \delta$, from~(\ref{eq:sitd}) we get
$p_i.\view(s_i) = p_i.\viewp(s_i) = v+1$. 
If $p_i.\viewp(s_i) > p_i.\prevvp(s_i)$, then 
the condition in line~\ref{line:send5} is true at $s_i$, and therefore,
$p_i$ sends $\WISH(p_i.\viewp(s_i)) = \WISH(v+1)$ to all processes at $s_i$.
Otherwise, $p_i.\viewp(s_i) = p_i.\prevvp(s_i)$, which implies
that $p_i$ sent $\WISH(v+1)$ when $p_i.\viewp$ first became
equal to $v+1$ sometime before $s_i$.
Thus, in both cases $p_i$ sends $\WISH(v+1)$ by $s_i$.
\end{itemize}

Thus, we get that for each correct process $p_i$, there exists a time $s_i\le T+\delta$
such that at $s_i$, $p_i$ sends $\WISH(v+1)$ to all processes.
By~(\ref{eq:sitd}), this implies that all correct processes send $\WISH(v+1)$
to all correct processes no later than $s = \max\{s_i\} \le T + \delta$. Then
Lemma~\ref{lem:simple-bound-generic} implies the required.\qed

\bigskip

Since by Lemma~\ref{lem:postgst1}, $\postGST(t)$ holds for all
$t > \GST + \rho$, from Lemma~\ref{lem:after-GST-bound-final-gen} we get
\begin{corollary}
For all views $v$, if a correct process enters $v$, 
$\tm{v} > \GST + \rho$, and $\timeout(v) > 2\delta$, then
all correct processes enter the view $v+1$ and
$\tl{v+1} \le \tl{v}+\timeout(v) + \delta$.
\label{lem:after-GST-bound-final}
\end{corollary}

\vspace{-\baselineskip} 

\begin{theorem}
  {\em \SYNC} satisfies
  Properties~\ref{prop:local-order}-\ref{prop:no-skip-sync:3} in
  Figure~\ref{fig:sync-properties} for $d = 2\delta$.
\label{th:main-app}
\end{theorem}

\paragraph{Proof.} 
Property~\ref{prop:local-order} is satisfied trivially. Let
$\B$ be the first view such that a correct process enters $\B$,
$\tm{\B}>\GST + \rho$ and $\timeout(\B)>2\delta$. 
Such a view exists by~(\ref{prop:increasing}) and Lemma~\ref{lem:gv-live}.
Since $\tm{\B}>\GST + \rho > \GST$, the view $\B$ satisfies
Property~\ref{prop:after-t}. By Lemmas~\ref{lem:gv-noskip}
and~\ref{lem:gv-live}, a correct process enters every view $v \ge \B$.
By Corollary~\ref{cor:ef-monotone},
\begin{equation}
\tm{v} \ge \tm{\B} > \GST.
\label{eq:v-after-gst}
\end{equation}
Since $\timeout$ is a non-decreasing function, $\timeout(v) > 2\delta$. 
Thus, by Corollary~\ref{lem:GST-bound2-final}, all correct processes enter $v$,
and $\tl{v} \le \tm{v} + 2\delta$, which validates 
Properties~\ref{prop:no-skip-sync:1} and~\ref{prop:no-skip-sync:2}.
To prove Property~\ref{prop:no-skip-sync:3}, 
fix a view $v\ge \B$. By~(\ref{eq:v-after-gst}),
$\tm{v} > \GST$, and therefore, by Corollary~\ref{lem:all-by-myself},  we
get $\tm{v+1} \ge \tm{v}+\timeout(v)$, which implies
Property~\ref{prop:no-skip-sync:3}.
\qed

\begin{theorem}
  Assume that $\tf \ge \GST$ and $F(1) > 2\delta$.
  Then {\em \SYNC}\/ satisfies
  Properties~\ref{prop:local-order}-\ref{prop:no-skip-sync:3},~\ref{prop:last-entry}
  and~\ref{prop:view1} in Figure~\ref{fig:sync-properties} for $\B = 1$ and
  $d = 2\delta$.
\label{th:main-prop:view1}
\end{theorem}

\paragraph{Proof.}
Property~\ref{prop:local-order} is satisfied trivially. Let
$\B = 1$. By Lemmas~\ref{lem:gv-noskip} and~\ref{lem:gv-live},
some correct process enters $\B$. 
To prove Property~\ref{prop:after-t},
let $p_i$ be a correct process that enters $\B = 1$
at $\tm{1}$. By Lemma~\ref{lem:desu-noto-main},
there exists a time $t < \tm{1}$ at which some correct process
attempts to advance from view $0$. 
Thus, by Lemma~\ref{lem:wish-start}, there exists a time
$s \le t < \tm{1}$ at which some correct process calls
$\start$. Since $s \ge \tf$, $\tm{1} > \tf \ge \GST$.
Thus, Property~\ref{prop:after-t} holds.

By Lemmas~\ref{lem:gv-noskip} and~\ref{lem:gv-live}, 
some correct process enters every view $v \ge 1 = \B$.
Thus, Corollary~\ref{cor:ef-monotone} implies that 
\begin{equation}
\tm{v} \ge \tm{1} \ge \tf \ge \GST.
\label{eq:v-after-gst-1}
\end{equation}
Then by Lemma~\ref{lem:postgst0},
$\postGST(\tm{v})$ holds. Since $\timeout$
is a non-decreasing function, $\timeout(v) \ge \timeout(1) > 2\delta$.
Thus, by Lemma~\ref{lem:GST-bound2}, all correct processes enter $v$, and
$\tl{v} \le \tm{v} + 2\delta$, which validates 
Properties~\ref{prop:no-skip-sync:1} and~\ref{prop:no-skip-sync:2} for $d=2\delta$.

To prove Properties~\ref{prop:no-skip-sync:3} and
\ref{prop:last-entry}, fix a view $v\ge \B$. By~(\ref{eq:v-after-gst-1}),
$\tm{v} > \GST$, and therefore, by Corollary~\ref{lem:all-by-myself},  we
get $\tm{v+1} \ge \tm{v}+\timeout(v)$, which implies
Property~\ref{prop:no-skip-sync:3}. Since by~(\ref{eq:v-after-gst-1}),
$\tm{v} \ge \tf$, by Lemma~\ref{lem:postgst0},
$\postGST(\tm{v})$ holds. We also have $\timeout(v) \ge \timeout(1) > 2\delta$.
Thus, by Corollary~\ref{lem:after-GST-bound-final}, 
$\tl{v+1} \le \tl{v} + \timeout(v) + \delta$, and therefore,
Property~\ref{prop:last-entry} holds.

To prove Property~\ref{prop:view1}, let
$$
\begin{array}{@{}l@{}}
t_1 = \max(\tfk{f+1}, \tf) = \tfk{f+1};
\ms
t_2 = \max(\ts, \tf) = \ts;
\ms
T = \max(\min(\tfk{f+1}, \ts - \delta), \tf).
\end{array}
$$

Suppose first that $T = \tf$. Then by~(\ref{eq:v-after-gst-1}),
$\GV{T} = \GV{\tf} = 0$. 
Since $\gvsym$ is non-decreasing, $\tm{\B} \ge T = \tf \ge \GST$. Thus, 
by Corollary~\ref{lem:upper-efirst}, 
no correct process can send $\WISH(v)$ for any $v > 1$
earlier than $T + \timeout(1) > T + 2\delta$.
Since by Lemma~\ref{lem:postgst0},
$\postGST(\tf)$ holds, by Lemma~\ref{lem:bound0},
$\tl{\B} \le \min(t_1+2\delta, t_2+\delta) \le t_2 + \delta = \ts + \delta$,
as needed. 

Suppose next that $T = \min(\tfk{f+1}, \ts - \delta) > \tf$. If
some correct process enters view $\B$ before $T$,
then by Lemma~\ref{lem:GST-bound2}, 
$$
\tl{1} \le T + 2\delta = 
\min(\tfk{f+1} + 2\delta, \ts+\delta) \le \ts + \delta,
$$
as needed. On the other hand, if no correct process
enters $\B = 1$ before $T$, then $\GV{T} = 0$. 
Since $\gvsym$ is non-decreasing, 
$\tm{\B} \ge T > \tf \ge \GST$. Thus, 
by Corollary~\ref{lem:upper-efirst}, 
no correct process can send $\WISH(v)$ for any $v > 1$
earlier than $T + \timeout(1) > T + 2\delta$.
Since by Lemma~\ref{lem:postgst0},
$\postGST(\tf)$ holds, by Lemma~\ref{lem:bound0} we have
$\tl{\B} \le \min(t_1+2\delta, t_2+\delta) \le t_2 + \delta = \ts + \delta$,
as needed. Thus, we proved that in all cases, 
$\tl{\B} \le \min(t_1+2\delta, t_2+\delta)$, which implies
the required.\qed

\begin{theorem}
  Let $\B = \GV{\GST + \rho}+1$ and $d = 2\delta$. Assume that
  $\tfk{f+1} \le \GST + \rho$ and $\timeout(\B) > 2\delta$.
  Then {\em \SYNC}\/ satisfies
  Properties~\ref{prop:local-order}-\ref{prop:no-skip-sync:3},~\ref{prop:last-entry}
  and~\ref{prop:convergence} in Figure~\ref{fig:sync-properties}.
\label{theorem:GST-bound-final}
\end{theorem}

\paragraph{Proof.} 
Property~\ref{prop:local-order} is satisfied trivially. Let
$\B = \GV{\GST + \rho}+1$. By Lemmas~\ref{lem:gv-noskip} and~\ref{lem:gv-live},
some correct process enters $\B$. 
By Lemma~\ref{lem:efirst-gv}, $\GV{\tm{\B}} = \B$. Since 
$\gvsym$ is non-decreasing, and $\B > \GV{\GST + \rho}$, we have
$\tm{\B} > \GST + \rho \ge \GST$. Hence, Property~\ref{prop:after-t} holds.
By Lemmas~\ref{lem:gv-noskip} and~\ref{lem:gv-live}, 
some correct process enters every view $v \ge \B$.
By Corollary~\ref{cor:ef-monotone}, $v \ge \B$ implies that
\begin{equation}
\tm{v} \ge \tm{\B} \ge \GST + \rho.
\label{eq:v-after-gst-8}
\end{equation}
Since $\timeout$
is a non-decreasing function, $\timeout(v) \ge \timeout(\B) > 2\delta$.
Thus, by Corollary~\ref{lem:GST-bound2-final}, all correct processes enter $v$ and
$\tl{v} \le \tm{v} + 2\delta$, which validates 
Properties~\ref{prop:no-skip-sync:1} and~\ref{prop:no-skip-sync:2}.

To prove Properties~\ref{prop:no-skip-sync:3} and
\ref{prop:last-entry}, fix a view $v\ge \B$. By~(\ref{eq:v-after-gst-8}),
$\tm{v} \ge \GST$, and therefore, by Corollary~\ref{lem:all-by-myself} we
get $\tm{v+1} \ge \tm{v}+\timeout(v)$, which implies
Property~\ref{prop:no-skip-sync:3}. Since by~(\ref{eq:v-after-gst-8}),
$\tm{v} \ge \GST + \rho$, by Lemma~\ref{lem:postgst1},
$\postGST(\tm{v})$ holds. We also have $\timeout(v) \ge \timeout(\B) > 2\delta$.
Thus, by Corollary~\ref{lem:after-GST-bound-final}, 
$\tl{v+1} \le \tl{v} + \timeout(v) + \delta$, and therefore,
Property~\ref{prop:last-entry} holds.

To prove Property~\ref{prop:convergence}, we consider two cases:
\begin{itemize}

\item $\GV{\GST + \rho} = 0$. Hence, $\B = 1$. Let
$$
\begin{array}{@{}l@{}}
t_1 = \max(\tfk{f+1}, \GST + \rho);
\ms
t_2 = \max(\ts, \GST + \rho);
\ms
T = \max(\min(\tfk{f+1}, \ts - \delta), \GST + \rho).
\end{array}
$$
Since $\tfk{f+1} \le \GST + \rho$ and 
$\min(\GST + \rho, \ts - \delta) \le \GST + \rho$, the above
can be re-written as follows:
$$
\begin{array}{@{}l@{}}
t_1 = \max(\tfk{f+1}, \GST + \rho) = \GST + \rho;
\ms
t_2 = \max(\ts, \GST + \rho);
\ms
T = \GST + \rho.
\end{array}
$$
Then $\GV{T} = 0$. Since $\gvsym$ is non-decreasing, $\tm{1} \ge \GST$. 
Thus, by Corollary~\ref{lem:upper-efirst}, 
no correct process can send $\WISH(v)$ for any $v > 1$
earlier than $T + \timeout(1) > T + 2\delta$.
Since by Lemma~\ref{lem:postgst1},
$\postGST(\GST + \rho)$ holds, by Lemma~\ref{lem:bound0},
$\tl{\B} \le \min(t_1+2\delta, t_2+\delta) \le t_1 + 2\delta = 
\GST + \rho + 2\delta$.
Since $\timeout(0) = 0$, we have
$$
\tl{\B} \le \GST + \rho + \timeout(\GV{\GST + \rho}) + 2\delta,
$$
which implies the upper bound stipulated 
by Property~\ref{prop:convergence}.

\item $\GV{\GST + \rho} > 0$.
Let $T = \GST + \rho +\timeout(\GV{\GST + \rho})+\delta$.
Suppose first that some correct process enters $\GV{\GST + \rho}+1$ 
before $T$. By Lemma~\ref{lem:efirst-gv}, 
$\GV{\tm{\GV{\GST + \rho}+1}} = \GV{\GST + \rho}+1$.
Since $\gvsym$ is non-decreasing, we have
$\tm{\GV{\GST + \rho}+1} > \GST + \rho$.
Thus, by Corollary~\ref{lem:GST-bound2-final},
all correct processes enter $\B$ by 
$\GST + \rho+\timeout(\GV{\GST + \rho})+3\delta$, as needed.
Suppose now that no correct processes enters $\B$ 
before $T$, so that $\tm{\B} \ge T \ge \GST$.
Then by Corollary~\ref{lem:upper-efirst},
\begin{equation}
\mbox{no correct process can send $\WISH(v)$ for any $v > \B$
earlier than $T + \timeout(\B) > T + 2\delta$.} 
\label{eq:tb2d}
\end{equation}
From Lemma~\ref{lem:postgst1}, $\postGST(\GST + \rho)$, and therefore,
by Lemma~\ref{lem:all-send-wish1-generic}, all correct
processes send $\WISH(\B)$ to all processes no later than $T +\delta$.
Since~(\ref{eq:tb2d}) holds, 
by Lemma~\ref{lem:simple-bound-generic},
all correct processes enter $\B$, and
$\tl{\B} \le T + 2\delta = \GST + \rho + \timeout(\GV{\GST + \rho}) + 3\delta$,
as needed.
\end{itemize}
\qed

\paragraph{Proof of Theorem~\ref{th:main}.} Follows from
Theorems~\ref{th:main-app}-\ref{theorem:GST-bound-final}.\qed

\clearpage

\section{Additional Material on Consensus Protocols}
\label{app:consensus}

\subsection{Safety Proof for Single-Shot HotStuff}
\label{sec:hotstuff-safety}

The protocol satisfies the Validity property, because deciding on a value
requires preparing it, and due to the validity check in $\ValidNewView$, any
prepared value is valid:
\begin{proposition}
\label{lemma:hotstuff:validityliveness}
$\forall v, C, \val .\, \accepted(C, v, \hash(\val)) \wedge \valid(C) {\implies} \validity(\val)$.
\end{proposition}
\paragraph{Proof. } Fix $v$, $C$ and $\val$ and assume
$\accepted(C, v, \hash(\val))$. Since $\accepted(C, v, \hash(\val))$, a quorum
$Q$ of processes sent $\PREPARED(v, \hash(\val))$. Then at least $f+1$ correct
processes checked the validity of $\val$ in the $\ValidNewView$ predicate, which
implies the required.\qed

\bigskip

Let
$
\committed(C, v, h)
\iff
\exists Q.\, 
\quorum(Q) \wedge C = \{\langle \COMMITTED(v, h) \rangle_j \mid p_j \in Q\}.
$

\begin{lemma}
\label{thm:hotstuff-main}
$\forall v, v', C, C', \val, \val'.\, \committed(C, v, \hash(\val)) \wedge \accepted(C', v', \hash(\val'))
\wedge {}$\\
\hspace*{2.25cm} $\valid(C) \wedge \valid(C') 
\wedge v < v' {\implies} \val = \val'$.
\end{lemma}
\paragraph{Proof.}
Fix $v$, $C$ and $\val$ and assume $\committed(C, v, \hash(\val))$. We prove by
induction on $v'$ that  
$$
\forall v', C', \val'.\, \accepted(C', v', \hash(\val')) \wedge \valid(C') \wedge
v < v' \implies \val = \val'.
$$
Assume this holds for all $v' < v^*$; we now prove it for $v' = v^*$. To this
end, assume $v < v'$ and $\accepted(C', v', \hash(\val'))$ for a well-formed
$C'$.

Since $\committed(C, v, \hash(\val))$, a quorum $Q$ of processes sent
$\COMMITTED(v, \hash(\val))$. Since $\accepted(C', v', \hash(\val'))$, a quorum
$Q'$ of processes sent $\PREPARED(v', \hash(\val'))$. The quorums $Q$ and $Q'$
have to intersect in some correct process $p_k$, which has thus sent both
$\COMMITTED(v, \hash(\val))$ and $\PREPARED(v', \hash(\val'))$. Since $v< v'$,
process $p_k$ must have sent $\COMMITTED(v, \hash(\val))$ before
$\PREPARED(v', \hash(\val'))$. Before sending $\COMMITTED(v, \hash(\val))$ the
process set $\lballot$ to $v$ (line~\ref{hotstuff:lock}) and had
$\cmd = \val$. 

Assume towards a contradiction that $\val \not= \val'$. Let $v''$ be the first
view after $v$ when $p_k$ prepared some proposal $\val'' \not= \val$, so that
$v'' \le v'$. When this happened, by Proposition~\ref{lemma:hotstuff:increase}
$p_k$ must have had $\cmd = \val$ and $\lballot \ge v$. Then by the
$\ValidNewView$ check (line~\ref{hotstuff:safety-check}), the leader of $v''$
provided a well-formed prepared certificate $C''$ such that
$\accepted(C'', v''', \hash(\val''))$ for $v'''$ such that
$v < v''' < v'' \le v'$. But then by induction hypothesis we have
$\val'' = \val$, and above we established $\val'' \not= \val$: a contradiction.
Hence, we must have $\val = \val'$, as required.\qed

\begin{proposition}
  \label{lemma:hotstuff:committed-prepared}
  $\forall v, C, h.\, \committed(C, v, h) \wedge \valid(C) {\implies} \exists C'.\, \accepted(C', v,
  h) \wedge \valid(C')$.
\end{proposition}

\begin{corollary}
Single-shot HotStuff satisfies Agreement.
\end{corollary}
\paragraph{Proof.} Assume two correct processes decide on values $\val$ and
$\val'$ in views $v$ and $v'$, respectively. Then
$\committed(C, v, \hash(\val))$ and $\committed(C', v', \hash(\val'))$ for some
well-formed $C$ and $C'$. By
Proposition~\ref{lemma:hotstuff:committed-prepared} we have
$\accepted(C_0, v, \hash(\val))$ and $\accepted(C'_0, v', \hash(\val'))$ for
some well-formed $C_0$ and $C'_0$. Without loss of generality assume $v \le v'$.
If $v = v'$, then $\val = \val'$ by
Proposition~\ref{lemma:hotstuff:singlecmd}. If $v < v'$, then $\val = \val'$
by Lemma~\ref{thm:hotstuff-main}.\qed

\clearpage

\subsection{Pseudocode of Two-Phase HotStuff}
\label{sec:2hs}

\begin{algorithm*}[H]
  \setcounter{AlgoLine}{0}

  \SubAlgo{\Upon $\newview(v)$\label{tendermint:newview}}{
    $\ballot \leftarrow v$\;
    $\voted \leftarrow \FALSE$\;
    $\stoptimer(\timernetwork)$\;
    \If{$p_i = \leader(\ballot)$}{
      $\starttimer(\timernetwork,\timeoutS(\ballot))$;
    }
    \Send $\langle \NEWLEADER(\ballot, \cballot, \cmd, \cert) \rangle_i$\\
    \quad \KwTo $\leader(\ballot)$\;\label{tendermint:send-newleader}
  }

  \smallskip
  \smallskip

  \SubAlgo{{\bf when $\timernetwork$ expired and}
    \hspace{6cm} {\bf received} $\{\langle \NEWLEADER(v, \vcballot_j, \vcmd_j, 
    \vcert_j) \rangle_j \mid p_j \in P\} = M$
  }{\label{tendermint:receive-newleader}
    \textbf{pre:} $\ballot = v \wedge p_i = \leader(v) \wedge (\forall m \in
    M.\, \ValidNewLeader(m))$\; 
    \uIf{$\exists j.\, \vcballot_j = \max\{\vcballot_{k} \mid p_k \in P\}
      \not= 0$}{
      \Send $\langle \PREPARE(v, \vcmd_j, \vcert_j) \rangle_i$ \KwTo
      \all\; 
    }\Else{
      \Send $\langle \PREPARE(v, \myproposal(), \bot) \rangle_i$ \KwTo
      \all\;    
    }     
    
  }

  \smallskip
  \smallskip

  \SubAlgo{\WhenReceived $\langle \PREPARE(v, \val, \_) \rangle_j =
    m$}{\label{tendermint:receive-prepare} 
    \textbf{pre:} $\ballot \,{=}\, v \wedge \voted \,{=}\, \FALSE \wedge \ValidNewView(m)$;\\
    \label{tendermint:safety-check}
    $\ctxn \leftarrow \val$\;
    $\voted \leftarrow \TRUE$\;
    \Send $\langle \PREPARED(v, \hash(\ctxn)) \rangle_i$ \KwTo \all\;
  }

  \smallskip
  \smallskip

  \SubAlgo{\WhenReceived $\{\langle \PREPARED(v, h)\rangle_j \mid p_j \in Q\} = C$ 
    {\bf for a quorum $Q$}}{\label{tendermint:receive-prepared}
    \textbf{pre:} $\ballot = v \wedge \voted = \TRUE \wedge \hash(\ctxn) = h$\;
    $\cmd \leftarrow \ctxn$\;
    $\cballot \leftarrow \ballot$\;
    $\lballot \leftarrow \ballot$\; \label{tendermint:lock}
    $\cert \leftarrow C$\;
    \Send $\langle \COMMITTED(v, h) \rangle_i$ \KwTo \all\;
  }

  \smallskip
  \smallskip

  \SubAlgo{\WhenReceived $\{\langle \COMMITTED(v, h)\rangle_j \mid p_j \in Q\}$
    {\bf for a quorum $Q$}}{\label{tendermint:receive-committed}
    \textbf{pre:} $\ballot = \lballot = v \wedge \hash(\ctxn) = h$\;
    ${\tt decide}(\ctxn)$;
  }

\end{algorithm*}

\bigskip

\noindent
The predicates $\accepted$, $\ValidNewLeader$ and $\ValidNewView$ are as defined
in Figure~\ref{fig:hotstuff}.

\clearpage

\subsection{Single-Shot All-to-All PBFT}
\label{sec:pbft}

\begin{algorithm*}[H]
  \setcounter{AlgoLine}{0}
  \SubAlgo{\Upon $\newview(v)$}{
    $\ballot \leftarrow v$\;
    $\voted \leftarrow \FALSE$\;
    \Send $\langle \NEWLEADER(\ballot, \lballot, \cmd, \cert) \rangle_i$\\
    \nonl \quad \KwTo $\leader(\ballot)$\; 
  }

  \smallskip
  \smallskip

  \SubAlgo{\WhenReceived $\{\langle \NEWLEADER(v, \vcballot_j, \vcmd_j, 
    \vcert_j) \rangle_j \mid p_j \in Q\} = M$ \hspace{3cm}
    {\bf for a quorum $Q$}}{\label{pbft:receive-newleader}
    \textbf{pre:} $\ballot = v \wedge p_i = \leader(v) \wedge (\forall m \in M.\, \ValidNewLeader(m))$\; 
    \uIf{$\exists j.\, \vcballot_j = \max\{\vcballot_{k} \mid p_k \in Q\}
      \not= 0$}{\Send $\langle \PREPARE(v,   \vcmd_j, M) \rangle_i$ \KwTo \all;}
    \Else{\Send $\langle \PREPARE(v,  \myproposal(), M) \rangle_i$ \KwTo \all;}
  }

  \smallskip
  \smallskip

  \SubAlgo{\WhenReceived $\langle \PREPARE(v, \val, \_) \rangle_j = m$}{\label{pbft:receive-prepare}
    \textbf{pre:}
    $\ballot = v \wedge \voted = \FALSE \wedge \ValidNewView(m)$\; \label{pbft:safety-check}
    $\ctxn \leftarrow \val$\;
    $\voted \leftarrow \TRUE$\;
    \Send $\langle \PREPARED(v, \hash(\ctxn)) \rangle_i$ \KwTo \all\; }

  \smallskip
  \smallskip

  \SubAlgo{\WhenReceived $\{\langle \PREPARED(v, h)\rangle_j \mid p_j \in Q\} = C$ 
    {\bf for a quorum $Q$}}{
    \textbf{pre:} $\ballot = v \wedge \voted = \TRUE \wedge \hash(\ctxn) = h$\;
    $\cmd \leftarrow \ctxn$\;
    $\lballot \leftarrow \ballot$\;\label{pbft:assign-cballot}
    $\cert \leftarrow C$\;
    \Send $\langle \COMMITTED(v, h) \rangle_i$ \KwTo \all\;
  }

  \smallskip
  \smallskip

  \SubAlgo{\WhenReceived $\{\langle \COMMITTED(v, h)\rangle_j \mid p_j \in Q\}$ 
    {\bf for a quorum $Q$}}{
    \textbf{pre:} $\ballot = \lballot = v \wedge \hash(\ctxn) = h$\;
    ${\tt decide}(\ctxn)$;
  }
\end{algorithm*}

\bigskip

$$
\accepted(C, v, h)
\iff
\exists Q.\, 
\quorum(Q) \wedge C = \{\langle \PREPARED(v, h) \rangle_j \mid p_j \in Q\}
$$

$$
\ValidNewLeader(\langle \NEWLEADER(v', v, \val, C) \rangle_{\_})
  \iff 
v < v' \wedge
({v \not= 0} {\implies} \accepted(C, v, \hash(\val)))
$$

$$
\begin{array}{@{}l@{}}
\ValidNewView(\langle \PREPARE(v, \val, M) \rangle_i) \iff {}
\ms
\quad  p_i = \leader(v) \wedge \validity(\val) \wedge {}
\ms
\quad \exists Q, \vcballot, \vcmd, \vcert.\, \quorum(Q) \wedge {}
\ms
\quad M = \{\langle \NEWLEADER(v, \vcballot_j, \vcmd_j,  \vcert_j) \rangle_j \mid p_j \in Q\} \wedge {}
\ms
\quad (\forall m \in M.\, \ValidNewLeader(m)) \wedge {}
\ms
\quad ((\exists j.\, \vcballot_j \not= 0) {\implies}
(\exists j.\, \vcballot_j = \max\{\vcballot_{k} \mid p_k \in Q\}
\wedge \val = \vcmd_j))
\end{array}
$$

\smallskip
\smallskip
\smallskip

\noindent
In view $1$ the leader can propose without waiting for $\NEWLEADER$ messages,
and processes can avoid sending these messages to this leader.

\clearpage

\subsubsection*{Safety}

\begin{proposition}
\label{lemma:pbft:validityliveness}
$\forall v, C, \val .\, \accepted(C, v, \hash(\val)) \wedge \valid(C) {\implies} \validity(\val)$.
\end{proposition}

This proposition implies that the protocol satisfies Validity. We next prove
Agreement.

\begin{proposition}
\label{lemma:pbft:view-increase}
The variables $\ballot$ and $\lballot$ at a correct process never decrease and
we always have $\lballot \le \ballot$.
\end{proposition}

\begin{proposition}
\label{lemma:pbft:singlecmd}
$\forall v, C, C', \val, \val '.\, \accepted(C, v, \hash(\val)) \wedge
\accepted(C', v, \hash(\val'))  \wedge {}$\\
\hspace*{2.9cm} $\valid(C) \wedge \valid(C') {\implies} \val = \val'$.
\end{proposition}

\begin{proposition}
\label{lemma:pbft:committed-prepared}
  $\forall v, C, h.\, \committed(C, v, h) \wedge \valid(C) \implies \exists C'.\, \accepted(C, v,
  h) \wedge \valid(C')$.
\end{proposition}

\begin{lemma}
\label{lemma:pbft:main}
$
\forall v, v', \val, \val', m.\, \committed(\_, v, \hash(\val)) \wedge
m = \langle \PREPARE(v', \val', \_)\rangle_{\_} \wedge {}$\\
\hspace*{2.25cm} $\valid(m) \wedge \ValidNewView(m) \implies \val' = \val
$.
\end{lemma}
\paragraph{Proof.}
Fix $v$, $C$ and $\val$ and assume $\committed(C, v, \hash(\val))$.  We prove by
induction on $v'$ that
$$
\forall m, v', \val'.\, 
m = \langle \PREPARE(v', \val', \_)\rangle_{\_} \wedge \valid(m) \wedge 
\ValidNewView(m) \implies \val' = \val.
$$
Assume this holds for all $v' < v^*$; we now prove it for $v' = v^*$. To this
end, assume $v < v'$ and $m = \langle\PREPARE(v', \val', M)\rangle_{\_}$ is a
sent message such that $\ValidNewView(m)$. Since a correct node only prepares
proposals satisfying $\ValidNewView$ (line~\ref{pbft:safety-check}), from the
induction hypothesis it follows that
$$
\forall C'', v'', \val''.\, v < v'' < v' \wedge \accepted(C'', v'',
\hash(\val'')) \wedge \valid(C'') {\implies} \val = \val''.
$$
Furthermore, by Propositions~\ref{lemma:pbft:singlecmd}
and~\ref{lemma:pbft:committed-prepared} we have
$$
\forall C'', \val''.\, 
\accepted(C'', v, \hash(\val'')) \wedge \valid(C'') {\implies} \val = \val'',
$$
so that overall we get
\begin{equation}\label{pbft:hyp2}
\forall C'', v'', \val''.\, v \le v'' < v' \wedge \accepted(C'', v'',
\hash(\val'')) \wedge \valid(C'') \implies \val = \val''.
\end{equation}

Let
$$
M = \{\langle \NEWLEADER(v', \vcballot_j, \vcmd_j, \vcert_j) \rangle_j \mid p_j \in
Q\}
$$
for some quorum $Q$. Since $\ValidNewView(m)$, we have
$\forall m' \in M.\, \ValidNewLeader(m')$, so that
$$
\forall p_j \in Q.\, \vcballot_j < v' \wedge ({\vcballot_j \not= 0}
{\implies} \accepted(\vcert_j, \vcballot_j, \hash(\vcmd_j)) \wedge
\valid(\vcert_j)).
$$
From this and~(\ref{pbft:hyp2}) we get that
\begin{equation}\label{pbft:hyp3}
  \forall p_j \in Q.\, {\vcballot_j \ge v} {\implies}  \vcmd_j = \val.
\end{equation}

Since $\committed(C, v, h)$, a quorum $Q'$ of processes sent $\COMMITTED(v, h)$.
The quorums $Q$ and $Q'$ have to intersect in some correct process $p_k$, which
has thus sent both $\COMMITTED(v, \hash(\val))$ and
$\NEWLEADER(v', \vcballot_k, \vcmd_k, \vcert_k)$. Since $v< v'$, this process
$p_k$ must have sent the $\COMMITTED$ message before the $\NEWLEADER$
message. Before sending $\COMMITTED(v, \hash(\val))$ the process set $\lballot$
to $v$ (line~\ref{pbft:assign-cballot}). Then by
Proposition~\ref{lemma:pbft:view-increase} process $p_k$ must have had
$\lballot \ge v$ when it sent the $\NEWLEADER$ message. Hence,
$\vcballot_k \ge v \not= 0$ and $\max\{\vcballot_j \mid p_j \in Q\} \ge v$.
Then from~(\ref{pbft:hyp3}) for any $j$ such that
$\vcballot_j = \max\{\vcballot_{k} \mid p_k \in Q\}$ we must have
$\vcmd_j = \val$. Since $\ValidNewView(m)$ holds, this implies
$\val' = \val$.\qed

\begin{corollary}
$\forall v, v', C, C', \val, \val'.\, \committed(C, v, \hash(\val)) \wedge \accepted(C', v', \hash(\val'))
\wedge {}$\\
\hspace*{2.5cm} $\valid(C) \wedge \valid(C') 
\wedge v < v' {\implies} \val = \val'$.
\end{corollary}

\begin{corollary}
PBFT satisfies Agreement.
\end{corollary}

\smallskip

\subsubsection*{Liveness}

\begin{theorem}\label{thm:pbft-latency}
  Let $v \ge \B$ be a view such that $\timeoutV(v) > 6\delta$ and $\leader(v)$ is
  correct. Then all correct processes decide at $v$ by
  $\tl{v}+4\delta$.
\end{theorem}
\paragraph{Proof.} 
By Property~\ref{prop:after-t}, we have $\tm{v} \ge \GST$, so that all messages
sent by correct processes after $\tm{v}$ get delivered to all correct processes
within $\delta$. Once a correct process enters $v$, it
sends its $\NEWLEADER$ message, so that $\leader(v)$ is guaranteed to receive a
quorum of such messages by $\tl{v} + \delta$. When this
happens, the leader will send its proposal in a $\PREPARE$ message, which
correct processes will receive by $\tl{v}+2\delta$. If they deem the proposal
safe, it takes them at most $2\delta$ to exchange the sequence of
$\PREPARED$ and $\COMMITTED$ messages leading to decisions. 
By~(\ref{overlap}), all correct processes will stay in $v$ until
at least $\tl{v}+(\timeout(v) - d) > \tl{v}+4\delta$.
Thus, the sequence of
message exchanges will complete before any of them exits view $v$, and all
correct processes will decide in this view by the time $\tl{v}+4\delta$.

It remains to show that the proposal $\val$ that $\leader(v)$ makes in view $v$
(line~\ref{pbft:receive-newleader}) will be deemed safe by all correct processes
according to the $\ValidNewView$ predicate
(line~\ref{pbft:receive-prepare}). All the conjuncts of $\ValidNewView$ except
for $\validity (\val)$ are trivially satisfied given that $\leader(v)$ is
correct.  If the leader is choosing its own proposal as $\val$, then it is valid
because correct processes propose valid values. Otherwise, from
$\ValidNewLeader$ we get that $\accepted(C, \_, \hash(\val))$ for a well-formed
$C$. Hence, by Proposition~\ref{lemma:pbft:validityliveness} we again have
$\validity(\val)$.\qed

\begin{corollary}
  Let $\timeoutV$ be such that~(\ref{prop:increasing}) holds. Then all correct
  processes eventually decide.
\end{corollary}

\subsubsection*{Latency under favorable conditions}

\begin{corollary}
  Assume that $\tf\geq\GST$ and $\timeout(1) > 6\delta$.  Then in the PBFT
  protocol all correct processes decide no later than
  $\ts+\sum_{k=1}^{f} (\timeoutV(k)+\delta) + 5\delta$.
\label{thm:pbft:latencyfromVequal1}
\end{corollary}

\begin{corollary}
  Assume that $\tf\geq\GST$, $\timeout(1) > 5\delta$ and $\leader(1)$ is
  correct. Then in the PBFT protocol all correct processes decide no later than
  $\ts + 4\delta$.
 \label{cor:pbft:Vequal1andcorrectleader}
\end{corollary}

\clearpage

\subsection{Single-Shot All-to-All SBFT}
\label{sec:sbft}

\begin{algorithm*}[H]
  \setcounter{AlgoLine}{0}
  \SubAlgo{\Upon $\newview(v)$}{
   $\ballot \leftarrow v$\;
    $\voted \leftarrow \FALSE$\;
    \Send $\langle \NEWLEADER(\ballot, \lballot, \cmd, \cert,
    \preballot, \ctxn) \rangle_i$\\
    \nonl \quad \KwTo $\leader(\ballot)$\; 
  }

  \smallskip
  \smallskip

  \SubAlgo{\WhenReceived $\{\langle \NEWLEADER(v, \vcballot_j, \vcmd_j, 
    \vcert_j, \vpreballot_j, \vctxn_j) \rangle_j \mid p_j \in Q\} = M$\hspace{2cm}
    {\bf for a quorum $Q$}}{\label{sbft:receive-newleader}
    \textbf{pre:} $\ballot = v \wedge p_i = \leader(v) \wedge (\forall m \in M.\, \ValidNewLeader(m))$\;
    {\bf let} $(\valslow, \vslow)\leftarrow(\bot, 0)$\label{comp-start}\;
    \lIf{$\exists j.\, \vcballot_j = \max\{\vcballot_{k} \mid p_k \in Q\}
      \not= 0$}{$(\valslow, \vslow)\leftarrow(\vcmd_j, \vcballot_j)$}
    {\bf let} $D\leftarrow \{(\val, v') 
    \mid \exists P \subseteq Q.\, |P|= f+1\wedge 
    (\forall p_j \in P.\, \vctxn_j = \val) \wedge {}$\label{sbft:D}
    \nonl \hspace*{2.8cm}$v' = \min\{\vpreballot_j \mid p_j \in P\}\}$\;
    {\bf let} $(\valfast, \vfast)\leftarrow(\bot, \max\{v'\mid (\_, v')\in D\})$\;
    \leIf{$ \exists !\, \val.\, (\val, \vfast) \in D$}{$\valfast\leftarrow \val$}
    {$\vfast=0$\label{comp-end}}
     \lIf{$\vslow\geq \vfast \wedge \vslow>0$}
     {\Send $\langle \PREPARE(v, \valslow, M) \rangle_i$ \KwTo \all}
     \lElseIf{$\vfast>\vslow$}{\Send $\langle \PREPARE(v, \valfast, M) \rangle_i$ \KwTo \all}
     \lElse{\Send $\langle \PREPARE(v, \myproposal(), M) \rangle_i$ \KwTo \all}
  }

  \smallskip
  \smallskip

\SubAlgo{\WhenReceived $\langle \PREPARE(v, \val, \_) \rangle_j = m$}{\label{sbft:receive-prepare}
    \textbf{pre:}
    $\ballot=v \wedge \voted = \FALSE \wedge \ValidNewView(m)$\;
    $\ctxn \leftarrow \val$\;
    $\preballot \leftarrow \ballot$\;
    $\voted \leftarrow \TRUE$\;
    $\stoptimer(\timersbft)$\;
    $\starttimer(\timersbft, \timeoutF(\ballot))$\;
    \Send $\langle \PREPARED(v, \hash(\ctxn)) \rangle_i$ \KwTo \all\; }

  \smallskip
  \smallskip

\SubAlgo{\WhenReceived 
  $\{\langle \PREPARED(v, h)\rangle_j \mid p_j \in \Proc\}$}{
    \textbf{pre:} $\ballot = v \wedge \voted=\TRUE \wedge \hash(\ctxn)=h$\;
    ${\tt decide}(\ctxn)$;
  }

\smallskip
\smallskip

\SubAlgo{{\bf when $\timersbft$ expired and}\hspace{7cm}
  {\bf received} $\{\langle \PREPARED(v, h)\rangle_j \mid p_j \in Q\} = C$ 
    {\bf for at least a quorum $Q$}}{\label{line:sbft-timer}
    \textbf{pre:} $\ballot = v \wedge \voted=\TRUE \wedge \hash(\ctxn)=h$\;
      $\cmd \leftarrow \ctxn$\;
    $\lballot \leftarrow \ballot$\;
    $\cert \leftarrow C$\;
      \Send $\langle \COMMITTED(v, h) \rangle_i$ \KwTo
      \all\;
}

  \smallskip
  \smallskip

  \SubAlgo{\WhenReceived $\{\langle \COMMITTED(v, h)\rangle_j \mid p_j \in Q\}$ 
    {\bf for a quorum $Q$}}{
    \textbf{pre:} $\ballot = \lballot = v \wedge \hash(\ctxn)=h$\;
    ${\tt decide}(\ctxn)$;
  }
\end{algorithm*}

\clearpage

$$
\accepted(C, v, h)
\iff
\exists Q.\, 
\quorum(Q) \wedge C = \{\langle \PREPARED(v, h) \rangle_j \mid p_j \in Q\}
$$

$$
\begin{array}{c}
\ValidNewLeader(\langle \NEWLEADER(v', v, \val, C, v_0, \val_0) \rangle_{\_})
  \iff {}
\\[2pt]
v \le v_0 < v' \wedge 
({v \not= 0} {\implies} \accepted(C, v, \hash(\val))) 
\end{array}
$$

$$
\begin{array}{@{}l@{}}
\ValidNewView(\langle \PREPARE(v, \val, M) \rangle_i) \iff {}
\ms
\quad  p_i = \leader(v) \wedge \validity(\val) \wedge {}
\ms
\quad \exists Q, \vcballot, \vcmd, \vcert, \vpreballot, \vctxn.\, 
\quorum(Q) \wedge {}
\ms
\quad M = \{\langle \NEWLEADER(v, \vcballot_j, \vcmd_j,  \vcert_j, 
  \vpreballot_j, \vctxn_j) \rangle_j \mid p_j \in Q\} \wedge {}
\ms
\quad (\forall m \in M.\, \ValidNewLeader(m)) \wedge {}
\ms
\quad \exists \vslow, \vfast, \valslow, \valfast.\,
\ms
\quad
(\vslow, \vfast, \valslow, \valfast
\mbox{ are computed 
from 
$ \vcballot, \vcmd,  \vcert, \vpreballot, \vctxn$
as in lines~\ref{comp-start}-\ref{comp-end}}) \wedge {}
\ms
\quad (\vslow\geq \vfast\wedge \vslow>0 {\implies}
\val = \valslow)\wedge 
(\vslow < \vfast {\implies} \val = \valfast)
\end{array} 
$$

\smallskip
\smallskip

\noindent
In view $1$ the leader can propose without waiting for $\NEWLEADER$ messages,
and processes can avoid sending these messages to this leader.

\subsubsection*{Safety}

\begin{proposition}
\label{lemma:sbft:validityliveness}
$\forall v, C, \val .\, \accepted(C, v, \hash(\val)) \wedge \valid(C) {\implies} \validity(\val)$.
\end{proposition}

This proposition implies that the protocol satisfies Validity. We next prove
Agreement.

\begin{proposition}
\label{prop:sbft:increase}
The variables $\ballot$, $\preballot$ and $\lballot$ at a correct process never
decrease and we always have $\lballot \le\preballot \le \ballot$.
\end{proposition}

\begin{proposition}
\label{lemma:sbft:singlecmd}
$$
\forall v, C, C', \val, \val '.\, \accepted(C, v, \hash(\val)) \wedge
\accepted(C', v, \hash(\val'))  \wedge \valid(C) \wedge
\valid(C')
{\implies} \val = \val'.
$$
\end{proposition}

Let
$$
\committedslow(C, v, h)
\iff
\exists Q.\, 
\quorum(Q) \wedge C = \{\langle \COMMITTED(v, h) \rangle_j \mid p_j \in Q\};
$$
$$
\committedfast(C, v, h)
\iff
C = \{\langle \PREPARED(v, h) \rangle_j \mid p_j \in \Proc\}.
$$

\begin{proposition}
\label{lemma:sbft:committedslow-prepared}
$$
\forall v, C, h.\, \committedslow(C, v, h) \wedge \valid(C) \implies
\exists C'.\, \accepted(C', v, h) \wedge \valid(C').
$$
\end{proposition}

\begin{proposition}
\label{lemma:sbft:committedfast-prepared}
$$
\forall v, C, C', h, h'.\, \committedfast(C, v, h) \wedge \valid(C) \wedge
\accepted(C', v, h') \wedge \valid(C') \implies h = h'.
$$
\end{proposition}

\begin{lemma}
\label{lemma:sbft:main}
$$
\begin{array}{l}
\forall v, v', \val, \val', m.\, 
(\committedslow(\_, v, \hash(\val)) \vee \committedfast(\_, v, \hash(\val)))
\wedge {}
\\
m = \langle \PREPARE(v', \val',
\_)\rangle_{\_} \wedge \valid(m) \wedge \ValidNewView(m)
\implies \val' = \val.
\end{array}
$$
\end{lemma}
\paragraph{Proof.}
Fix $v$, $C$ and $\val$ and assume 
$$
\committedslow(C, v, \hash(\val)) \vee \committedfast(C, v, \hash(\val)).
$$
We prove by induction on $v'$ that 
$$
\forall m, v', \val'.\, 
m = \langle \PREPARE(v', \val', \_)\rangle_{\_} \wedge \valid(m) \wedge 
\ValidNewView(m) \implies \val' = \val.
$$
Assume this holds for all $v' < v^*$; we now prove it for $v' = v^*$. To this
end, assume $v < v'$ and $m = \langle\PREPARE(v', \val', M)\rangle_{\_}$ is a
sent message such that $\ValidNewView(m)$. Note that from the induction
hypothesis it follows that
$$
\forall C'', v'', \val''.\, v < v'' < v' \wedge \accepted(C'', v'',
\hash(\val'')) \wedge \valid(C'') {\implies} \val = \val''.
$$
Furthermore, by
Propositions~\ref{lemma:sbft:singlecmd},~\ref{lemma:sbft:committedslow-prepared},
and~\ref{lemma:sbft:committedfast-prepared} we have
$$
\forall C'', \val''.\, 
\accepted(C'', v, \hash(\val'')) \wedge \valid(C'') {\implies} \val = \val'',
$$
so that overall we get
\begin{equation}\label{sbft:hyp2}
\forall C'', v'', \val''.\, v \le v'' < v' \wedge \accepted(C'', v'',
\hash(\val'')) \wedge \valid(C'') \implies \val = \val''.
\end{equation}

Let
$$
M = \{\langle \NEWLEADER(v', \vcballot_j, \vcmd_j, \vcert_j, \vpreballot_j,
\vctxn_j) \rangle_j \mid p_j \in Q\}.
$$
for a quorum $Q$. Since $\ValidNewView(m)$, we have
$\forall m' \in M.\, \ValidNewLeader(m')$, so that 
\begin{multline}\label{sbft:validnewleader}
\forall p_j \in Q.\, \vcballot_j \le \vpreballot_j < v' \wedge {}
\\
({\vcballot_j \not= 0} {\implies} \accepted(\vcert_j, \vcballot_j,
\hash(\vcmd_j)) \wedge \valid(\vcert_j)).
\end{multline}
From this and~(\ref{sbft:hyp2}) we get that
\begin{equation}\label{sbft:hyp3}
  \forall p_j \in Q.\, {\vcballot_j \ge v} {\implies}  \vcmd_j = \val.
\end{equation}

Let $R$ be the set of correct processes in $Q$. By induction hypothesis, for any
$v''$ such that $v < v'' < v'$, a process in $R$ cannot accept a
$\PREPARE(v'', h'', \_)$ message for $h'' \not= \hash(\val)$. Then,
since~(\ref{sbft:validnewleader}) implies $\vpreballot_j < v'$, we get
\begin{equation}\label{sbft:preballot}
\forall p_j \in R.\, \vpreballot_j > v \implies \vctxn_j = \val.
\end{equation}

Let $\vslow, \vfast, \valslow, \valfast$ be computed from
$\vcballot, \vcmd, \vcert, \vpreballot, \vctxn$ as in
lines~\ref{comp-start}-\ref{comp-end}. Then $\ValidNewView(m)$ implies
\begin{equation}\label{sbft:val-defn}
(\vslow\geq \vfast\wedge \vslow>0 {\implies} \val' = \valslow) \wedge 
(\vslow < \vfast {\implies} \val' = \valfast).
\end{equation}

We now consider two cases, depending on whether
$\committedslow(C, v, \hash(\val))$ or $\committedfast(C, v, \hash(\val))$.

1. $\committedslow(C, v, \hash(\val))$. In this case a quorum $Q'$ of processes
sent $\COMMITTED(v, \hash(\val))$.  The quorums $Q$ and $Q'$ have to intersect
in some correct process $p_k$, which has thus sent both
$\COMMITTED(v, \hash(\val))$ and
$\NEWLEADER(v', \vcballot_k, \vcmd_k, \vcert_k, \vpreballot_j, \vctxn_j)$. 
Since $v< v'$, this process 
$p_k$ must have sent the $\COMMITTED$ message before the $\NEWLEADER$
message. Before sending $\COMMITTED(v, \hash(\val))$ the process set $\lballot$
to $v$ (line~\ref{pbft:assign-cballot}). Then by
Proposition~\ref{prop:sbft:increase} process $p_k$ must have had
$\lballot \ge v$ when it sent the $\NEWLEADER$ message. Hence,
$\vcballot_k \ge v$ and $\max\{\vcballot_j \mid p_j \in Q\} \ge v$.  Then
from~(\ref{sbft:hyp3}) for any $j$ such that
$\vcballot_j = \max\{\vcballot_{k} \mid p_k \in Q\}$ we must have
$\vcmd_j = \val$, so that
\begin{equation}\label{sbft:slow}
\vslow \ge v \wedge \valslow = \val.
\end{equation}

Since $|R| \ge |Q|-f$, from~(\ref{sbft:preballot}) for the $D$ defined in
line~\ref{sbft:D} we get
$$
\forall (\val'', v'') \in D.\, v'' > v \implies \val'' = \val.
$$
Hence,
$$
\vfast > v \implies \valfast = \val.
$$
From this,~(\ref{sbft:slow}) and~(\ref{sbft:val-defn}) we get $\val' = \val$, as
required.

2. $\committedfast(C, v, \hash(\val))$. Then each process in $R$ sent
$\PREPARED(v, \hash(\val))$, and this must have happened before it sent the
$\NEWLEADER$ message for view $v'$. Hence, by
Proposition~\ref{prop:sbft:increase}, when each process in $R$ sent its
$\NEWLEADER$ message for view $v'$, it had $\preballot \ge v$. Hence,
\begin{equation}\label{sbft:preballot-v}
\forall p_j \in R.\, \vpreballot_j \ge v.
\end{equation}
Since a correct process can accept only a single proposal in a view, we have
$$
\forall p_j \in R.\, \vpreballot_j = v \implies \vctxn_j = \val.
$$
Together with~(\ref{sbft:preballot}), this yields
$$
\forall p_j \in R.\, \vpreballot_j \ge v \implies \vctxn_j = \val.
$$
This and~(\ref{sbft:preballot-v}) give $\forall p_j \in R.\, \vctxn_j =
\val$. Then for the $D$ defined in line~\ref{sbft:D}, we get
$$
\forall (\val'', v'') \in D.\, v'' \ge v \wedge \val'' = \val.
$$
Furthermore, since $|R| \ge |Q|-f \ge f+1$, we also have $D \not= \emptyset$, so
that
\begin{equation}\label{sbft:fast2}
\vfast \ge v \wedge \valfast = \val.
\end{equation}
Finally,~(\ref{sbft:hyp3}) implies
$$
\vslow \ge v \implies \valslow = \val.
$$
From this,~(\ref{sbft:fast2}) and~(\ref{sbft:val-defn}) we get $\val' = \val$,
as required.\qed

\begin{corollary}
SBFT satisfies Agreement.
\end{corollary}

\subsubsection*{Liveness}

\begin{theorem}
  Assume all processes are correct and let $v \ge \B$ be a view such that
  $\timeoutV(v) > 5\delta$. Then all processes decide at $v$ by
  $\tl{v}+3\delta$.
\label{thm:sbft-fast-latency}
\end{theorem}

\paragraph{Proof.} 
By Property~\ref{prop:after-t}, we have $\tm{v} \ge \GST$, so that all messages
sent by correct processes after $\tm{v}$ get delivered to all correct processes
within $\delta$. Once a correct process enters $v$, it sends its $\NEWLEADER$
message, so that $\leader(v)$ is guaranteed to receive a quorum of $\NEWLEADER$
messages by $\tl{v} + \delta$. When this happens, the leader will send its
proposal in a $\PREPARE$ message, which correct processes will receive by
$\tl{v}+2\delta$. If all processes deem the proposal safe, then by
$\tl{v}+3\delta$, every process will receive $3f+1$ matching $\PREPARED$
messages and decide.
By~(\ref{overlap}), all correct processes will stay in $v$ until
at least $\tl{v}+(\timeout(v) - d) > \tl{v}+3\delta$.
Thus, the
above sequence of message exchanges will complete before any of them exits view
$v$, and all correct processes will decide in this view by the time
$\tl{v}+3\delta$.

It remains to show that the proposal $\leader(v)$ makes in view $v$
(line~\ref{sbft:receive-newleader}) will be deemed safe by all 
processes according to the $\ValidNewView$ predicate
(line~\ref{sbft:receive-prepare}). All the conjuncts of
$\ValidNewView$ except for $\validity(\val)$ are trivially satisfied
given that $\leader(v)$ is correct. If the leader is choosing its own
proposal as $\val$, then it is valid because correct processes propose
valid values. If the leader is choosing $\valslow$ as $\val$, then from
$\ValidNewLeader$ we get that $\accepted(C, \_, \hash(\valslow))$ for
a well-formed $C$. Hence, by Proposition~\ref{lemma:sbft:validityliveness} we again have
$\validity(\val)$. Finally, if the leader is choosing $\valfast$ as $\val$, then
$f+1$ processes sent $\val$ in their $\NEWLEADER$
message. Thus, at least one process has checked the
validity of $\val$. Hence, we again have
$\validity(\val)$.\qed

\begin{theorem}
  Let $v \ge \B$ be a view such that $\timeoutF(v) > 2\delta$,
  $\timeoutV(v) - \timeoutF(v) > 5\delta$ (so that $\timeoutV(v) > 7\delta$) and
  $\leader(v)$ is correct. Then all correct processes decide at $v$ by
  $\tl{v}+\timeoutF(v)+3\delta$.
\end{theorem}

\paragraph{Proof.} 
By Property~\ref{prop:after-t}, we have $\tm{v} \ge \GST$, so that all messages
sent by correct processes after $\tm{v}$ get delivered to all correct processes
within $\delta$. Once a correct process enters $v$, it sends its $\NEWLEADER$
message, so that $\leader(v)$ is guaranteed to receive a quorum of $\NEWLEADER$
messages by $\tl{v} + \delta$. When this happens, the leader will send its
proposal in a $\PREPARE$ message, which correct processes will receive by
$\tl{v}+2\delta$. As in the proof of Theorem~\ref{thm:sbft-fast-latency}, we can
show that all correct processes will deem the proposal safe. Fix a correct
process $p_i$ and let $t_1\leq \tl{v}+2\delta$ be the time when $p_i$ receives
the $\PREPARE$ message. Then every correct process will receive the leader's
$\PREPARE$ message and send its $\PREPARED$ message by $t_1 +\delta$. The
process $p_i$ will thus receive all $\PREPARED$ messages sent by correct
processes by $t_1+2\delta$. The process $p_i$ starts $\timersbft$ at time $t_1$,
and $\timeoutF(v) > 2\delta$. Thus, by the time $\timersbft$ expires at $p_i$,
it will have received a quorum of $\PREPARED$ messages. Since $p_i$ was picked
arbitrarily, this holds for any correct process. A correct process starts its
timer no later than $\tl{v}+2\delta$, so that every correct process will send
$\COMMITTED$ by $\tl{v}+2\delta+\timeoutF(v)$. It then takes at most $\delta$ to
exchange $\COMMITTED$ messages, leading to decisions. By
By~(\ref{overlap}), all correct processes will stay in $v$ until
at least $\tl{v}+(\timeout(v) - d) > \tl{v}+\timeoutF(v)+3\delta$.
Thus, the above sequence of message exchanges
will complete before any of them exits view $v$, and all correct processes will
decide in this view by the time $\tl{v}+\timeoutF(v)+3\delta$. \qed

\begin{corollary}
  Let $\timeoutV$ and $\timeoutF$ be such that~(\ref{prop:increasing}) holds and
  $\forall \theta.\,\exists v.\,\forall v'.\, v' \ge v {\implies}
  \timeoutV(v')-\timeoutF(v')>\theta$. Then all correct processes eventually decide.
\end{corollary}

\subsubsection*{Latency under favorable conditions}

\begin{corollary}
  Assume all processes are correct, $\tf\geq\GST$, $\timeout(1) > 4\delta$ and
  $\leader(1)$ is correct. Then in the SBFT protocol all correct processes
  decide no later than $\ts + 3\delta$.
 \label{cor:sbft:Vequal1andcorrectleader}
\end{corollary}

\subsubsection*{SBFT without the extra timer}

We can dispense with $\timersbft$ in the SBFT protocol. In this variant, as soon
as a process receives a quorum of $\PREPARED$ messages at
line~\ref{line:sbft-timer}, it sends the $\COMMITTED$ messages. This reduces the
latency when the protocol decides through the slow path, and leaves the
fast-path latency unchanged.

\begin{theorem}
  Assume that all processes are correct and let $v \ge \B$ be a view such that
  $\timeoutV(v) > 5\delta$. Then in the modified SBFT all processes decide at
  $v$ by $\tl{v}+3\delta$.
\end{theorem}

\begin{theorem}
  Let $v \ge \B$ be a view such that $\timeoutV(v) > 6\delta$ and $\leader(v)$
  is correct. Then in the modified SBFT all correct processes decide at $v$ by
  $\tl{v}+4\delta$.
\end{theorem}

The lower slow-path latency comes at the expense of a higher message complexity,
since a process sends a $\COMMITTED$ message even if in the end it decides on
the fast path. In contrast, a process running the previously presented version
of SBFT may decide through the fast path before $\timersbft$ expires, and thus
avoid sending $\COMMITTED$.

\clearpage

\subsection{Single-Shot Tendermint}
\label{sec:original-tendermint}

\begin{algorithm}[H]
  \setcounter{AlgoLine}{0}

  \SubAlgo{\Upon $\newview(v)$}{
    $\ballot \leftarrow v$\;
    $\voted \leftarrow \FALSE$\;
    $\stoptimer(\timertendermint)$\;
    $\starttimer(\timertendermint, \timeoutL(\ballot))$\;
    \If{$p_i=\leader(\ballot)$}{
      \uIf{$\cballot\not= 0$}{\Broadcast $\langle \PREPARE(v, \cmd,
        \cballot) \rangle_i$;}
      \Else{\Broadcast $\langle \PREPARE(v, \myproposal(), 0) \rangle_i$;}
      }
  }

  \smallskip
  \smallskip

\SubAlgo{\WhenReceived $\langle \PREPARE(v, \val, \_) \rangle_j =
    m$}{
    \textbf{pre:}
    $\ballot = v \wedge \voted = \FALSE \wedge \ValidNewView(m)$\;
    $\ctxn \leftarrow \val$\;
    $\voted \leftarrow \TRUE$\;
    \Broadcast $\langle \PREPARED(v, \hash(\ctxn)) \rangle_i$\;
  }

 \smallskip
  \smallskip

\SubAlgo{\WhenReceived $\{\langle \PREPARED(v, h)\rangle_j \mid p_j \in Q\}$ 
    {\bf for a quorum $Q$}}{
    \textbf{pre:} $\ballot = v \wedge \voted = \TRUE \wedge\hash(\ctxn) = h$\;
    $\cmd \leftarrow \ctxn$\;
    $\cballot \leftarrow v$\;
    \If{$\timertendermint\text{\rm{} has not expired}$}{
    $\lballot \leftarrow \ballot$\;
    $\lcmd \leftarrow \ctxn$\;
    \Broadcast $\langle \COMMITTED(v, h) \rangle_i$\;
    }
  }
  \smallskip
  \smallskip
  \SubAlgo{\WhenReceived $\{\langle \COMMITTED(v, h)\rangle_j \mid p_j \in Q\}$
    {\bf for a quorum $Q$}}{
    \textbf{pre:} $\ballot = \lballot = v \wedge \hash(\ctxn) = h$\;
    ${\tt decide}(\ctxn)$;
  }
\end{algorithm}

\bigskip

$$
\begin{array}{@{}l@{}}
\accepted(v, h)
\iff
\exists Q.\, 
\quorum(Q) \wedge ({\bf received}\,\{\langle \PREPARED(v, h) \rangle_j \mid p_j \in Q\})
\\[20pt]
\ValidNewView(\langle \PREPARE(v, \val, v') \rangle_i) \iff 
p_i = \leader(v) \wedge \validity(\val) \wedge {}
\\[2pt]
(\lballot \not \not= 0 {\implies}
\val = \lcmd \vee 
(\exists v'.\, v > v'>\lballot \wedge \accepted(v', \hash(\val))))
\end{array}
$$

\bigskip
\bigskip

\noindent
In Tendermint processes exchange messages using a reliable broadcast primitive
($\Broadcast$). The primitive guarantees that, if a correct process receives a
message $m$ by a time $t$, then all correct processes will receive $m$ by
$\max\{t, \GST\}+\Theta$.

\clearpage

\subsubsection*{Safety}

The proof of safety is virtually identical to the one for single-shot HotStuff
(\S\ref{sec:hotstuff-safety}) and is omitted.

\subsubsection*{Liveness}

\begin{proposition}
\label{lemma:tendermintoriginal:increase}
The variables $\lballot$, $\cballot$ and $\ballot$ at a correct process never
decrease and we always have $\lballot \le \cballot \le \ballot$.
\end{proposition}

\begin{proposition}
\label{lemma:tendermintoriginal:singlecmd}
For any $v$, $\val$, and $\val'$, if $\accepted(v, \hash(\val))$ and
$\accepted(v, \hash(\val'))$ at any two correct processes, then $\val = \val'$.
\end{proposition}

\begin{proposition}
\label{lemma:tendermintoriginal:validityliveness}
For any $v$ and $\val$, if $\accepted(v, \hash(\val))$ at some correct process,
then $\validity(\val)$.
\end{proposition}

\begin{lemma}
  Let $v \ge \B$ be a view such that $\timeoutL(v) > 2\delta+2\Theta$,
  $\timeoutV(v) > 2\delta + 3\Theta$ and $\leader(v)$ is correct. Let
  $\leader(v).\cballot=v_0$ when $\leader(v)$ enters view $v$. Assume that for
  each correct process $p_j$ we have $p_j.\lballot\le v_0$ when $p_j$ receives
  the leader's proposal in $v$. Then all correct processes decide in view $v$ by
  $\tl{v}+3\Theta$.
\label{lemma:tendermintoriginal:safe}
\end{lemma}
\paragraph{Proof}
By Property~\ref{prop:after-t} we have $\tm{v} \ge \GST$, so that all messages
broadcast by correct processes after $\tm{v}$ get delivered to all correct
processes within $\Theta$. When $\leader(v)$ enters view $v$ (no later than
$\tl{v}$), it will broadcast its proposal $x$ in a $\PREPARE(v, \val, v_0)$
message, which correct processes will receive by $\tl{v}+\Theta$. We first prove
that no later than $\tl{v}+\Theta$ this message will satisfy the $\ValidNewView$
predicate at all correct processes.

Assume first that $v_0=0$, so that the leader is proposing its own value, which
must be valid because the leader is correct. Then every correct process has
$\lballot=0$ when receiving the leader's proposal, and thus the proposal
satisfies the last conjunct of $\ValidNewView$. Hence, every correct process
will deem the proposal safe upon its receipt. Assume now that $v_0>0$. Since the
leader is correct, we have $\accepted(v_0, \hash(\val))$ at the leader when it
makes the proposal. Hence, by
Proposition~\ref{lemma:tendermintoriginal:validityliveness} we have
$\validity(\val)$. Furthermore, since the messages comprising the quorum of
$\PREPARED(v_0, \hash(\val))$ messages received by the leader were sent by
reliable broadcast and $\tm{v} \ge \GST$, all correct processes will satisfy
$\accepted(v_0, \hash(\val))$ by $\tl{v}+\Theta$. By the assumption of the
lemma, each correct process $p_j$ has $p_j.\lballot=v'\leq v_0$ when receiving
the leader's proposal. If $v'<v_0$, then by $\tl{v}+\Theta$ the leader's
proposal will satisfy the last conjunct of $\ValidNewView$ and will thus be
deemed safe by $p_j$. If $v'=v_0$, then by
Proposition~\ref{lemma:tendermintoriginal:singlecmd} the process $p_j$ has
$\lcmd = \val$ when it receives the leader's proposal. Then the leader's
proposal again satisfies the last conjunct of $\ValidNewView$.

Hence, by $\tl{v}+\Theta$ each correct process will receive the leader's
proposal and will deem it safe. It will then send a $\PREPARED(v, \hash(\val))$
message, so that all correct processes will receive a quorum of such messages by
$\tl{v}+2\Theta$. When a process enters a view, it starts $\timertendermint$,
which is set to $\timeoutL(v)>2\delta+2\Theta$. By
Property~\ref{prop:no-skip-sync:2}, we have $\tm{v} \ge \tl{v} - 2\delta$, so
that the $\timertendermint$ of any correct process cannot expire before
$\tl{v}+2\Theta$. Hence, every correct process will receive a quorum of
$\PREPARED(v, \hash(\val))$ messages before its $\timertendermint$ expires and
will thus send the corresponding $\COMMITTED$ message by
$\tl{v}+2\Theta$. Afterwards, it takes correct processes at most $\Theta$ to
exchange the $\COMMITTED$ messages, leading to decisions. 
By~(\ref{overlap}), all correct processes will stay in $v$ until
at least $\tl{v}+(\timeout(v) - d) > \tl{v}+3\Theta$.
Thus, the above sequence of message exchange will complete
before any of them exits view $v$, and all correct processes will decide in this
view by $\tl{v}+3\Theta$. \qed

\begin{lemma}
  Let $v \ge \B$ be a view such that $\timeoutL(v) > 2\delta+2\Theta$,
  $\timeoutV(v) - \timeoutL(v) > 2\delta+\Theta$ (so that
  $\timeoutV(v) > 4\delta + 3\Theta$) and $\leader(v)$ is correct. If a correct
  process locks a value in $v$, then all correct processes will have $\cballot=v$
  when leaving $v$.
\label{lemma:tendermintoriginal:validall}
\end{lemma}

\paragraph{Proof}
Let $t\geq \tm{v} \ge \GST$ be the time when a correct process $p_i$ locks a
value in view $v$. When a correct process enters $v$, it starts its
$\timertendermint$, which defines the period of time during which a process can
lock a value. Thus, the latest time $p_i$ can lock a value in $v$ is
$\tl{v}+\timeoutL(v)$, i.e., $t\leq\tl{v}+\timeoutL(v)$. To lock a value, $p_i$
has to receive a corresponding $\PREPARE$ message and a quorum of $\PREPARED$
messages. Since these messages are sent via reliable broadcast, all correct
processes are guaranteed to receive them by $\tl{v}+\timeoutL(v)+\Theta$. By
Property~\ref{prop:no-skip-sync:3} all correct processes will stay in $v$ until
at least $\tm{v}+\timeoutV(v) > \tm{v}+\timeoutL(v)+2\delta+\Theta$. By
Property~\ref{prop:no-skip-sync:2} we have $\tm{v}\geq \tl{v} - 2\delta$. Hence,
all correct processes will stay in $v$ until at least
$\tl{v}+\timeoutL(v)+\Theta$. Then each correct process will receive the
$\PREPARE$ message and the quorum of $\PREPARED$ messages while still in view
$v$, and will thus set its $\cballot$ to $v$ before exiting the view, as
required.\qed

\begin{theorem}
  Let $v \ge \B$ be a view such that $\timeoutL(v) > 2\delta+2\Theta$,
  $\timeoutV(v) - \timeoutL(v) > 2\delta+\Theta$ (so that
  $\timeoutV(v) > 4\delta + 3\Theta$) and $\leader(v)$ is correct. Then all
  correct processes decide in a view no later than $v+3f$ by
  $\tl{v}+\sum_{k=v}^{v+3f-1} (\timeoutV(k)+\delta)+3\Theta$.
\label{thm:livetendermintoriginal}
\end{theorem}

\paragraph{Proof}
Consider first the case when each correct process has $\lballot=0$ when
receiving the leader's proposal in $v$. Then by
Lemma~\ref{lemma:tendermintoriginal:safe} all correct processes decide at $v$ by
$\tl{v}+3\Theta$. We now consider the case when some correct process is locked
on a value at time $\tl{v}$. Let $p_i$ be a process that is locked on the
highest view among correct processes, and let this view be $v_0 < v$. Assume
first that $\leader(v).\cballot\geq v_0$ when $\leader(v)$ enters view $v$. Then
by Lemma~\ref{lemma:tendermintoriginal:safe}, all correct processes decide at
$v$ by $\tl{v}+3\Theta$.

Assume now that that $\leader(v).\cballot< v_0$ when $\leader(v)$ enters $v$, so
that $p_i \not= \leader(v)$. By the definition of $\leader()$, leaders rotate
round-robin, so that for some view $v'\leq v+3f$ we have $\leader(v') =
p_i$. Let $v_1\geq v_0$ be the highest view locked among all correct processes
at time when they receive the leader's proposal in $v'$. We prove that
$p_i.\cballot\geq v_1$ when $p_i$ enters $v'$. We know that $p_i$ was locked at
the highest view $v_0<v$ among all correct processes when these received the
leader's proposal in $v$. Then by
Proposition~\ref{lemma:tendermintoriginal:increase}, $p_i.\cballot\geq v_0$ when
$p_i$ enters $v'$. Thus, if no correct process locks a value between views $v$
and $v'$, then $v_1=v_0$, so that $p_i.\cballot\geq v_1$ when $p_i$ enters
$v'$. On the other hand, if a correct process locks a value between views $v$
and $v'$, then $v_0 < v \leq v_1 < v'$ and by
Lemma~\ref{lemma:tendermintoriginal:validall} and
Proposition~\ref{lemma:tendermintoriginal:increase}, $p_i$ has
$p_i.\cballot\geq v_1$ when it enters $v'$. Thus, in all cases we have
$p_i.\cballot\geq v_1$ when $p_i$ enters $v'$. Then by
Lemma~\ref{lemma:tendermintoriginal:safe} all correct processes decide in view
$v'$ by $\tl{v'}+3\Theta$. By Proposition~\ref{lem:last-entry-sum} we have
$\tl{v'} \le \tl{v}+\sum_{k=v}^{v'-1} (\timeoutV(k)+\delta)$, so that all
correct processes decide in view $v'$ by
$\tl{v}+\sum_{k=v}^{v'-1} (\timeoutV(k)+\delta)+3\Theta \le
\tl{v}+\sum_{k=v}^{v+3f-1} (\timeoutV(k)+\delta)+3\Theta$, as required. \qed

\begin{corollary}
  Let $\timeoutV$ and $\timeoutL$ be such that~(\ref{prop:increasing}) holds and
  $\forall \Theta.\,\exists v.\,\forall v'.\, v' \ge v {\implies}
  \timeoutV(v')-\timeoutL(v')>\Theta$. Then all correct processes eventually
  decide.
\end{corollary}

\subsubsection*{Latency under favorable conditions}

\begin{corollary}
  Assume that $\tf\geq\GST$, $\timeoutL(v) > 2\delta+2\Theta$ and
  $\timeoutV(v) - \timeoutL(v) > 2\delta+\Theta$. Then all correct processes
  decide no later than
  $\ts+\sum_{k=1}^{f} (\timeoutV(k)+\delta) + \delta+ 3\Theta$.
\end{corollary}

\begin{corollary}
  Assume that $\tf \geq\GST$, $\timeoutL(v) > 2\delta+2\Theta$,
  $\timeoutV(v) > 2\delta + 3\Theta$ and $\leader(1)$ is correct. Then all
  correct processes decide no later than $\ts + \delta + 3\Theta$.
\end{corollary}

\clearpage

\section{Linear Versions of Consensus Protocols}
\label{sec:linear}

\subsection{Threshold Signatures}

The linear versions of the consensus protocols make use of threshold
signatures\footnote{D. Boneh, B. Lynn, and H. Shacham. Short signatures from the
  Weil pairing. {\em J. Cryptology}, 17(4):297–319, 2004.}. A $k$-out-of-$n$
threshold signature scheme is a protocol that allows any subset of $k$ processes
out of $n$ to generate a digital signature, but that disallows the creation of a
valid signature if fewer than $k$ processes participate in the protocol. A
process $p_i$ participates by \emph{partially} signing a message $m$ using its
private key. A process that gathers a set $M$ of at least $k$ partial signatures
for a message $m$ can combine them into a single compact signature using
$\Combine(M)$. Any process can then verify the signature using a public key
shared by all processes.

We use two threshold signature schemes with $k=n$ (denoted $\sigma$) and
$k=2f+1$ (denoted $\tau$). We denote by $\langle m \rangle_{\sigma_i}$ a message
$m$ partially signed by process $p_i$ using the $\sigma$ scheme, and denote by
$\langle m \rangle_{\sigma}$ a combined signature on a message $m$. We use
similar notation for the $\tau$ scheme.

\clearpage

\subsection{Single-Shot Linear HotStuff}
\label{sec:hotstuff-linear}

\begin{algorithm*}[H]
  \setcounter{AlgoLine}{0}
  \SubAlgo{\Upon $\newview(v)$}{
    $\ballot \leftarrow v$\;
    $\voted\leftarrow\FALSE$\;
    \Send $\langle \NEWLEADER(\ballot, \cballot, \cmd, \cert) \rangle_i$\\
    \nonl \quad \KwTo $\leader(\ballot)$\; 
  }

  \smallskip

  \SubAlgo{\WhenReceived $\{\langle \NEWLEADER(b, \vcballot_j,\vcmd_j, 
    \vcert_j)\rangle_j \mid p_j \in Q\} = M$\hspace{3cm}
     {\bf for a quorum $Q$}}{
    \textbf{pre:} $\ballot = v \wedge p_i = \leader(v) \wedge 
    (\forall m \in M.\, \ValidNewLeader(m))$\; \label{safety-newleader}
    \uIf{$\exists j.\, \vcballot_j = \max\{\vcballot_k \mid p_k \in Q\}
      \not= 0$}{\Send $\langle \PREPARE(v, \vcmd_j, \vcert_j) \rangle_i$ \KwTo
      \all}
    \Else{\Send $\langle \PREPARE(v, \myproposal(), \bot) \rangle_i$ \KwTo \all
    }
    
  }

  \smallskip

  \SubAlgo{\WhenReceived $\langle \PREPARE(v, \val, \_) \rangle_j = m$}{
    \textbf{pre:}
    $\ballot = v \wedge \vote = \FALSE \wedge \ValidNewView(m)$\;
    $\ctxn \leftarrow \val$\;
    \Send $\partialthreshold{\PREPARED(v, \hash(\val))}{\tau_i}$ \KwTo $\leader(\ballot)$\; }

  \smallskip

  \SubAlgo{\WhenReceived $\{\partialthreshold{\PREPARED(v, h)}{\tau_j} \mid p_j \in Q\} = M$ 
    {\bf for a quorum $Q$}}{
    \textbf{pre:} $\ballot = v \wedge  p_i = \leader(v) \wedge
    \voted = \TRUE \wedge \hash(\ctxn) = h$\;
    \Send $\langle \PRECOMMIT(v, \Combine(M)) \rangle_i$ \KwTo \all\;
  }

  \smallskip

  \SubAlgo{\WhenReceived $\langle \PRECOMMIT(v, \thcert) \rangle_{j}$}{
    \textbf{pre:} $\ballot = v \wedge p_j = \leader(v) \wedge
    \voted = \TRUE \wedge {}$\\
    \nonl \hspace{0.81cm}$\thcert = \threshold{\PREPARED}{v}{\hash(\ctxn)}{\tau}$\;
    $\cmd \leftarrow \ctxn$\;
    $\cballot \leftarrow \ballot$\;\label{ballot-assignment}
    $\cert \leftarrow \thcert$\;
    \Send $\partialthreshold{\PRECOMMITTED(v, \hash(\cmd))}{\tau_i}$ \KwTo $\leader(\ballot)$\;
  }

  \smallskip

  \SubAlgo{\WhenReceived $\{\partialthreshold{\PRECOMMITTED(v, h)}{\tau_j} \mid p_j \in Q\} = M$ 
    {\bf for a quorum $Q$}}{
    \textbf{pre:} $\ballot = \cballot = v \wedge p_i = \leader(v) \wedge
    \hash(\ctxn) = h$\;
     \Send $\langle \COMMIT(b, \Combine(M)) \rangle_i$ \KwTo \all\;
  }

  \smallskip

  \SubAlgo{\WhenReceived $\langle \COMMIT(v, \thcert) \rangle_{j}$}{
    \textbf{pre:} $ \ballot = \cballot = v \wedge p_j = \leader(v) 
    \wedge{}$\\
    \nonl \hspace{0.81cm}$\thcert = \threshold{\PRECOMMITTED}{v}{\hash(\ctxn)}{\tau}$\;
    $\lballot \leftarrow \cballot$\;
    \Send $\partialthreshold{\COMMITTED(v, \hash(\cmd))}{\tau_i}$ \KwTo $\leader(\cballot)$\;
  }

  \smallskip

  \SubAlgo{\WhenReceived $\{\partialthreshold{\COMMITTED(v, h)}{\tau_j} \mid p_j \in Q\} =
    M$ {\bf for a quorum $Q$}}{
    \textbf{pre:} $\ballot = \lballot = v \wedge p_i = \leader(v) \wedge
    \hash(\ctxn) = h$\;
    \Send $\langle \DECIDE(v, \Combine(M)) \rangle_i$ \KwTo \all\;
  }

  \smallskip

  \SubAlgo{\WhenReceived $\langle \DECIDE(v, \thcert) \rangle_{j}$}{
    \textbf{pre:} $\ballot = \lballot = v \wedge p_j = \leader(v) \wedge {}$\\
    \nonl \hspace{0.81cm}$\thcert = \threshold{\COMMITTED}{v}{\hash(\ctxn)}{\tau}$\;
    ${\tt decide}(\ctxn)$;
  }
\end{algorithm*}

\clearpage

$$
\begin{array}{l}
\ValidNewLeader(\langle \NEWLEADER(v', v, \val, \thcert) \rangle_{\_})
  \iff 
\ms
\quad v < v' \wedge
({v \not= 0} {\implies}\thcert = \threshold{\PREPARED}{v}{\hash(\val)}{\tau})
\end{array}
$$

$$
\begin{array}{l}
\ValidNewView(\langle \PREPARE(v, \val, \thcert) \rangle_i) \iff
p_i = \leader(v) \wedge \validity(\val)\wedge{}
\ms
\quad (\lballot \not \not= 0 {\implies}
\val = \cmd \vee {}
\ms
\quad (\exists v'.\, v>v'>\lballot \wedge \thcert = \threshold{\PREPARED}{v'}{\hash(\val)}{\tau}))
\end{array}
$$

\bigskip

\noindent
In view $1$ the leader can propose without waiting for $\NEWLEADER$ messages,
and processes can avoid sending these messages to this leader.

\subsubsection*{Liveness}

\begin{theorem}
  Let $v \ge \B$ be a view such that $\timeout(v) > 10\delta$ and $\leader(v)$
  is correct. Then all correct processes decide in view $v$ by $\tl{v}+8\delta$.
\label{thm:livelinearhotstuff}
\end{theorem}
\begin{corollary}
  Let $\timeoutV$ be such that~(\ref{prop:increasing}) holds. Then all correct
  processes eventually decide.
\end{corollary}

\subsubsection*{Latency under favorable conditions}

\begin{corollary}
  Assume that $\tf\geq\GST$ and $\timeout(1) > 10\delta$. Then in the linear
  HotStuff protocol all correct processes decide no later than
  $\ts+\sum_{k=1}^{f} (\timeoutV(k)+\delta) + 8\delta$.
\label{thm:hslinear:latencyfromVequal1}
\end{corollary}

\begin{corollary}
  Assume that $\tf\geq\GST$, $\timeout(1) > 9\delta$ and $\leader(1)$ is
  correct. Then in the linear HotStuff protocol all correct processes decide no
  later than $\ts + 8\delta$.
 \label{cor:hslinear:Vequal1andcorrectleader}
\end{corollary}
\clearpage

\subsection{Single-Shot Linear SBFT}

\begin{algorithm*}[H]
  \setcounter{AlgoLine}{0}
  \SubAlgo{\Upon $\newview(v)$}{
   $\ballot \leftarrow v$\;
    $\voted \leftarrow \FALSE$\;
    \Send $\langle \NEWLEADER(\ballot, \lballot, \cmd, \cert,
    \preballot, \ctxn) \rangle_i$\\
    \nonl \quad \KwTo $\leader(\ballot)$\; 
  }

  \smallskip
  \smallskip

  \SubAlgo{\WhenReceived $\{\langle \NEWLEADER(v, \vcballot_j, \vcmd_j, 
    \vcert_j, \vpreballot_j, \vctxn_j) \rangle_j \mid p_j \in Q\} = M$\hspace{3cm}
    {\bf for a quorum $Q$}}{
    \textbf{pre:} $\ballot = v \wedge p_i = \leader(v) \wedge (\forall m \in M.\, \ValidNewLeader(m))$\;
    {\bf let} $(\val_{\rm slow}, v_{\rm slow})\leftarrow(\bot, 0)$\label{sbft-linear:comp-start}\;
    \lIf{$\exists j.\, \vcballot_j = \max\{\vcballot_{k} \mid p_k \in Q\}
      \not= 0$}{$(\val_{\rm slow}, v_{\rm slow})\leftarrow(\vcmd_j, \vcballot_j)$}
    {\bf let} $D\leftarrow \{\langle \val, v' \rangle 
    \mid \exists P \subseteq Q.\, |P|= f+1\wedge 
    (\forall p_j \in P.\, \vctxn_j = \val) \wedge {}$\\
    \nonl \hspace*{2.8cm}$v' = \min\{\vpreballot_j \mid p_j \in P\}\}$\;
    {\bf let} $(\val_{\rm fast}, v_{\rm fast})\leftarrow(\bot, \max\{v'\mid \langle \_,
    v'\rangle\in D\})$\;
    \leIf{$ \exists !\, \val.\, \langle \val, v_{\rm fast}\rangle\in D$}{$\val_{\rm fast}\leftarrow \val$}
    {$v_{\rm fast}=0$\label{sbft-linear:comp-end}}
    $\stoptimer(\timersbft)$\;
    $\starttimer(\timersbft, \timeoutF(\ballot))$\;
     \uIf{$v_{\rm slow}\geq v_{\rm fast} \wedge v_{\rm slow}>0$}
     {\Send $\langle \PREPARE(v, \val_{\rm slow}, M) \rangle_i$ \KwTo \all\;}
     \uElseIf{$v_{\rm fast}>v_{\rm slow}$}{\Send $\langle \PREPARE(v, \val_{\rm fast}, M) \rangle_i$ \KwTo \all\;}
     \Else{\Send $\langle \PREPARE(v, \myproposal(), M) \rangle_i$ \KwTo \all\;}
  }

  \smallskip
  \smallskip

  \SubAlgo{\WhenReceived $\langle \PREPARE(v, \val, \_) \rangle_j = m$}{
    \textbf{pre:}
    $\ballot=v \wedge \voted = \FALSE \wedge \ValidNewView(m)$\;
    $\ctxn \leftarrow \val$\;
    $\preballot \leftarrow \ballot$\;
    $\voted \leftarrow \TRUE$\;
    \Send $\partialthreshold{\PREPARED(v, \hash(\ctxn))}{\tau_i}$ \KwTo $\leader(\ballot)$\; 
    \Send $\partialthreshold{\PREPARED(v, \hash(\ctxn))}{\sigma_i}$ \KwTo $\leader(\ballot)$\; }

\smallskip
\smallskip

\SubAlgo{\WhenReceived 
  $\{\partialthreshold{\PREPARED(v, h)}{\sigma_i} \mid p_j \in \Proc\} = M$}{
    \textbf{pre:} $\ballot = v \wedge p_i = \leader(v) \wedge \voted = \TRUE
    \wedge \hash(\ctxn) = h$\;
    \Send $\langle \DECIDEFAST(v, \Combine(M)) \rangle_i$ \KwTo \all\;
  }

  \smallskip
  \smallskip

  \SubAlgo{\WhenReceived $\langle \DECIDEFAST(v, \thcert) \rangle_{j}$}{
    \textbf{pre:} $\ballot = v \wedge p_j = \leader(b) \wedge \voted = \TRUE
    \wedge{}$\\
    \nonl \hspace{0.81cm}$\thcert = \threshold{\PREPARED}{v}{\hash(\ctxn)}{\sigma}$\;
    ${\tt decide}(\ctxn)$;
  }

\end{algorithm*}

\clearpage

\begin{algorithm*}[H]
\SubAlgo{ {\bf when $\timersbft$ expired and received}
  $\{\partialthreshold{\PREPARED(v, h)}{\tau_j} \mid p_j \in Q\} = M$ 
    {\bf for at least  quorum $Q$}}{\label{line:linear-sbft-timer}
    \textbf{pre:} $\ballot = v \wedge p_i = \leader(v) \wedge \voted = \TRUE
    \wedge \hash(\ctxn) = h$\;
    \Send $\langle \COMMIT(v, \Combine(M)) \rangle_i$ \KwTo \all\;
  }

  \smallskip
  \smallskip

  \SubAlgo{\WhenReceived $\langle \COMMIT(v, \thcert) \rangle_{j}$}{
    \textbf{pre:} $\ballot = v \wedge p_j = \leader(v) \wedge 
    \voted=\TRUE \wedge{}$\\
    \nonl \hspace{0.81cm}$\thcert = \threshold{\PREPARED}{v}{\hash(\ctxn)}{\tau}$\;
    $\cmd \leftarrow \ctxn$\;
    $\lballot \leftarrow \ballot$\;
    $\cert \leftarrow \thcert$\;
    \Send $\partialthreshold{\COMMITTED(v, \hash(\cmd))}{\tau_i}$ \KwTo $\leader(\ballot)$\;
  }

  \smallskip
  \smallskip

  \SubAlgo{\WhenReceived $\{\partialthreshold{\COMMITTED(v, h)}{\tau_j} \mid p_j \in Q\} =
    M$ {\bf for a quorum $Q$}}{
    \textbf{pre:} $\ballot = \lballot = v \wedge p_i = \leader(v) \wedge
    \hash(\ctxn) = h$\;
    \Send $\langle \DECIDESLOW(v, \Combine(M)) \rangle_i$ \KwTo \all\;
  }

  \smallskip
  \smallskip

  \SubAlgo{\WhenReceived $\langle \DECIDESLOW(v, \thcert) \rangle_{j}$}{
    \textbf{pre:} $\ballot = \lballot = v \wedge p_j = \leader(b) \wedge {}$\\
    \nonl \hspace{0.81cm}$\thcert = \threshold{\COMMITTED}{v}{\hash(\ctxn)}{\tau}$\;
    ${\tt decide}(\ctxn)$;
  }

\end{algorithm*}

\bigskip

$$
\begin{array}{l}
\ValidNewLeader(\langle \NEWLEADER(v', v, \val, C, v_0, \val_0) \rangle_{\_})
  \iff
\ms
\quad v \le v_0 < v' \wedge 
({v \not= 0} {\implies}\thcert = \threshold{\PREPARED}{v}{\hash(\val)}{\tau})
\end{array}
$$

\smallskip
\smallskip

$$
\begin{array}{@{}l@{}}
\ValidNewView(\langle \PREPARE(v, \val, M) \rangle_i) \iff {}
\ms
\quad  p_i = \leader(v) \wedge \validity(\val) \wedge {}
\ms
\quad \exists Q, \vcballot, \vcmd, \vcert, \vpreballot, \vctxn.\, 
\quorum(Q) \wedge {}
\ms
\quad M = \{\langle \NEWLEADER(v, \vcballot_j, \vcmd_j,  \vcert_j, 
  \vpreballot_j, \vctxn_j) \rangle_j \mid p_j \in Q\} \wedge {}
\ms
\quad (\forall m \in M.\, \ValidNewLeader(m)) \wedge {}
\ms
\quad \exists \vslow, \vfast, \valslow, \valfast.\,
\ms
\quad
(\vslow, \vfast, \valslow, \valfast
\mbox{ are computed 
from 
$ \vcballot, \vcmd,  \vcert, \vpreballot, \vctxn$
as in lines~\ref{sbft-linear:comp-start}-\ref{sbft-linear:comp-end}}) \wedge {}
\ms
\quad (\vslow\geq \vfast\wedge \vslow>0 {\implies}
\val = \valslow)\wedge 
(\vslow < \vfast {\implies} \val = \valfast)
\end{array} 
$$

\bigskip

\noindent
In view $1$ the leader can propose without waiting for $\NEWLEADER$ messages,
and processes can avoid sending these messages to this leader.

\subsubsection*{Liveness}

\begin{theorem}
  Assume that all processes are correct and let $v \ge \B$ be a view such that
  $\timeoutV(v) > 6\delta$. Then all processes decide at $v$ by
  $\tl{v}+4\delta$.
\end{theorem}
\begin{theorem}
  Let $v \ge \B$ be a view such that $\timeoutF(v) > 2\delta$,
  $\timeoutV(v) - \timeoutF(v) > 6\delta$ (so that $\timeoutV(v) > 8\delta$) and
  $\leader(v)$ is correct. Then all correct processes decide at $v$ by
  $\tl{v}+\timeoutF(v)+4\delta$.
\end{theorem}
\begin{corollary}
  Let $\timeoutV$ and $\timeoutF$ be such that~(\ref{prop:increasing}) holds and
  $\forall \theta.\,\exists v.\,\forall v'.\, v' \ge v {\implies}
  \timeoutV(v')-\timeoutF(v')>\theta$. Then all correct processes eventually decide.
\end{corollary}

\subsubsection*{Latency under favorable conditions}

\begin{corollary}
  Assume all processes are correct, $\tf\geq\GST$, $\timeout(1) > 5\delta$ and
  $\leader(1)$ is correct. Then in the linear SBFT protocol all correct
  processes decide no later than $\ts + 4\delta$.
 \label{cor:sbftlinear:Vequal1andcorrectleader}
\end{corollary}

\subsubsection*{Linear SBFT without the extra timer}

Like with the all-to-all SBFT protocol of \S\ref{sec:sbft}, we can dispense with
$\timersbft$ in the linear SBFT protocol. In this variant, as soon as the leader
receives a quorum of $\PREPARED$ messages at line~\ref{line:linear-sbft-timer},
it sends the $\COMMIT$ messages. This again reduces the slow-path latency at the
expense of a higher message complexity.

\begin{theorem}
  Assume that all processes are correct and let $v \ge \B$ be a view such that
  $\timeoutV(v) > 6\delta$. Then in the modified linear SBFT all processes
  decide at $v$ by $\tl{v}+4\delta$.
\end{theorem}

\begin{theorem}
  Let $v \ge \B$ be a view such that $\timeoutV(v) > 8\delta$ and $\leader(v)$
  is correct. Then in the modified linear SBFT all correct processes decide at
  $v$ by $\tl{v}+6\delta$.
\end{theorem}

\clearpage

\fi

\end{document}